\documentclass[10pt,preprint]{elsarticle}
\usepackage[textwidth=6.5in,textheight=9in]{geometry}
\usepackage{amssymb,graphicx,longtable,multirow,amsmath}
\usepackage{wasysym}
\usepackage{color}
\usepackage{epsfig}
\usepackage{epstopdf}
\usepackage[]{natbib}
\usepackage{enumitem}
\usepackage{xargs}
\usepackage[bottom]{footmisc}
\usepackage[super]{nth}
\usepackage{hyperref}
\usepackage[capitalize]{cleveref}

\newcommand{\be}{\begin{equation}}
\newcommand{\ba}{\begin{align}}
\newcommand{\ee}{\end{equation}}
\newcommand{\ea}{\end{align}}
\newcommand{\half}{\frac{1}{2}}

\newcommand{\Wi}{\ensuremath{W_i}}
\newcommand{\Wj}{\ensuremath{W_j}}
\newcommand{\Wij}{\ensuremath{W_{ij}}}
\newcommand{\Wji}{\ensuremath{W_{ji}}}
\newcommand{\Wri}{\ensuremath{\mathcal{W}_i^R}}
\newcommand{\Wrj}{\ensuremath{\mathcal{W}_j^R}}
\newcommand{\Wrij}{\ensuremath{\mathcal{W}_{ij}^R}}
\newcommand{\Wrji}{\ensuremath{\mathcal{W}_{ji}^R}}
\newcommand{\xij}{\ensuremath{x_{ij}}}
\newcommand{\vij}{\ensuremath{v_{ij}}}
\newcommand{\vhatij}{\ensuremath{\hat{v}_{ij}}}
\newcommand{\paralpha}{\ensuremath{\partial_\alpha}}
\newcommand{\pargam}{\ensuremath{\partial_\gamma}}

\newcommand{\alphatmp}{\ensuremath{{\alpha_{\text{tmp}}}}}
\newcommand{\alphamin}{\ensuremath{\alpha_{\min}}}
\newcommand{\alphamax}{\ensuremath{\alpha_{\max}}}
\newcommand{\alphaz}{\ensuremath{{\alpha_0}}}
\newcommand{\etacrit}{\ensuremath{\eta_{\text{crit}}}}
\newcommand{\etafold}{\ensuremath{\eta_{\text{fold}}}}
\newcommand{\fkern}{\ensuremath{f_{\text{kern}}}}
\newcommand{\vsig}{{\ensuremath{v_{\text{sig}}}}}
\newcommand{\qacc}{\ensuremath{{q^\alpha_{\text{acc}}}_{ij}}}
\newcommand{\DE}{\ensuremath{\Delta E^{\text{thermal}}}}
\newcommand{\DEij}{\ensuremath{\Delta E_{ij}^{\text{thermal}}}}
\newcommand{\smin}{\ensuremath{s_{\min}}}
\newcommand{\smax}{\ensuremath{s_{\max}}}
\newcommand{\bb}[1]{\textbf{#1}}
\newcommand{\rinner}{\ensuremath{r_{\text{inner}}}}
\newcommand{\router}{\ensuremath{r_{\text{outer}}}}
\newcommand{\rhoinner}{\ensuremath{\rho_{\text{inner}}}}
\newcommand{\rhoouter}{\ensuremath{\rho_{\text{outer}}}}
\newcommand{\Pinner}{\ensuremath{P_{\text{inner}}}}
\newcommand{\Pouter}{\ensuremath{P_{\text{outer}}}}
\newcommand{\sshell}{\ensuremath{s_{\text{shell}}}}
\newcommand{\Interp}[1]{\ensuremath{\left\langle #1 \right\rangle}}
\newcommandx*\sigmaab[2][1=\alpha,2=\beta]{\ensuremath{\sigma^{#1#2}}}
\newcommand{\Uvec}{\ensuremath{\mathbf{U}}}
\newcommand{\Fvec}{\ensuremath{\mathbf{F}}}

\DeclareMathOperator{\sgn}{sgn}

\newcommand{\tauKH}{\ensuremath{\tau_{\text{KH}}}}
\newcommand{\rcloud}{\ensuremath{r_{\text{cloud}}}}
\newcommand{\rhocloud}{\ensuremath{\rho_{\text{cloud}}}}
\newcommand{\rhoam}{\ensuremath{\rho_{\text{ambient}}}}
\newcommand{\tauCR}{\ensuremath{\tau_{\text{crush}}}}

\newcommand{\rhoI}{\ensuremath{\rho_{\text{I}}}}
\newcommand{\rhoII}{\ensuremath{\rho_{\text{II}}}}
\newcommand{\rhoIII}{\ensuremath{\rho_{\text{III}}}}
\newcommand{\PI}{\ensuremath{P_{\text{I}}}}
\newcommand{\PII}{\ensuremath{P_{\text{II}}}}
\newcommand{\PIII}{\ensuremath{P_{\text{III}}}}

\begin{document}

\begin{frontmatter}
\title{CRKSPH - A Conservative Reproducing Kernel Smoothed Particle Hydrodynamics Scheme}
\author[add1,add2]{Nicholas Frontiere}
\ead{nfrontiere@uchicago.edu}
\author[add3]{Cody D.~Raskin}
\ead{raskin1@llnl.gov}
\author[add3]{J.~Michael Owen}
\ead{mikeowen@llnl.gov}

\address[add1]{Department of Physics, University of Chicago, Chicago, IL, USA, 60637}
\address[add2]{High Energy Physics Division, Argonne National Laboratory, Lemont, IL, USA, 60439}
\address[add3]{Lawrence Livermore National Laboratory, P.O. Box 383, L-38, Livermore, CA, USA, 94550} 

\begin{abstract}

We present a formulation of smoothed particle hydrodynamics (SPH) that utilizes a first-order consistent reproducing kernel, a smoothing function that exactly interpolates linear fields with particle tracers. Previous formulations using reproducing kernel (RK) interpolation have had difficulties maintaining conservation of momentum due to the fact the RK kernels are not, in general, spatially symmetric. Here, we utilize a reformulation of the fluid equations such that mass, linear momentum, and energy are all rigorously conserved without any assumption about kernel symmetries, while additionally maintaining approximate angular momentum conservation. Our approach starts from a rigorously consistent interpolation theory, where we derive the evolution equations to enforce the appropriate conservation properties, at the sacrifice of full consistency in the momentum equation. Additionally, by exploiting the increased accuracy of the RK method's gradient, we formulate a simple limiter for the artificial viscosity that reduces the excess diffusion normally incurred by the ordinary SPH artificial viscosity. Collectively, we call our suite of modifications to the traditional SPH scheme Conservative Reproducing Kernel SPH, or CRKSPH. CRKSPH retains many benefits of traditional SPH methods (such as preserving Galilean invariance and manifest conservation of mass, momentum, and energy) while improving on many of the shortcomings of SPH, particularly the overly aggressive artificial viscosity and zeroth-order inaccuracy. We compare CRKSPH to two different modern SPH formulations (pressure based SPH and compatibly differenced SPH), demonstrating the advantages of our new formulation when modeling fluid mixing, strong shock, and adiabatic phenomena.
\end{abstract}

\end{frontmatter}
\section{Introduction}
Originally derived in \cite{Lucy1977, Gingold1977}, Smoothed Particle Hydrodynamics (SPH) is a meshfree technique for simulating fluid dynamics, where particles (or ``nodes'') serve as interpolation points carrying the fluid properties.
SPH is a Lagrangian method, i.e., it discretizes the hydrodynamic equations using particles that move with the fluid velocity. The continuum fluid limit is represented by convolving the discrete particle properties (such as mass, momentum, and energy) with an interpolation kernel, generally denoted by $W$.  The functional form of $W$ is a free parameter, though in practice kernels with compact support such as the spline functions outlined in \cite{Schoenberg1969} are preferred.
SPH has many desirable properties for a hydrodynamic method: (1) it is Galilean invariant, which has a tremendous advantage in many astrophysical
applications with arbitrary gravitational potentials, (2) its Lagrangian nature allows the resolution to follow the mass (as opposed to prescribed heuristics for refinement
utilized in mesh codes), (3) it is agnostic to the particular geometries or material surface boundaries of a given problem, whereas grid imprinting can be a
concern in mesh-based methods, (4) it can be formulated to inherently obey the conservation laws of mass, momentum, and energy at machine precision,
(5) particle connectivity is mutable, enabling more accurate modeling of extreme material deformations, (6) it is easily extensible to multiple
dimensions, and (7) it is easily parallelizable.

While SPH has been successfully applied in many areas, most traditional SPH methodologies exhibit some known weaknesses. The most serious of these is SPH's lack of zeroth-order consistency, the so-called ``E0-error'' \cite{Agertz2007,Dilts1999,Morris1996,Read2010}. In other words, there is no guarantee that a constant pressure field (let alone a field with a more complicated, higher-order shape), for example, is interpolated correctly for non-uniform point distributions.
In the presence of a density discontinuity (and the attendant variation in the point distribution/weighting), this lack of zeroth-order consistency leads to errors that mimic a numerical surface tension.  This can drastically reduce accuracy in various fluid mixing problems \cite{Agertz2007,Okamoto2003}. 
Another common problem with SPH is the formulation of the artificial viscosity required to properly capture shock hydrodynamics;
for instance, the standard viscosity formalism of Monaghan and Gingold \cite{Monaghan:1983dn} introduces dissipation in any smooth convergent flow (regardless of the presence of a shock), which can result in over-damped solutions \cite{Cullen2010}.
Various corrections to the SPH viscosity have been proposed, e.g. \cite{Balsara1995, Morris1997, Cullen2010,Read2010}, and, in general, the viscosity treatment in SPH remains an active area of research. 

A number of studies have been designed to redress these SPH deficiencies.  One approach to the zeroth-order consistency problem is to replace the ordinary density or volume weighting of SPH with functions of the pressure \cite{Saitoh2013,Hopkins2012}, leading to Density Independent SPH (DISPH) or Pressure based SPH (PSPH).  This effort is motivated by noting that many classic test cases of fluid mixing (such as Rayleigh-Taylor, Kelvin-Helmholtz, etc.) involve discontinuous density fields but continuous (or even constant) pressure fields.  By converting the weighting of the SPH interpolations from being a function of the discontinuous variable (density) to functions of the continuous pressure, one can, to some extent, sidestep the zeroth-order errors of SPH.  However, the E0-error is in fact still present in these PSPH formulations -- a constant function on a set of disordered points will still not be interpolated correctly with PSPH.

Other investigators \cite{Price2008} have suggested introducing artificial thermal energy conduction akin to the suggestions of \cite{Noh1987}, noting that the SPH formalism is derived assuming continuous underlying fluid properties, and the discontinuous methods by which many classical mixing test cases are established (notably Kelvin-Helmholtz and Rayleigh-Taylor studies) often involve initially discontinuous fluid distributions.  
This point is well taken, but there are many concerns regarding artificial heat conduction such as unphysical transport of entropy. 
A strict lack of entropy diffusion is one of the strengths of a Lagrangian hydrodynamic formulation which we are loath to sacrifice.

Other approaches to the zeroth-order error problem replace the ordinary SPH interpolation methodology with interpolation bases that are intrinsically more accurate, allowing reproduction of fields to arbitrary order.  Two interesting examples are Reproducing Kernel (RK) methodologies \cite{Monaghan1985, Liu1995,Liu1998,Bonet2000} and the Moving Least Squares (MLSPH) approach of \cite{Dilts1999,Dilts2000}.  The RK methodology enhances the ordinary SPH interpolation kernel with additional terms/degrees of freedom that are recomputed for each new configuration of the points in order to exactly reproduce functions to any desired order.  This eliminates the zeroth-order error of SPH, but introduces a complication in that each point now has unique values for these additional terms in the kernel, and, thereby, makes the kernels between points non-symmetric.  This breaks the assumptions that are traditionally used to enforce conservation of linear momentum in SPH \cite{Monaghan2005}, and, to date, RK methodologies have accepted this lack of conservation and relied on the improved accuracy of RK differencing to keep this error in check.  This approach has worked reasonably well for low-deformation problems involving solids modeling \cite{Liu1998,Jun1998,Bonet2000,Bonet2002}, but this loss of strict conservation is a problem for fluid calculations involving large deformations and shock hydrodynamics.

The MLSPH method of \cite{Dilts1999,Dilts2000} goes further in altering the underlying interpolation basis functions, using a least-squares approach with arbitrary polynomial basis sets.  Unlike existing RK methods, MLSPH is formulated in a conservative manner, which is an important strength of this approach for fluid and shock hydrodynamics.  However, MLSPH represents a further departure from an ordinary SPH methodology, due to choices of the basis sets and how the sampling volumes are shaped/chosen.  MLSPH is an intriguing and promising technique, but for the purposes of this paper, it is too far afield from traditional SPH formalism. Moreover, we wish to develop an alternative approach that more closely leverages experience with applying SPH to many problems of interest to astrophysicists - an area where traditional SPH has seen its greatest use.

More recently, \cite{Hopkins2015} has proposed some very interesting techniques in the Meshless Finite Mass (MFM) and Meshless Finite Volume (MFV) algorithms.  These approaches also seek to rectify the interpolation inaccuracies of SPH by reformulating the underlying numerical differencing, using Riemann solvers to evaluate point to point interactions and a least-squares gradient operator to project the fluid values passed to the solver.  Although they are in fact meshfree, these methods in some ways bear more resemblance to moving mesh methods such as AREPO \cite{Springel2010}.  One concern with the MFV method of this pair is that it involves a mass advection term that may introduce mixed material complications in multi-material problems.  We will not consider direct comparisons with MFM or MFV, as these techniques are larger departures from ordinary SPH than we are seeking in this work;  such comparisons may be quite interesting in future studies, however.

Another recent less radical modification of SPH is proposed in \cite{Rosswog2015}, where the author investigates the use of modified forms of the SPH gradient operators that are more accurate than the standard SPH choices.  The motivation to improve the underlying numerical estimators of SPH is similar to our own in this investigation, and again comparisons of the methods outlined in \cite{Rosswog2015} with our approach would be interesting.

Another issue we wish to address beyond SPH's interpolation accuracy is the overly dissipative nature of the standard SPH artificial viscosity formulations.  The traditional SPH viscosity due to Monaghan and Gingold  \cite{Monaghan:1983dn} is activated by any convergent flow, which is not always appropriate.  Additionally, the pair-wise nature of this viscosity cannot distinguish a shearing from a converging flow, leading to overly diffuse solutions in fluid shearing and mixing phenomena.  This effect leads to some of the more notoriously studied issues such as unphysical damping in the Kelvin-Helmholtz instability, and particularly troubling for astrophysicists, unphysical transport of angular momentum in gravitationally bound rotating disks such as the classic Keplerian disk \cite{Balsara1995,Maddison1995,Cullen2010}.  Several studies aimed at correcting these issues have been pusblished,  ranging from efforts to solely solve the shearing problem by detecting and removing the viscosity from shears \cite{Balsara1995}, to more complicated/sophisticated efforts seeking to time evolve the components of the viscosity itself with physically-based sources from the hydrodynamic flow \cite{Morris1997,Cullen2010}.  Other studies have sought to improve the SPH viscosity by carefully choosing the type of interpolation kernel while adjusting the number of neighbors sampled by each point \cite{Read2010,WalterAly2012}.  Each of these solutions tends to treat specific problems (such as the overactivity of the viscosity in specific shearing test problems) to varying degrees of success, often with compromises in other situations (such as loss of sufficient dissipation in some shock scenarios).  Many of these approaches are also quite sophisticated and complex -- in this study we seek a simpler solution to these issues encountered in the viscosity treatment.

Our philosophy in this paper is to develop a method as closely related to ordinary SPH as possible while addressing what we see as SPH's greatest weaknesses: the zeroth-order interpolation errors and artificial viscosity formulation.  We choose to use reproducing kernel theory as the numerical basis of our interpolation and gradient operators.  RK interpolation is a direct extension of the corresponding SPH operators, adding only sufficient additional terms to allow reproduction of functions to the desired fidelity; specifically, we use linear reproducing kernels, i.e., smoothing functions that interpolate fields exactly up to linear terms. We utilize the mathematical framework of \cite{Dilts1999} to rederive the hydrodynamic equation for momentum based on RK interpolation, such that linear momentum is preserved exactly to machine precision, albeit relinquishing rigorous consistency (as investigated in \cref{sec:consist}). Although we also derive the time evolution equations for both specific thermal energy as well as total energy, we rely on the so-called ``compatible'' discretization ideas of \cite{Owen2014} to advance the specific thermal energy. The compatible energy methodology ensures total energy conservation is met exactly while favorable adiabatic evolution is maintained. We note that, unlike ordinary SPH,  angular momentum is only approximately conserved in our formalism, common in RK schemes as the pair-wise forces are no longer radially aligned when using non-symmetric kernels. However, \cref{sec:angularMomentum} illustrates that the affect is sub-percent in rotating problems, where proper angular momentum treatment is crucial; \cite{Raskin2016} further illustrated that the increased accuracy in simulated angular momentum transport of our formalism demonstrates significantly improved solutions in rotating phenomena when compared to SPH, regardless of the fact that angular momentum is not strictly conserved.  

The final novel ingredient in our scheme is derived from exploiting the accurate gradient operator of the RK formulation, where we construct a high-order limiter applicable to the standard SPH artificial viscosity \cite{Monaghan:1983dn} in the spirit of \cite{Christensen1990}.  This new limiter, while a minor modification of the SPH viscosity, greatly reduces the overly aggressive dissipation of the unlimited SPH viscosity.  Taken together, we call the resulting method Conservative Reproducing Kernel Smoothed Particle Hydrodynamics (CRKSPH).

The outline of the paper is as follows. We reprise the reproducing kernel interpolation methodology in \cref{sec:rk}. \Cref{sec:crksph} describes the three major alterations to the SPH formalism utilized in the CRKSPH framework: (a) linear reproducing kernels, (b) conservative dynamic equations (including the compatible energy update), and (c) limited artificial viscosity. Results from standard hydrodynamic tests using the CRKSPH scheme along with selected comparisons to other SPH methods are presented in \cref{sec:crktests}. Finally, in \cref{sec:conclusion} we conclude with discussion and future work.

We briefly take a moment to collect our notation. Throughout this paper, Latin subscripts denote node indices, while Greek superscripts denote dimensional components (e.g.~$x_i^\alpha$ is the $\alpha$-component of the positional vector for node $i$). We employ summation notation for repeated superscripts, such as $v_i^\alpha v_j^\alpha= \mathbf{v}_i \cdot \mathbf{v}_j$, and we succinctly write spatial gradients as $\paralpha f \equiv \partial f / \partial x^\alpha$.

\section{Reproducing Kernels}
\label{sec:rk}
To address the inability of SPH to adequately reconstitute fields of a desired order, \cite{Liu1995,Bonet2000} suggest adding terms to the traditional SPH interpolation kernel
that allow for the exact reproduction of constant, linear, or higher order fields. This results in an enhanced particle interpolation method referred to as reproducing kernel methods (RPKM). 

In ordinary SPH interpolation \citep{Monaghan2005}, an arbitrary function $\psi(\bb{r})$ can be approximated through convolution with a kernel $W(\bb{r},h)$
\begin{align}
  \label{eq:eqnKern}
  \psi(\bb{r})=\int \psi(\bb{r}')W(\bb{r}-\bb{r}',h)d\bb{r}',
\end{align}
which can then be discretized for particle interpolants via
\begin{align}
  \label{eq:disc}
  \psi_i=\sum_j  \psi_jW(|r_i-r_j|,h)V_j=\sum_j m_j \frac{\psi_j}{\rho_j}W(|r_i-r_j|,h)
\end{align}
where $V_j, \rho_j,$ and $m_j$ are the volume, density, and mass of the $j^{\text{th}}$ fluid parcel; note $V_j=m_j/\rho_j$ in this relation. As a concrete example, substituting $\rho$ for $\psi$ yields the SPH density estimate equation:
\begin{align}
 \label{eq:SPHSum}
  \rho_i = \sum_j  m_jW(|r_i-r_j|,h)
\end{align}
illustrating how the interpolants' mass is ``smoothed" to approximate particle density, and hence the designation Smoothed Particle Hydrodynamics.
  The choice of interpolation kernel $W(r,h)$ is arbitrary for the method. As previously described, it is desirable that $W(r,h)$ be approximately Gaussian, yet have compact support (i.e., a finite extent $r_c$ beyond which $W(r>r_c,h)=0$).  A common choice for the SPH kernel function is the cubic spline \cite{Schoenberg1969,Monaghan2005}
\begin{align}
  \label{eq:eqnW3}
  W_3(\bb{x},h)=W_3(\eta)=A_d \begin{cases}  1-\frac{3}{2}\eta^2+\frac{3}{4}\eta^3, \> &0\le \eta \le 1 \\ \frac{1}{4}(2-\eta)^3, &1 < \eta \le 2 \\ 0, &\eta > 2 \end{cases} 
\end{align}
where $\eta^\alpha=x^\alpha/h$, and $A_d$ is a normalization constant in $d$-dimensions. The cubic spline conforms to the following \bb{required conditions} 
of an SPH kernel: (i) $W_3$ approaches a delta function as the smoothing scale $h\rightarrow 0$, (ii)$\int W_3 d\bb{r} = 1$, here enforced by the normalization $A_d$, and (iii) has compact support. 

The RPKM is derived by re-examining \cref{eq:eqnKern}. Plugging the Taylor expansion of $\psi(\bb{r}')$ into the equation yields
\begin{align}
  \label{eq:UpOrder}
  \psi(\bb{r})= \; &\psi(\bb{r})\int W(\bb{r}-\bb{r}',h)d\bb{r}' + \psi '(\bb{r})\int W(\bb{r}-\bb{r}',h) (\bb{r-r}')d\bb{r}' \nonumber \\
&+ \frac{1}{2} \psi ''(\bb{r})\int W(\bb{r}-\bb{r}',h) (\bb{r-r}')^2d\bb{r}' + \; \cdots
\end{align}
Thus, if the following ``consistency equations'' are satisfied
\begin{align}
  \label{eq:FirstCons}
  \int W(\bb{r}-\bb{r}',h)d\bb{r}' &=1 \\
  \label{eq:RestCons}
  \int W(\bb{r}-\bb{r}',h) (\bb{r-r}')^{\otimes m} d\bb{r}' &=0, \; m \in [1,n]
\end{align}
where $(\bb{r-r}')^{\otimes m}$ represents the outer $m$th product of the vector $\bb{r-r}'$, 
we are left with a kernel that is $n$th-order accurate, i.e., will exactly reproduce polynomial fields of order $n$. We note that \cref{eq:FirstCons}
was the second criteria listed above for any SPH kernel, but \cref{eq:RestCons} is a stronger constraint that is not satisfied by the simple cubic
spline $W_3$ as stated. We also recognize that the consistency equations are satisfied by a delta function for any order $n$, as one would expect
since such a kernel is exactly reproducing. In the next section we explicitly construct a linear-order reproducing kernel, which will be the basis of our method in this paper.

\subsection{Linear-order Reproducing Kernels}
\label{sec:linearrk}
The reproducing kernel formulation can be extended to any order of consistency; in this paper we focus on RK of linear order.  We denote the reproducing kernel by $\mathcal{W}^R$, as distinguished from (and built upon) the unmodified SPH kernel $W$.  The linearly corrected reproducing kernel and its gradient are given as
\begin{align}
  \label{eq:wR}
  \Wrij(\xij) \equiv& A_i \left(1+B_i^\alpha \xij^\alpha\right)\Wij(\xij) \\
  \label{eq:gradwR}
  \pargam\Wrij(\xij) =& A_i \left(1+B_i^\alpha \xij^\alpha\right)\pargam \Wij(\xij) \nonumber \\
  &+ \pargam A_i\left(1+B_i^\alpha x_{ij}^\alpha\right)\Wij(\xij) \nonumber\\
  &+ A_i\left(\pargam B_i^\alpha \xij^\alpha + B_i^\gamma\right)\Wij(\xij)
\end{align}
where we have used the notational shorthand $\xij^\alpha \equiv x_i^\alpha - x_j^\alpha$, $\Wij(\xij) \equiv [W_i(\xij, h_i) + W_j(\xij, h_j)]/2$, and $h_i$ and $h_j$ are the individual smoothing scales of points $i$ and $j$.
The constants $A_i$ and $B_i^\alpha$ are determined by the discrete form of the consistency relations \cref{eq:FirstCons,eq:RestCons} with $n=1$
\begin{align}
  \label{eq:Cons1}
  \sum_j V_j \Wrij = 1 \\
  \label{eq:Cons2}
  \sum_j V_j x_{ij}^\alpha \Wrij = 0.
\end{align}
It is important to note the distinction that these conditions are met in the discrete case, i.e., only in the continuous (infinite resolution) domain does the ordinary SPH interpolation exactly meet the restriction of \cref{eq:FirstCons}, which is how the normalization constants for the SPH kernel (such as $A_d$ in \cref{eq:eqnW3}) are determined. In the discrete case of \cref{eq:Cons1}, the SPH interpolation only approximately meets this criterion, whereas the RK interpolation enforces it.

Imposing the discrete constraints of \cref{eq:Cons1,eq:Cons2} on \cref{eq:wR}, shown explicitly in \cref{sec:RKDerv}, leads to the following solution for $(A_i, B_i^\alpha)$:
\begin{align}
  \label{eq:coefA}
  A_i &= \left[m_0-\left(m_2^{-1}\right)^{\alpha\beta}m_1^\beta m_1^\alpha\right]^{-1} \\
  \label{eq:coefB}
  B_i^\alpha &= -\left(m_2^{-1}\right)^{\alpha\beta}m_1^\beta,
\end{align}
with derivatives obtained by repeated application of the chain-rule,
\begin{align}
  \label{eq:gradA}
  \pargam A_i = &-A_i^2\{\pargam m_0 - \left(m_2^{-1}\right)^{\alpha\beta}m_1^\beta\pargam m_1^\alpha-\left(m_2^{-1}\right)^{\alpha\beta}\pargam m_1^\beta m_1^\alpha \nonumber\\
  &+\left(m_2^{-1}\right)^{\alpha\phi}\pargam m_2^{\phi\psi}\left(m_2^{-1}\right)^{\psi\beta}m_1^\beta m_1^\alpha\}\\
  \label{eq:gradB}
  \pargam B_i^\alpha = & -\left(m_2^{-1}\right)^{\alpha\beta}\pargam m_1^\beta + \left(m_2^{-1}\right)^{\alpha\phi}\pargam m_2^{\phi\psi}\left(m_2^{-1}\right)^{\psi\beta}m_1^\beta,
\end{align}
where we have defined geometric moments (and their derivatives) as
\begin{align}
  \label{eq:m0}
  m_0 &\equiv \sum_j V_j \Wij \\
  \label{eq:m1}
  m_1^\alpha &\equiv \sum_j x_{ij}^\alpha V_j\Wij \\
  \label{eq:m2}
  m_2^{\alpha\beta} &\equiv \sum_j x_{ij}^\alpha x_{ij}^\beta V_j\Wij \\
  \label{eq:gradm0}
  \pargam m_0 &= \sum_j V_j \pargam\Wij \\
  \label{eq:gradm1}
  \pargam m_1^\alpha &= \sum_j V_j\left(x_{ij}^\alpha \pargam\Wij+\delta^{\alpha\gamma}\Wij\right) \\
  \label{eq:gradm2}
  \pargam m_2^{\alpha\beta} &= \sum_j V_j\left[x_{ij}^\alpha x_{ij}^\beta \pargam\Wij+\left(x_{ij}^\alpha\delta^{\beta\gamma}+\delta^{\alpha\gamma}x_{ij}^\beta\right)\Wij\right].
\end{align}
In addition to satisfying \cref{eq:Cons1,eq:Cons2} exactly, $\mathcal{W}^R$ maintains the compact support of $W$ as well as the delta function limiting behavior.
$\mathcal{W}^R$, therefore, satisfies all three of the above listed criteria for utilization as an SPH kernel. Moreover, the linear order formulation allows for a kernel that reconstructs linear fields to machine precision.

It is evident from \crefrange{eq:coefA}{eq:gradm2} that the reproducing kernel and its gradient are only dependent on the geometric moments of the underlying kernel and the point weight $V_j$: this dependence implies it is only necessary to recompute the point-wise kernel enhancements $(A_i, B_i^\alpha)$ when the points move.  Until such geometric changes occur, the computed values of the kernel corrections can be reused as necessary.

Finally, given the relations for $\Wrij$ and $\pargam\Wri$ in \cref{eq:wR,eq:gradwR}, we can express the RK formulation for the interpolation and gradient of a general field $F(x^\alpha)$ as
\begin{align}
  \label{eq:rkinterp}
  \Interp{F(x_i^\alpha)} &= \sum_j V_j F_j \Wrij \\
  \label{eq:rkgrad}
  \Interp{\pargam F(x_i^\alpha)} &= \sum V_j F_j \pargam\Wrij.
\end{align}
These relations are nearly identical to their SPH counterparts, simply substituting the enhanced RK kernel for the SPH one, $\Wij \to \Wrij$.  However, these expressions are now exact for any linear field by construction, and therefore avoid the zeroth-order consistency error of ordinary SPH interpolation. For further inspection, \cref{fig:1dInterp} in \cref{sec:RKDerv} illustrates interpolating a linear field to machine precision using the RK method.  

\section{CRKSPH Formalism}
\label{sec:crksph}

\subsection{Conservative Fluid Equations}
\label{sec:crkmomderiv}
The reproducing kernel interpolation and gradient (\cref{eq:rkinterp,eq:rkgrad}) are, by construction, free of the zeroth-order error of ordinary SPH, and in general, provide more accurate results.  However, for use in constructing the hydrodynamic evolution equations, they pose a serious challenge: conservation of linear momentum.  The derivation of the ordinary SPH momentum equation (\cref{eq:SPHvel}, see \cite{Monaghan2005,Owen2014} for a complete explanation) depends on the spatial symmetry of the kernel function (i.e. $\paralpha \Wij = -\paralpha \Wji$) to make the SPH pair-wise forces anti-symmetric, which manifestly conserves linear momentum.  As the additional terms $(A_i, B_i^\alpha)$ in the RK kernel function vary from point to point, $\paralpha \Wrij \ne -\paralpha \Wrji$ in general, and the usual method of deriving a conservative momentum equation does not work. 

Fortunately, it is possible to construct a conservative differencing of the hydrodynamic evolution equations using reproducing kernels.  
We utilize the formalism outlined in the derivation of MLSPH, and refer the reader to the thorough exposition and excellent discussion in \cite{Dilts1999} for more detail.

We begin by assuming interpolation using a generic kernel $\psi(x),$ which must satisfy the constraints
\begin{equation}
  \label{eq:Interp}
  \sum_j \psi_j \equiv 1 \; \Rightarrow \; \sum_j \nabla \psi_j=0.
\end{equation}
In our case, $\psi_j=V_j \Wrij$, which satisfies this condition by \cref{eq:Cons1}.
The dynamic fluid equations can be written in conservative form as
\begin{equation}
  \label{eq:hydro}
  \rho \frac{D\Uvec}{Dt}=\paralpha \Fvec(\Uvec),
\end{equation}
where the conserved density $\Uvec$ and flux $\Fvec$ are defined as
\begin{equation}
  \label{eq:consvar}
  \Uvec = \begin{bmatrix}
    1/\rho \\
    v^\alpha  \\
    u + v^2/2
  \end{bmatrix}, \;\; F(\Uvec)=\begin{bmatrix}
  v^\alpha \\
  \sigmaab  \\
  \sigmaab v^\beta
  \end{bmatrix}, \;\; \sigmaab = -P \delta^{\alpha \beta} + \tau^{\alpha \beta},
\end{equation}
where $\rho$ is the mass density, $v^\alpha$ velocity, $u$ the specific thermal energy, $\sigmaab$ the stress tensor, $P$ the pressure, $\delta^{\alpha\beta}$ the Kronecker delta, and $\tau^{\alpha\beta}$ the viscous deviatoric stress tensor which is zero for the ideal fluid case of interest here.  
Multiplying \cref{eq:hydro} by the basis function $\psi$ and taking the volumetric integral we have
\begin{equation}
  \label{eq:volform}
  \int_V \psi \rho \frac{D\Uvec}{Dt} = \int_V \psi \paralpha \Fvec(\Uvec).
\end{equation}
We approximate the flux $\Fvec$ with our interpolants $\psi$ as $\Fvec \approx \sum_j \Fvec_j \psi_j$. We also assume that for any smooth function $f,$ we can apply the one-point quadrature approximation
\begin{equation}
  \int f \psi \approx V_i f_i.
\end{equation}
Plugging these two relations into \cref{eq:volform} and using the definition of mass $m_i= \rho_i V_i$ we arrive at the approximate dynamic equation
\begin{equation}
  \label{eq:firstform}
  m_i \frac{D\Uvec_i}{Dt}=\sum_j \Fvec_j \int_V \psi_i \paralpha \psi_j.
\end{equation}
Performing integration by parts
\begin{equation}
  \label{eq:secondform}
  m_i \frac{D\Uvec_i}{Dt} = \sum_j \Fvec_j \left( \oint_{\partial V} \psi_i \psi_j \hat{n}^\alpha - \int_V \paralpha \psi_i \psi_j \right).
\end{equation}
Note, $\oint_{\partial V}$ represents the surface integral on the bounding surface of $V$, with $\hat{n}^\alpha$ the local surface normal to $\partial V$.
Both momentum equations are not yet in conservative form, but making use of the identities from \cref{eq:Interp}
\begin{align}
  \sum_j \int_V \psi_i \paralpha \psi_j = \int_V \psi_i \sum_j \paralpha \psi_j  &= 0 \nonumber \\
  \label{eq:id1}
  \Rightarrow \; \sum_j \Fvec_i \int_V \psi_i \paralpha \psi_j &= 0 \\
  \sum_j \int_V \psi_i \paralpha \psi_j = \sum_j \left(\oint_{\partial V} \psi_i \psi_j \hat{n}^\gamma - \int_V  \psi_j \paralpha \psi_i \right) &= 0 \nonumber \\
  \label{eq:id2}
  \Rightarrow \; \sum_j \Fvec_i \left (\oint_{\partial V} \psi_i \psi_j \hat{n}^\gamma - \int_V \psi_j \paralpha \psi_i \right) &= 0,
\end{align}
we can sum \crefrange{eq:firstform}{eq:id2} and arrive at a third approximate dynamic equation
\begin{equation}
  \label{eq:thirdform}
  2m_i \frac{D\Uvec_i}{Dt} = \sum_j(\Fvec_i + \Fvec_j)\left( \int_V \psi_i \paralpha \psi_j -  \int_V \psi_j \paralpha \psi_i + \oint_{\partial V} \psi_i \psi_j \hat{n}^\gamma  \right),
\end{equation}
where we note that the RHS of the equation is anti-symmetric in indices $i,j$. For the fluid material problems we consider in this paper, the total fluid volume $V$ does not have any rigid boundaries $\partial V$ (consistent with the equivalent assumptions for SPH),
so the boundary term in \cref{eq:thirdform} can be dropped. Applying the one-point quadrature approximation (as the interpolant functions are smooth), we can approximate the volume integrals using
\begin{equation}
  \label{eq:approx}
  \int_V \psi_i \paralpha \psi_j \approx V_i \paralpha \psi_j.
\end{equation}
Using \cref{eq:approx} and dropping the boundary terms in \cref{eq:thirdform} we arrive at a discretized evolution equation
\begin{equation}
  \label{eq:master}
  m_i \frac{D\Uvec_i}{Dt} = \half \sum_j(\Fvec_i + \Fvec_j) \left( V_i \paralpha \psi_j - V_j \paralpha \psi_i \right),
\end{equation}
from which we can directly obtain the dynamic momentum and energy equations using \cref{eq:consvar}
\begin{align}
  \label{eq:momeq}
  m_i \frac{Dv^\alpha_i}{Dt} &= \half \sum_j(\sigmaab_i + \sigmaab_j) \left( V_i \partial_\beta \psi_j - V_j \partial_\beta \psi_i \right) \\
  m_i \frac{Du_i}{Dt} + m_i v^\alpha_i \frac{Dv^\alpha_i}{Dt} &= \half \sum_j (\sigmaab_i v^\beta_i + \sigmaab_j v^\beta_j) \left( V_i \partial_\alpha \psi_j - V_j \partial_\alpha \psi_i \right) \nonumber \\
  \label{eq:engeq}
  \Rightarrow \; m_i \frac{Du_i}{Dt} &= \half \sum_j \sigmaab_j (v^\beta_j - v^\beta_i) \left( V_i \partial_\alpha \psi_j - V_j \partial_\alpha \psi_i \right).
\end{align}
Finally, imposing the ideal fluid stress tensor $\sigmaab = -P \delta^{\alpha\beta}$ and our choice of basis function $\psi_j = V_j \Wrij$, the resulting evolution equations are
\begin{align}
  \label{eq:momeq2}
  m_i \frac{Dv^\alpha_i}{Dt} &= -\half \sum_j V_i V_j (P_i + P_j) \left( \partial_\alpha \Wrij  - \partial_\alpha \Wrji \right) \\
  \label{eq:engeq2}
  m_i \frac{Du_i}{Dt} &= \half \sum_j V_i V_j P_j \left( v^\alpha_i - v^\alpha_j \right) \left( \partial_\alpha \Wrij - \partial_\alpha \Wrji \right).
\end{align}

To explicitly illustrate conservation invariance, we examine the pair-wise forces due to \cref{eq:momeq2}.  We can see the force from point $j$ upon $i$ is
\begin{equation}
  \label{eq:aij}
  m_i a_{ij}^\alpha = -\half V_i V_j (P_i + P_j) \left( \partial_\alpha \Wrij  - \partial_\alpha \Wrji \right),
\end{equation}
where $a_{ij}$ is the acceleration on $i$ due to $j$.  Reversing the indices, the force due to node $i$ on point $j$ is
\begin{equation}
  \label{eq:aji}
  m_j a_{ji}^\alpha = -\half V_j V_i (P_j + P_i) \left( \partial_\alpha \Wrji  - \partial_\alpha \Wrij \right).
\end{equation}
Examination of the right-hand sides of \cref{eq:aij,eq:aji} show they are equal up to the terms $\left( \partial_\alpha \Wrij  - \partial_\alpha \Wrji \right) = -\left( \partial_\alpha \Wrji  - \partial_\alpha \Wrij \right)$, which imply
\begin{equation}
  \label{eq:antisym}
  m_i a_{ij}^\alpha = -m_j a_{ji}^\alpha.
\end{equation}
\Cref{eq:antisym} demonstrates that the pair-wise forces due to \cref{eq:momeq2} are anti-symmetric, and therefore, using this relation as the momentum equation with the reproducing kernel formalism enforces exact linear momentum conservation. 

Although the pairwise forces due to \cref{eq:momeq} are equal and opposite, it is important to note that those forces are not guaranteed to be radially oriented between the interacting points (at least for RK corrections beyond zeroth-order, i.e., with more than the $A_i$ correction in \cref{eq:wR}), contrary to ordinary SPH. As a result, angular momentum is not exactly conserved in our formalism, unlike linear momentum. In \cref{sec:angularMomentum}, we examine the magnitude of this loss of exact angular momentum conservation in a rotating spherical collapse test case and find it to be at the sub-percent level. It is worth considering that the treatment of angular momentum plays a key role in certain applications, such as gravitationally bound rotating disks. However, the quality of the simulation depends on more than the exact total conservation of angular momentum; one important, and often overlooked, complication of SPH is the fact that the inaccuracies due to either interpolation error or overactivity of the artificial viscosity can result in incorrect angular momentum transport, depite rigorous total conservation. In \cite{Raskin2016} we examine a family of generalized Keplerian disk problems, where we find the inaccuracies of the angular momentum transport of SPH (using modern viscosity prescriptions to minimize artificial transport), result in significantly degraded solutions compared to our conservative CRK formalism; these tests illustrate the key role of accurately modeling angular momentum transport in these astrophysically relevant scenarios.  Nonetheless, we do not wish to downplay the importance of angular momentum conservation: as discussed in \cref{sec:angularMomentum}, we find the deviation from exact angular momentum conservation is at the sub-percent level and converges rapidly toward zero with increasing resolution. In rotating problems this quantity should be monitored just as energy should be in non-energy conserving methods, and in fact a variety of numerical effects can contribute larger errors to the angular momentum such as approximate gravitational solvers (tree or particle-mesh for instance), non-radial physical forces such as material strength, etc.  If desired, total angular momentum conservation of the CRK method can be restored by dropping to zeroth-order consistency in the RK formalism (forcing $B_i^\alpha=0$ in \cref{eq:wR}), at the cost of reduced accuracy in the interpolation method.

We conclude this derivation discussion by noting two features of \cref{eq:master}. First, enforcing linear momentum conservation as is done in \cref{eq:approx} renders the dynamic equations inconsistent, i.e., no longer exactly reproducing. As shown rigorously in section 3.7 of \cite{Dilts1999}, inserting a simple polynomial field into \cref{eq:master} will result in dynamic equations that do not precisely evaluate the field gradient using an arbitrary polynomial basis. Thus, although linear RKs are used as basis functions in this paper, the evolution due to a linear pressure field via \cref{eq:momeq2} will in general not exactly reproduce the expected constant acceleration field, as opposed to achieving exact reproducibility. In other words, the inconsistency is a compromise for achieving locally conservative equations that we find to be necessary for problems involving compressible hydrodynamics in extreme flows, such as systems involving strong shock phenomena. An investigation of the inconsistency error for linear problems can be found in \cref{sec:consist}. 

Second, with regards to the validity of approximations in our formalism compared to SPH, it can be shown that the derivation of \cref{eq:master} is merely a generalized representation of the traditional SPH fluid equations. In fact, as shown in \cite{Dilts1999}, the inconsistency of the quadrature approximation can be removed given a discretized constraint (Equation 34 in \cite{Dilts1999}), which is sufficiently satisfied when the kernel is symmetric and the boundary terms assumed to be zero (as is true in nominal SPH). Substituting the SPH kernel $W$ into \cref{eq:momeq2}, as opposed the RK basis $\mathcal{W}^R$, yields a common form of the symmetric SPH acceleration equation
\begin{align}
 m_i \frac{Dv^\alpha_i}{Dt} &= -\half \sum_j V_i V_j (P_i + P_j) \left( \partial_\alpha W_{ij}  - \partial_\alpha W_{ji} \right) \\
 \Rightarrow \frac{Dv^\alpha_i}{Dt}  &= - \sum_j \frac{m_j}{\rho_i \rho_j} (P_i+P_j) \partial_\alpha W_{ij}
\end{align}
illustrating the validity of the quadrature approximations and the return to consistency.

At this point, we have the fundamentals to form a fully conservative hydrodynamics method based on RK theory. However, before we put together the full formalism, we consider two further ingredients in the CRKSPH prescription: a limited form of artificial viscosity in \cref{sec:limitedq}, and an improved energy update in \cref{sec:compwork}.

\subsection{Limited Artificial Viscosity} 
\label{sec:limitedq}
We begin with the standard SPH viscosity attributed to Monaghan and Gingold \cite{Monaghan:1983dn}, which approximates the classic bulk (linear) and Von Neumann-Richtmyer (quadratic) viscosity via
\begin{align}
  \label{eq:visc}
  Q_i &= \rho_i \left(-C_l c_i \mu_i + C_q \mu_i^2 \right) \\
  \label{eq:unlimmu}
  \mu_i &= \min\left(0, \frac{\vij^\alpha \eta_i^\alpha}{\eta_i^\alpha \eta_i^\alpha + \epsilon^2}\right) \\
  \eta_i^\alpha &= \xij^\alpha/h_i
\end{align}
where $Q_i$ is the artificial viscous pressure, $\vij^\alpha \equiv v_i^\alpha - v_j^\alpha$, $\xij^\alpha \equiv x_i^\alpha - x_j^\alpha$, $(C_l, C_q)$ are the viscous linear and quadratic coefficients, $c_i$ is the sound speed, and $\epsilon$ is a small number to avoid division by zero.  Note we have used the convention in these relations that the subscript $i$ denotes the choice of which smoothing scale is used between the pair $(i,j)$, in this case $h_i$, with $\eta_i^\alpha = (x_i^\alpha - x_j^\alpha)/h_i$. Using our subscript convention, $Q_j$ is obtained by using $h_j$ in the above relations. We also note that $Q$ is defined as an artificial viscous pressure in \cref{eq:visc}, rendering its implementation trivial in the evolution relations (\crefrange{eq:momeq2}{eq:engeq2}) by replacing the pressure $P$ with $P + Q$. 

This formulation of the viscosity does an excellent job of capturing one-dimensional shock phenomena. It is inspired by the classic bulk and Von Neumann-Richtmyer viscosities, but replaces the full velocity divergence $\paralpha v^\alpha$ with the pair-wise approximation $\mu_i$ of \cref{eq:unlimmu}. The success of the Monaghan-Gingold viscosity can be attributed to this pair-wise formulation. The definition of $\mu_i$ allows $Q_i$ to respond to individual velocity jumps between points and to dissipate extreme values effectively -- efforts to directly use the SPH interpolation for $\Interp{\paralpha v^\alpha}$ (such as in the earliest work by \cite{Lucy1977,Monaghan1977}) can fail to pick up extreme pair-wise velocities, allowing noise in the velocity field to grow at or below the resolution scale.  

While this pair-wise formulation is key to the success of Monaghan-Gingold viscosity, there are two serious issues with this definition for $\mu_i$, both related to the fact that the viscosity will activate and add dissipation whenever the dot product $\vij^\alpha \xij^\alpha < 0$ in \cref{eq:unlimmu}. First, consider two points in a purely shearing flow.  In this pair-wise definition, there will be times when $\vij^\alpha \xij^\alpha < 0$ even though there is actually no compression, and the viscosity will be triggered.  The full velocity divergence $\paralpha v^\alpha$ would correctly detect the lack of compression in such pure shears, but in multiple dimensions there simply is not enough information from two point-wise velocities to distinguish shear from compression.  This is the source of errors noted in shearing flows such as the classic Keplerian disk problem or models of the Kelvin-Helmholtz instability, and has inspired increasingly sophisticated corrections suggested by authors such as \cite{Balsara1995,Morris1997,Cullen2010} in an effort to add back information from the full velocity gradient.

The second issue with \cref{eq:unlimmu} is that, even in simple one-dimensional flows, not all compressions should necessarily trigger dissipation: modeling the propagation of acoustic waves, isentropic (adiabatic) compressions, such as in the pre-shock flow of the classic Noh implosion test \cite{Noh1987}, certain regimes in laser driven implosions as found in, e.g., inertial confinement fusion experiments \cite{Kidder1976}, or pre-shock gaseous inflow in astrophysical scenarios, are all examples of phenomena that can suffer from artificial dissipation.  Loss of proper adiabatic behavior can seriously impact the usefulness of a numerical model, and few of the efforts to limit the SPH viscosity have dealt with this issue.

Here we develop a simple limiter formulation for $\mu_i$ inspired by the ideas of \cite{Christensen1990}.  This approach is based on the observation that domains with a linear velocity field, indicating smooth flow rather than the presence of a shock, should have vanishing artificial viscosity. We accomplish this by replacing the computed pair-wise velocity jump $\vij^\alpha$ in \cref{eq:unlimmu}, with a limited value projected to the midpoint position between points $i$ and $j$.  We compute the linearly extrapolated velocity jump $\vhatij^\alpha$ as
\begin{align}
  \label{eq:viLim}
  \hat{v}_i^\alpha &\equiv v_i^\alpha + \half \phi_{ij} \partial_\beta v_i^\alpha (x_j^\beta-x_i^\beta)  \\ 
  \label{eq:vjLim}
  \hat{v}_j^\alpha &\equiv v_j^\alpha + \half \phi_{ji} \partial_\beta v_j^\alpha (x_i^\beta-x_j^\beta)  \\ 
  \label{eq:vDiff}
  \vhatij^\alpha &\equiv \hat{v}_i^\alpha - \hat{v}_j^\alpha.
\end{align}
The term $\phi_{ij} \in [0,1]$ is a pair-wise limiter designed to allow a high-order solution ($\phi=1$) in a smooth field, while rolling over to a low-order ($\phi=0$) evaluation for discontinuous or extreme values.  We use the classic van Leer limiter \cite{vanLeer1974,Toro1989} familiar from the world of mesh-based hydrodynamics as the basis of our limiter, with a modification at small separations.  Our modified limiter is given as
\begin{align}
  \label{eq:phi}
  \phi_{ij} &= \max\left[0, \min\left[1, \frac{4 r_{ij}}{(1 + r_{ij})^2}\right]\right] \times
  \left\{ \begin{array}{l@{\quad}l}
    \exp\left\{-\left((\eta_{ij} - \etacrit)/\etafold\right)^2\right\}, & \eta_{ij} < \etacrit \\
    1, & \eta_{ij} \ge \etacrit \\
  \end{array} \right. \\
  \label{eq:rij}
  r_{ij} &\equiv \frac{\left(\partial_\beta v_i^\alpha \right) \xij^\alpha \xij^\beta}
                     {\left(\partial_\beta v_j^\alpha \right) \xij^\alpha \xij^\beta} \\
  \eta_{ij} &\equiv \min(\eta_i, \eta_j) = \min\left( (\xij^\alpha \xij^\alpha)^{1/2}/h_i, (\xij^\alpha \xij^\alpha)^{1/2}/h_j \right).
\end{align}
Note in \cref{eq:phi}, the limiter is symmetric: $\phi_{ij} = \phi_{ji}$.  $r_{ij}$ is formed from the projected velocity jump computed from the velocity gradients $\partial_\beta v_i^\alpha$ and $\partial_\beta v_j^\alpha$, serving analogously to the ratio of the forward and backward solution differences in mesh-based limiters.  We use the ordinary RK gradient operator to find this velocity gradient as
\begin{equation}
  \partial_\beta v_i^\alpha = -\sum_j V_j \vij^\alpha \partial_\beta \Wrij.
\end{equation}
The first piece of \cref{eq:phi} is simply the ordinary van Leer limiter; the second term activated for $\eta_{ij} < \etacrit$, forces the limiter to zero as points are driven close together.  We choose $\etacrit$ such that this term only comes into effect for points that are getting closer together than we would expect based on physics.  We parameterize the evolution of the smoothing scale in terms of the desired number of points per smoothing scale, denoted by $n_h$.  In this parlance we choose $(\etacrit, \etafold) = (1/n_h, 0.2)$, so that in ordinary smooth regions this second multiplier should never activate.  We find this correction helps by dissipating small-scale noise as it arises in calculations.

To conclude, the limited form of the artificial viscosity we use in CRKSPH is merely the ordinary SPH Monaghan-Gingold viscosity of \cref{eq:visc}, except now $\mu_i$ is computed using our monotonically extrapolated velocity jump
\begin{equation}
  \label{eq:limmu}
  \mu_i = \min\left(0, \frac{\vhatij^\alpha \eta_i^\alpha}{\eta_i^\alpha \eta_i^\alpha + \epsilon^2}\right).
\end{equation}

\subsection{The Compatible Energy Discretization}
\label{sec:compwork}
Although we reference the evolution equations for both the specific thermal energy and total energy in \cref{eq:CRKDuDt,eq:CRKDEDt} for completeness, in this paper, we use the ``compatible'' discretization described in \cite{Owen2014} to advance the specific thermal energy in our CRKSPH examples.  The major advantage of this method is that it both enforces exact energy conservation (yielding good results for strong shock problems) while maintaining favorable behavior on adiabatic problems, a property typically sacrificed by total energy algorithms.  We demonstrate the advantage of the compatible discretization for adiabatic evolution using an isentropic test case in \cref{sec:KidderTest}.  In the following, we briefly summarize the compatible energy update methodology, including an improvement to the pair-wise distribution of the work not described in the original algorithm of \cite{Owen2014}. For a more thorough discussion we refer the reader to \cite{Owen2014}.

The essence of this idea is that we will exactly account for the pair-wise discrete work implied by the discretized momentum equation, regardless of the details of how the momentum equation is derived, i.e., SPH, PSPH, CRKSPH, etc. We begin by writing down the total energy of the discretized system (ignoring any external sources or sinks of energy) as
\begin{equation}
  E = \sum_i m_i \left( \frac{1}{2} v_i^2 + u_i \right).
\end{equation}
Note here we have adopted the convenient notational contraction $v_i^2 = v_i^\alpha v_i^\alpha$, i.e, the square of the vector magnitude.
The total energy change across a timestep (denoting the beginning of timestep values by superscript 0 and end of timestep values by superscript 1) is
\begin{equation}
  \label{eq:DeltaE}
  E^1 - E^0 = \sum_i m_i \left[ \frac{1}{2} \left( v_i^1 \right)^2 + u_i^1 - \frac{1}{2} \left( v_i^0 \right)^2 - u_i^0 \right].
\end{equation}
Total energy conservation is enforced by setting $E^1 - E^0 = 0$; we use $(v_i^\alpha)^1 = (v_i^\alpha)^0 + (a_i^\alpha)^0 \Delta t$ to rewrite \cref{eq:DeltaE} as
\begin{align}
  0 &= \sum_i m_i \left[ 
                  \left( (v_i^\alpha)^0 + \frac{1}{2} (a_i^\alpha)^0 \Delta t \right)
                  (a_i^\alpha)^0 \Delta t + u_i^1 - u_i^0 
                  \right] \\
    \label{eq:TotalEBalance}
    &= \sum_i m_i \left[
                  (v_i^\alpha)^{1/2} (a_i^\alpha)^0 \Delta t + \Delta u_i
                  \right]
\end{align}
where $\Delta t$ is the timestep, $a_i^\alpha$ is the total acceleration on node $i$, $(v_i^\alpha)^{1/2} = (v_i^\alpha)^0 + (a_i^\alpha)^0 \Delta t/2$ is the half-timestep velocity, and $\Delta u_i = u_i^1 - u_i^0$ is the specific thermal energy change.  There are a number of possibilities we could choose for how to construct $\Delta u_i$ such that \cref{eq:TotalEBalance} is met.  One natural approach is to consider the pair-wise work contribution between any interacting pair of nodes $i$ and $j$.  We can express the desired total thermal energy change of the system in terms of the pair-wise interactions as
\begin{align}
  \DE &= \sum_i m_i \Delta u_i 
       =  -\sum_i m_i (v_i^\alpha)^{1/2} (a_i^\alpha)^0 \Delta t \nonumber \\
      &= -\sum_i m_i (v_i^\alpha)^{1/2} \left( \sum_j (a_{ij}^\alpha)^0 \right) \Delta t, \nonumber
\end{align}
where $a_{ij}^\alpha$ represents the pair-wise contribution to the acceleration of node $i$ due to node $j$.  The corresponding pair-wise contribution to the total work is
\begin{align}
  \label{eq:PairWork}
  \DEij &= m_i \Delta u_{ij} + m_j \Delta u_{ji} \\
        &= -\left( m_i (v_i^\alpha)^{1/2} (a_{ij}^\alpha)^0 \Delta t +
                m_j (v_j^\alpha)^{1/2} (a_{ji}^\alpha)^0 \Delta t \right) \nonumber \\
        &= m_i \left[ (v_j^\alpha)^{1/2} - (v_i^\alpha)^{1/2} \right]
           (a_{ij}^\alpha)^0 \Delta t, \nonumber
\end{align}
where $\Delta u_{ij}$ represents the specific thermal energy change of node $i$ due to its interaction with node $j$.  Note that in \cref{eq:PairWork} we have explicitly used the fact that pair-wise forces are anti-symmetric ($m_i a_{ij}^\alpha = -m_j a_{ji}^\alpha$).  This is not a required property to derive the compatible energy equation, it simply removes the necessity of referring to both $a_{ij}^\alpha$ and $a_{ji}^\alpha$ in the equation for node $i$.  Since both the SPH and CRKSPH formalisms we consider in this paper are symmetric in the pair-wise forces, we will use this simplification.

\Cref{eq:PairWork} represents the exact discrete pair-wise work due to the interaction of nodes $i$ and $j$, however, we still have to decide how to partition this work between these two nodes.  We parameterize this choice via $f_{ij}$
\begin{equation}
  \label{eq:uij}
  \Delta u_{ij} = f_{ij} \DEij/m_i
                = f_{ij} \left[ (v_j^\alpha)^{1/2} - (v_i^\alpha)^{1/2} \right] (a_{ij}^\alpha)^0 \Delta t,
\end{equation}
and exact conservation of the energy is guaranteed so long as $f_{ij} + f_{ji} = 1$.  In \cite{Owen2014} we present several choices for $f_{ij}$ and settle on a somewhat arbitrary form that tends to reduce the variation of energy between points as work is done.  We have since moved to a more physical form based on a function of the specific entropy per point
\begin{align}
  \label{eq:pointentropy}
  s_i &= P_i/\rho_i^\gamma, \;\;\;
  \smin = \min(|s_i|, |s_j|), \;\;\;
  \smax = \max(|s_i|, |s_j|) \\
  \label{eq:compfij}
  f_{ij} &= \left\{ \begin{array}{l@{\quad}l}
    1/2                   & |s_i - s_j| = 0, \;\text{otherwise} \\
    \smin/(\smin + \smax) & \Delta u_{ij} \ge 0 \;\text{and}\; s_i \ge s_j, \;\text{or}\; \Delta u_{ij} < 0 \;\text{and}\; s_i < s_j  \\
    \smax/(\smin + \smax) & \Delta u_{ij} \ge 0 \;\text{and}\; s_i < s_j, \;\text{or}\; \Delta u_{ij} < 0 \;\text{and}\; s_i \ge s_j  \\
  \end{array} \right.
\end{align}
This form of $f_{ij}$ tends to heat the cooler point, or cool the warmer point, depending on the sign of the work, sympathetically to our prior preferred method (Eq.~(28) in \cite{Owen2014}).  However, the new definition in \cref{eq:compfij} now bases the discrepancy in the heating based on how different the entropies of points $i$ and $j$ are, rather than our prior more ad-hoc approximations.  \Cref{eq:compfij} has proven more reliable for studying adiabatic problems, which are one of the main motivations when considering this compatible differencing approach vs.~simply evolving the total energy.

\Crefrange{eq:uij}{eq:compfij} completely define the new specific thermal energy $u_i$ for each point at the end of a time step.  We still use the derivative energy equation (\cref{eq:CRKDuDt}) to predict interim values of $u_i$ during a time advancement cycle, but the final energy per point is updated in this compatible manner in our default CRKSPH methodology.  The only complication to this approach is that it requires knowledge of the pair-wise accelerations $a_{ij}^\alpha$ as well as the mid-timestep velocity difference $(v_j^\alpha)^{1/2} - (v_i^\alpha)^{1/2}$ when updating the energy.  Thus, we need to either retain the pair-wise accelerations (i.e., extra memory) or recompute them (extra computation) when updating the energy.  In our current implementation, we choose to retain the pair-wise accelerations and burn the memory for the sake of computational speed on multi-core MPI distributed architectures.  It is possible that the more computationally demanding second choice of recomputing the pair-wise accelerations may see benefits on architectures such as GPU accelerated machines with limited memory and significant FLOPs to burn. 

\subsection{Summary and Additional Ingredients}
\label{sec:crkeqs}
We now have the major ingredients to construct a fully conservative differencing method based on reproducing kernels.  For clarity, we succinctly summarize the evolution equations derived above:
\begin{align}
  \label{eq:CRKDvDt}
  \frac{Dv^\alpha_i}{Dt} &= -\frac{1}{2 m_i} \sum_j V_i V_j (P_i + P_j + Q_i + Q_j) \left( \partial_\alpha \Wrij  - \partial_\alpha \Wrji \right) \\
  \label{eq:CRKcompE}
  u_i(t + \Delta t) &= u_i(t) + \sum_j \Delta u_{ij} \Delta t \\
  \label{eq:CRKcompE2}
  \Delta u_{ij} &= \frac{f_{ij}}{2} \left[ v_j^\alpha(t) + v_j^\alpha(t + \Delta t) - v_i^\alpha(t) - v_i^\alpha(t + \Delta t) \right] \frac{Dv_{ij}^\alpha}{Dt} \\
  \label{eq:CRKfij}
  f_{ij} &= \left\{ \begin{array}{l@{\quad}l}
    1/2                   & |s_i - s_j| = 0, \;\text{otherwise} \\
    \smin/(\smin + \smax) & \Delta u_{ij} \ge 0 \;\text{and}\; s_i \ge s_j, \;\text{or}\; \Delta u_{ij} < 0 \;\text{and}\; s_i < s_j  \\
    \smax/(\smin + \smax) & \Delta u_{ij} \ge 0 \;\text{and}\; s_i < s_j, \;\text{or}\; \Delta u_{ij} < 0 \;\text{and}\; s_i \ge s_j  \\
  \end{array} \right. \\
  \label{eq:sminmax}
  \smin &= \min(|s_i|, |s_j|), \;\;\;
  \smax = \max(|s_i|, |s_j|) \\
  \label{eq:CRKvisc}
  Q_i &= \rho_i \left(-C_l c_i \mu_i + C_q \mu_i^2 \right) \\
  \label{eq:CRKmu}
  \mu_i &= \min\left(0, \frac{\vhatij^\alpha \eta_i^\alpha}{\eta_i^\alpha \eta_i^\alpha + \epsilon^2}\right) \\
  \label{eq:CRKvhat}
  \vhatij^\alpha &= v_i^\alpha - v_j^\alpha - \frac{\phi_{ij}}{2}\left(\partial_\beta v_i^\alpha + \partial_\beta v_j^\alpha  \right) \xij^\beta  \\ 
  \phi_{ij} &= \max\left[0, \min\left[1, \frac{4 r_{ij}}{(1 + r_{ij})^2}\right]\right] \times
  \left\{ \begin{array}{l@{\quad}l}
    \exp\left\{-\left((\eta_{ij} - \etacrit)/\etafold\right)^2\right\}, & \eta_{ij} < \etacrit \\
    1, & \eta_{ij} \ge \etacrit \\
  \end{array} \right. \\
  \label{eq:CRKrij}
  r_{ij} &= \frac{\partial_\beta v_i^\alpha \xij^\alpha \xij^\beta}{\partial_\beta v_j^\alpha \xij^\alpha \xij^\beta}, \; \; \eta_{ij} = \min(\eta_i, \eta_j) \\
  \label{eq:CRKDvDx}
  \partial_\beta v_i^\alpha &= -\sum_j V_j \vij^\alpha \partial_\beta \Wrij \\
  \label{eq:CRKvol}
  V_i^{-1} &= \sum_j \Wi \\
  \label{eq:CRKrho}
  \rho_i &= \frac{\sum_j m_{ij} V_j \Wrij}{\sum_j V_j^2 \Wrij}, \; \;
    m_{ij} \equiv \left\{ \begin{array}{l@{\quad}l}
      m_j, & \text{$i$ and $j$ same material} \\
      m_i, & \text{$i$ and $j$ different materials} \\
    \end{array} \right.
\end{align}
where \cref{eq:CRKDvDt} is the conservative momentum relation derived in \cref{sec:crkmomderiv}, \crefrange{eq:CRKcompE}{eq:sminmax} the compatible energy update from \cref{sec:compwork}, and \crefrange{eq:CRKvisc}{eq:CRKDvDx} the limited artificial viscosity of \cref{sec:limitedq}.  Note in \cref{eq:CRKvhat} we explicitly use the fact our chosen $\phi_{ij}$ is symmetric with respect to $i$ and $j$ for simplification; if a non-symmetric limiter is chosen, the full expressions of \crefrange{eq:viLim}{eq:vDiff} must be used instead.  $s_{i} = P_i/\rho_i^\gamma$ is the entropic function based on the point-wise specific entropy, used in the definition of the work sharing term $f_{ij}$ (\cref{eq:CRKfij}).  Our standard set of parameters for the viscosity used throughout this paper are $C_l=2$, $C_q=1$, $\epsilon^2=10^{-2}$, $\etacrit=1/n_h$ (where $n_h$ is the expected number of nodes per smoothing scale in one dimension), and $\etafold=0.2$. Furthermore, unless otherwise specified, the equation of state assumed for our tests is that of an ideal gas, viz.
\begin{align}
\label{eq:EOS}
P = (\gamma-1)\rho u,
\end{align}
where $\gamma$ is the specific heat ratio. 

We have introduced two aspects of the algorithm left as free parameters to this point: how to define the volume per particle (\cref{eq:CRKvol}) and the mass density update (\cref{eq:CRKrho}).  These choices are not necessarily independent or unique -- our primary desire was to come up with a summation form for the mass density that takes advantage of the improved accuracy of our RK kernel basis.  The usual cancellation of $\rho$ from inside the SPH interpolation to get the ordinary SPH density equation, $\Interp{\rho} = \sum_j (m_j/\rho_j) \rho_j \Wij = \sum m_j \Wij$, does not occur with CRKSPH since we are not necessarily using $V_j = m_j/\rho_j$.  However, we can easily enough adapt the interpolated mass divided by interpolated volume definition of \cref{eq:CRKrho} based on an earlier investigation of alternative SPH mass density forms from \cite{Owen2011}.  Much hinges upon how we define the volume per point $V_i$; this is critical for both the mass density definition as well as determining how the RK weighting will be established.  The volume relation of \cref{eq:CRKvol} is the inverse of the SPH number density.  We do not want a function of the mass density in the volume definition or \cref{eq:CRKrho} would become iterative. That being said, \Cref{eq:CRKvol} certainly has its shortcomings; it will suffer the ordinary SPH oscillations and errors, with its largest errors near surfaces.  Likely, we will need improved relations for examining multiple-materials and solids interacting across/with surfaces, but for the continuous fluid problems we examine in this paper these relations suffice.  One multiple-material aspect of \cref{eq:CRKrho} should be pointed out, however. The equation will exactly interpolate the density of a static multi-material boundary when the particle spacing is uniform -- see the hydrostatic box problem in \cref{sec:HydroBox} for an example.  Aside from this one aspect, we leave the proper treatment of multiple-material problems for future work.

Although not utilized as part of the primary state update in CRKSPH, we note a few equations that are useful for alternative testing and calculating mid-step estimates required for multi-step time integration methods.  For such mid-step estimates of the thermal energy and mass density, we use the relations
\begin{align}
  \label{eq:CRKDuDt}
  \frac{Du_i}{Dt} &= \frac{1}{2 m_i} \sum_j V_i V_j (P_j + Q_j) \vij^\alpha \left( \partial_\alpha \Wrij - \partial_\alpha \Wrji \right) \\
   \label{eq:contEqn}
  \frac{D\rho_i}{Dt} &= -\rho_i \paralpha v_i^\alpha.
\end{align}
It is also possible to derive a relation for the total energy evolution should we desire to replace the compatible energy update of \crefrange{eq:CRKcompE}{eq:sminmax} with a total energy method
\begin{equation}
  \label{eq:CRKDEDt}
  \frac{DE_i}{Dt} = m_i v_i^\alpha \frac{Dv_i^\alpha}{Dt} + \half \sum_j V_i V_j (P_j + Q_j) \vij^\alpha \left( \partial_\alpha \Wrij - \partial_\alpha \Wrji \right).
\end{equation}
We do not use \cref{eq:CRKDEDt} in this paper other than for one comparison in \cref{sec:KidderTest} to demonstrate why we prefer the compatible energy update. 

Lastly, we have not yet specified our choice for the interpolation kernel basis $W$.  Just as in ordinary SPH this is an arbitrary parameter, and we have experimented with several forms.  We have settled on a member of the B-spline family due to \cite{Schoenberg1969}, in this case the seventh-order kernel $W_7$.  The general family of B-splines of order $k$ can be compactly written as
\begin{align}
  \label{eq:nbspline}
  W_k(\eta) &= A_\nu h^{-\nu} (k!)^{-1} \sum_{i=0}^{k+1} (-1)^i \binom{k+1}{i} \left(\eta - i + \frac{k + 1}{2}\right)_+^k \\
  \label{eq:AW7}
  k &= 7 \;\; \Rightarrow \;\; A_\nu = \left\{\begin{array}{l@{\quad}l}
  1               & \nu = 1 \\
  2268/(1487 \pi) & \nu = 2 \\
  3/(4 \pi)       & \nu = 3 \\
  \end{array}\right.
\end{align}
where $(\ldots)_+^k$ is the so-called ``one-sided power function,'' implying for arguments less than zero the result is zero, but for positive values the exponent power $k$ is applied. The sampling radius (i.e., the radius at which the kernel falls to zero) of a B-spline of order $k$ is $\eta_{\max}=(k+1)/2$; therefore, our chosen seventh-order kernel has a compact support radius of $\eta_{\max}=4$.  $A_\nu$ is the SPH normalization such that the volume integral $\int_V dV \, W=1$ in $\nu$ dimensions; for the case of the our implemented kernel $W_7,$ \cref{eq:AW7} lists these constants.  Note that for $k=3$ \cref{eq:nbspline} also generates the cubic B-spline which has long been popular in SPH \cite{Monaghan2005}, while $k=5$ generates a quintic kernel that more recent researchers have found useful for reasons such as accuracy and resistance to artificial clumping of the SPH points.  We have successfully used both cubic and quintic B-splines with CRKSPH (as well as the Wendland kernels described in \cite{WalterAly2012}), but overall we have found that the $k=7$ seventh-order kernel provides the best results, and use it throughout this paper. The choice of kernel is a free parameter in CRKSPH, and the results are not strongly dependent on this selection. Truly any reasonable kernel could be used for this method, and we compare alternative choices in \cref{sec:Wchoice}.

\section{Evaluation}
\label{sec:crktests}
We organize the evaluation of the CRKSPH framework by the physical mechanisms at play in the various tests we examine. Our analysis is designed to investigate convergent behavior, numerical robustness, retention of SPH conservation properties, reduction of inherent SPH errors, and improvements gained with our viscosity treatment. All of our tests are drawn from examples in the literature for their applicability to a wide range of physical problems across multiple fields within physics. When possible these tests are constructed in a manner consistent with their presentation in prior methods papers in order to facilitate comparison with those works.  We also present results using standard SPH implementations, if and where appropriate, to elucidate improvements or comparable performance. In the following discussion, PSPH refers to the pressure-discretized SPH formulation of \cite{Hopkins2012,Hopkins2015,Saitoh2013}. CompSPH refers to the standard formulation of SPH with the addition of the compatible energy evolution described above and in \cite{Owen2014}. Where applicable, we also present results using compSPH with the Cullen viscosity prescription \cite{Cullen2010} referred to in plots and text as ``compSPH+Cullen''. We summarize the equations for compSPH, the Cullen viscosity, and PSPH in \cref{sec:compSPH}, \cref{sec:CDHvisc},  and \cref{sec:PSPH}, respectively. 

For the most part, we present compSPH and CRKSPH comparisons using similar numerical parameters.  The places where they differ are in the viscosity coefficients of \cref{eq:visc,eq:CRKvisc} ($C_l=C_q=1$ for compSPH vs.~$C_l=2, C_q=1$ for CRKSPH) and choice of the base interpolation kernel $W$ (the fifth-order spline $W_5$ from \cref{eq:nbspline} for compSPH vs.~seventh-order $W_7$ for CRKSPH). The kernel choice for CRKSPH does not have a large effect for most problems (see discussion in \cref{sec:Wchoice}), owing to the fact that any arbitrary kernel can be made first-order accurate in the RK framework -- $W_7$ has merely proven to yield the overall best results, as does $W_5$ in the compSPH schema.  For PSPH, we strive to exactly reproduce the method outlined in appendix F2 of \cite{Hopkins2015} (repeated in our \cref{sec:PSPH}); of particular note, is the quintic kernel given in \cref{eq:Wfive}, which is distinct from the quintic form of \cref{eq:nbspline} due to \cite{Schoenberg1969} only in the renormalization of the extent to fall to zero at $\eta=1$.  To preserve the same number of neighbors sampled per point between all methods, we adjust the resolution scale $h$ such that each method maintains a constant radial number of 4 neighbors -- thereby equalizing resolution and computational expense. 
This choice evaluates to a total of 268 neighbors per point in 3D, and 50 neighbors in 2D.  In our code, this radial number of neighbors is parameterized as the effective number of points per smoothing scale, or $n_h$.  The quintic form of \cref{eq:nbspline} has a maximum extent $\eta_{\max}=3$, corresponding to $n_h=4/3$ for compSPH.  For the seventh-order kernel used in CRKSPH, $\eta_{\max}=4$, coinciding with $n_h=1$.  The quintic kernel applied in PSPH given by \cref{eq:Wfive} has $\eta_{\max}=1$, resulting in $n_h=4$.  The details of how the smoothing scale $h$ is updated in our implementation may be found in \cite{Owen2010}; in brief, our methodology strives to optimize the total kernel weight sampled at each point rather than maintain a strict number of neighbors, so the expected neighbor counts quoted here are approximate.
For time-stepping in all solvers, we use a modified second-order Runge-Kutta advance with a Courant-Friedrichs-Lewy (CFL) coefficient of 0.25.



\subsubsection{The Kidder Isentropic Compression Test Case}
\label{sec:KidderTest}
In order to demonstrate the utility of the compatible energy update outlined in \cref{sec:compwork}, it is informative to compare it against the result of evolving the total energy (\cref{eq:CRKDEDt}) on an adiabatic test problem.  In this section, we examine an idealized isentropic implosion described in \cite{Kidder1976, Maire:2009dr}.  This test consists of an isentropic (shockless) implosion of an ideal gas shell, which allows us to obtain an analytic solution at all times and radii, as well.  There are solutions for planar, cylindrical, and spherical geometries corresponding to our 1D, 2D, and 3D CRKSPH methodologies; here we consider the 1D planar case.

This test problem has a self-similar solution we can describe as follows.  Consider a spherical shell initially in the radial range $r \in [\rinner, \router]$.  Assume $(\rhoinner, \Pinner)$ are initial density and pressure at the inner radius $\rinner$, and $(\rhoouter, \Pouter)$ the corresponding initial values at the outer radius $\router$.  We assume an isentropic compression with fixed entropy $\sshell = \Pinner/\rhoinner^\gamma$, implying $\rhoinner = (\Pouter/\Pinner)^{1/\gamma} \rhoouter$.  We also assume an adiabatic constant $\gamma = 1 + 2/\nu$ where $\nu$ is dimensionality such that $\nu = 1$ corresponds to the planar solution, $\nu = 2$ the cylindrical, and $\nu = 3$ spherical.  The final self-similar solution describing the evolution of the shell is given as
\begin{align}
  R(r,t) &= a(t) r \\
  \rho(R(r,t), t) &= a(t)^{-2/(\gamma - 1)} \rho_0(r) \frac{R(r,t)}{a(t)} \\
  P\left(R(r,t), t\right) &= a(t)^{-2\gamma/(\gamma - 1)} P_0(r) \frac{R(r,t)}{a(t)} \\
  v(R(r,t), t) &= \partial_t a(t)  \frac{R(r,t)}{a(t)} 
\end{align}
for points at radius $R(r,t)$ (with initial radius $r$) and time $t \in [0, \tau]$, where $\tau$ is the focusing time of the shell
\begin{equation}
  \tau = \sqrt{\frac{\gamma - 1}{2 \sshell^2 \gamma^2}\frac{\router^2 - \rinner^2}{\rhoouter^{2(\gamma - 1)} - \rhoinner^{2(\gamma - 1)}}},
\end{equation}
and the scaling function
\begin{equation}
  a(t) = \sqrt{1 - \left (\frac{t}{\tau} \right)^2},
\end{equation}
which is valid for $t \in [0, \tau]$.  The initial density, pressure, and velocity profiles are
\begin{align}
  \rho_0(r) &= \left(\frac{\router^2 - r^2}{\router^2 - \rinner^2} \rhoinner^{\gamma - 1} +
                     \frac{r^2 - \rinner^2}{\router^2 - \rinner^2} \rhoouter^{\gamma - 1} \right)^{1/(\gamma - 1)} \\
  P_0 &= s \left( \rho_0(r) \right)^\gamma \\
  v_0 &= 0.
\end{align}

The main difficulty of this problem for SPH-like calculations is that the isentropic solution requires time varying pressures at the inner and outer surfaces:
\begin{align}
  P(R(\rinner, t)) &= \Pinner a(t)^{-2\gamma/(\gamma - 1)} \\
  P(R(\router, t)) &= \Pouter a(t)^{-2\gamma/(\gamma - 1)}.
\end{align}
Our solution to this issue is to enforce the analytic solution on a sufficient set of nodes on each end of the shell (we choose a rind of 10 such points from each end) such that the points free to evolve never interact with the boundaries of the shell.

\begin{figure}[ht]
\centering
\includegraphics[width=0.45\textwidth]{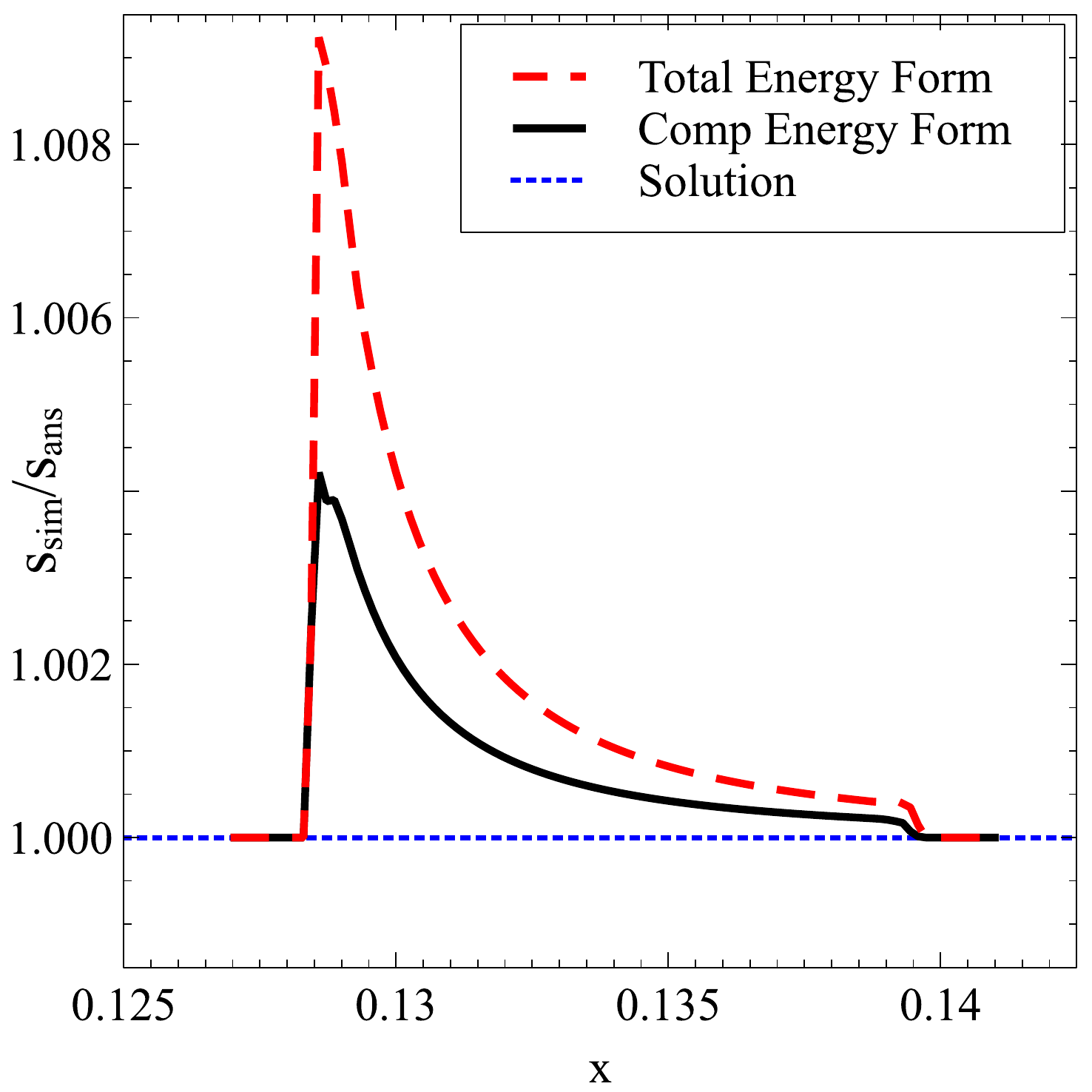}
\caption{Ratio of the simulated to expected entropy across the shell for the planar Kidder isentropic implosion at $t = 0.99\tau$, using $N=100$ particles.  Both results are CRKSPH using different methods of updating the energy: red (dashed) evolves the total energy relation $DE_i/Dt$ using \cref{eq:CRKDEDt}; black (solid) uses the compatible energy update (\cref{eq:compfij,eq:uij}).  The thin blue (dotted) line along $s_{\text{sim}}/s_{\text{ans}}=1$ shows the analytic solution. The compatible update is roughly more accurate by a factor of 2 when compared to differencing the total energy for this problem.}
\label{fig:KidderEntropy}
\end{figure}
Following the example shown in \cite{Maire:2009dr}, we start with $(\rinner, \router) = (0.9, 1)$, $(\Pinner, \Pouter) = (0.1, 10)$, and $\rhoouter = 0.01$.  We advance to $t = 0.99\tau$, yielding a compression of just over a factor of seven.  We compare two models using CRKSPH: one evolving the total energy $DE_i/Dt$ using \cref{eq:CRKDEDt}, and the other using the compatible energy update (\cref{eq:compfij,eq:uij}).  Each is modeled with 100 points (of which the inner and outer 10 points are used to enforce the boundary conditions).  \Cref{fig:KidderEntropy} plots radial profiles of the ratio of simulated to expected entropy at the final time.  It is evident that the compatible energy advance is more accurate by roughly a factor of 2 in maintaining the proper entropy: the minimum to maximum entropy error for the total energy mode is roughly 0.9\% vs.~0.4\% for the compatible model.  The CPU time is nearly identical between the two approaches due to our choice of storing the pair-wise accelerations for use in the compatible update.

Why does the compatible energy update fare better on these adiabatic problems compared with the total energy model?  Both methods conserve total energy to machine precision, so at first blush one might think they should be nearly identical.  The critical difference is that the compatible methodology uses more information about the pair-wise work.  \Cref{eq:PairWork} is philosophically similar to the energy evolution one would get by directly using the specific thermal energy update relation (\cref{eq:CRKDuDt}), save the following distinction: instead of differencing the continuous equation for $Du/Dt,$ thereby allowing the discretization error to creep into the energy evolution,  we are precisely accounting for the work done by the discretized pair-forces evaluated from the momentum relation in \cref{eq:CRKDvDt}.  Thus, the thermal evolution will benefit from the compatible differencing formalism when considering adiabatic problems.

Viewed in another way, consider that inferring the thermal energy as the difference between the total and kinetic energies allows the error of both those relations to be exacerbated in the thermal evolution. This problem will be at its most egregious when the thermal energy represents a small fraction of the total energy, and therefore, finding the thermal energy translates to finding a small number as the difference of two large ones. Henceforth, all further tests of CRKSPH utilize the compatible update for the energy evolution. 

\subsection{Acoustic Wave}
\label{sec:acoustic}
\begin{figure}[ht]
\centering
\includegraphics[width=0.45\textwidth]{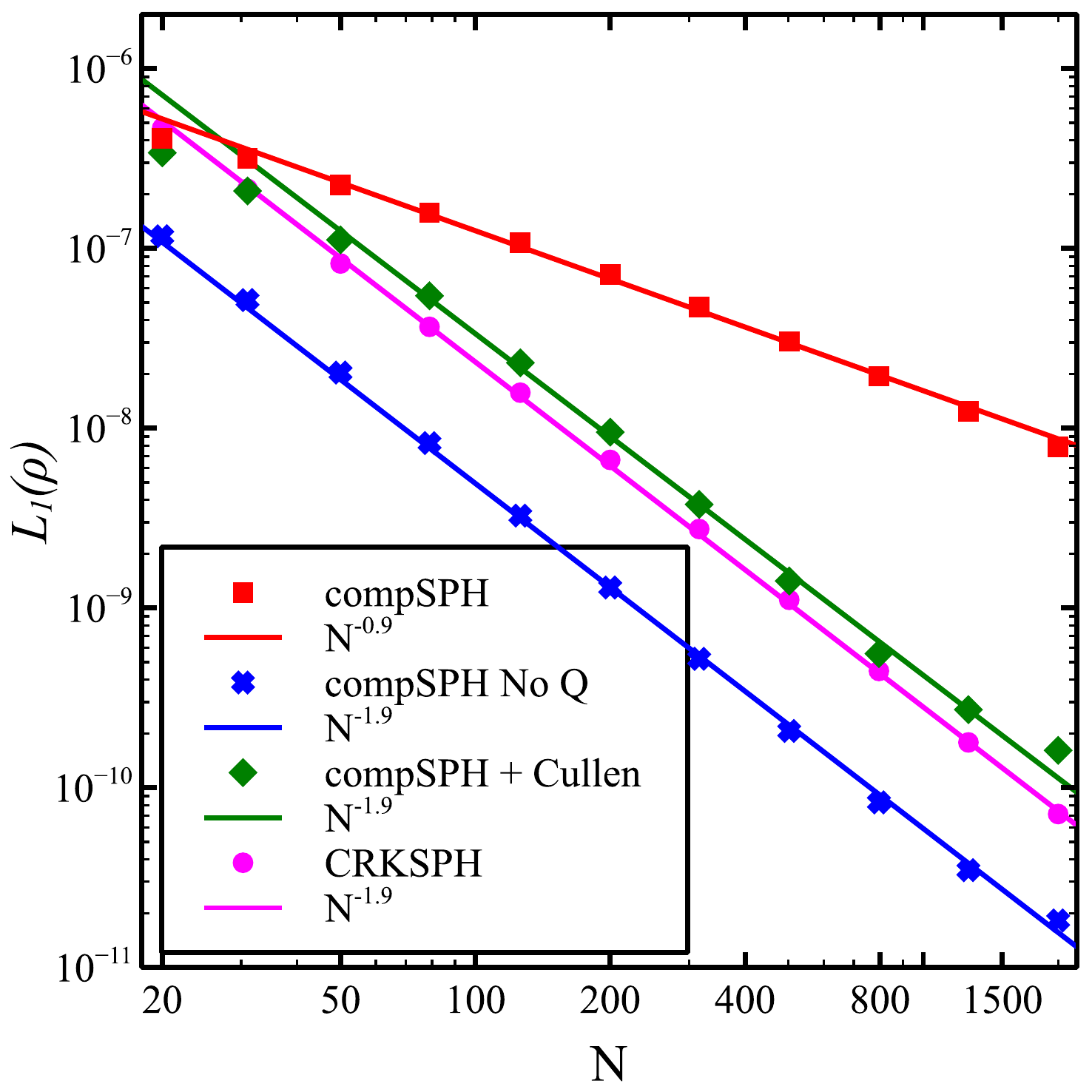}
\caption{The convergence of the density $L_1$ norm as a function of resolution $N$ in the 1D acoustic wave test for compSPH with Monaghan and Gingold viscosity, Cullen viscosity, and no viscosity at all, compared to the CRKSPH convergence rate. All methods demonstrate second-order convergence besides the Monaghan viscosity variation, which does not contain any explicit viscosity suppression, rendering it first-order. Thus, the simple viscosity limiter in CRKSPH correctly deactivates the viscosity in smooth-flow, achieving the proper theoretical convergence rate for this solver. }
\label{fig:acoustic}
\end{figure}
The evolution of a sound wave is a popular \cite{Stone2008, Springel2010, Hopkins2015} test for hydrodynamic solvers, owing to its smooth continuous solution that should demonstrate convergence at the theoretical maximum rate of a given method. Here, we examine the propagation of an acoustic wave for a ($\gamma=5/3$) gas inside a 1D periodic unit box, with unit density ($\rho_0 = 1$) and sound speed ($c_s=1$); a sinusoidal perturbation of unit wavelength ($\lambda = 1$) with amplitude $A=10^{-6}$ defines the sound wave, viz.
\begin{align}
 \label{eq:awaveqn}
 \rho_i = \rho_0 + \delta_i, \; v_i = c_s \delta_i, \; P_i = P_0 + \delta_i, \; \delta_i =  A \sin \left(\frac{2\pi x_i }{\lambda}\right)
\end{align}
where the background pressure $P_0$ is determined by the ideal gas equation of state, namely $P_0 = c^2_s \rho_0 / \gamma = 3/5$.
We model this scenario using a variety of particle counts $N \in (20, 31, 50, 59, 126, 200, 316, 502, 796, 1261, 2000)$ (chosen for roughly equal logarithmic steps) in order to examine the convergence with spatial resolution.  Analytically, the fluid evolution is simple -- the propagating wave returns to its initial condition after each period. We measure numerical convergence in the $L_1$ norm of the density as
\begin{align}
  \label{eq:L1}
  L_1(\rho) = \frac{1}{N} \sum_i |\rho_i - \rho(x_i)|,
\end{align}
where N is the total number of particles, $\rho_i$ is the density of the $i^{th}$ node, and $\rho(x)$ is the analytical solution. \Cref{fig:acoustic} shows this norm as a function of N for a variety of models at $t=5$, i.e., after the wave has propagated around our periodic volume five times. We examine four methods: ordinary compSPH, compSPH with zero viscosity ($C_l=C_q=0$ in \cref{eq:visc}), compSPH with the Cullen-Dehnen viscosity modification (\cref{sec:CDHvisc}), and CRKSPH. Considering that both compSPH and CRKSPH are nominally second-order in space, we would expect to achieve second-order convergence for this problem. As we can see in \cref{fig:acoustic}, three of our tests do achieve the predicted convergence rate: compSPH with zero viscosity, compSPH with the Cullen modified viscosity, and CRKSPH. However, compSPH without explicit viscosity suppression is limited to first-order. This is attributed to the fact that the ordinary Monaghan-Gingold viscosity of \cref{eq:visc} is activated for any compressing flow.  The sound-wave has compressional regions where the viscosity is erroneously triggered, despite the absence of a shock, yielding extra dissipation that cuts the convergence rate to first-order. Reassuringly, explicitly suppressing the viscosity in compSPH by setting the viscous coefficients to zero, or applying the sophisticated viscosity limiters of Cullen \& Dehnen, recovers second-order convergence in this problem, supporting this interpretation. This problem also demonstrates our simple viscosity limiter in CRKSPH correctly eliminates the viscosity in this scenario, and allows CRKSPH to achieve second-order convergence as one would expect.

\subsection{Hydrostatic Box}
\label{sec:HydroBox}
To illustrate the SPH error incurred near density jumps, we analyze an idealized square contact discontinuity consisting of a high-density box of gas in pressure equilibrium with a low-density background \cite{Hess2010,Saitoh2013,Hopkins2015}. Our domain consists of a unit length periodic box, filled with a $\gamma=1.5$ gas at unit initial pressure and zero velocity. The innermost volume $(x,y) \in ([1/4,3/4], [1/4,3/4])$ is created with an initial density $\rho_{\text{box}}=4$, while the surrounding gas is initialized at a lower value of $\rho_{\text{medium}}=1$.  Since the problem is in pressure equilibrium we expect no evolution.

\begin{figure}[ht]
\centering
\includegraphics[width=0.95\textwidth]{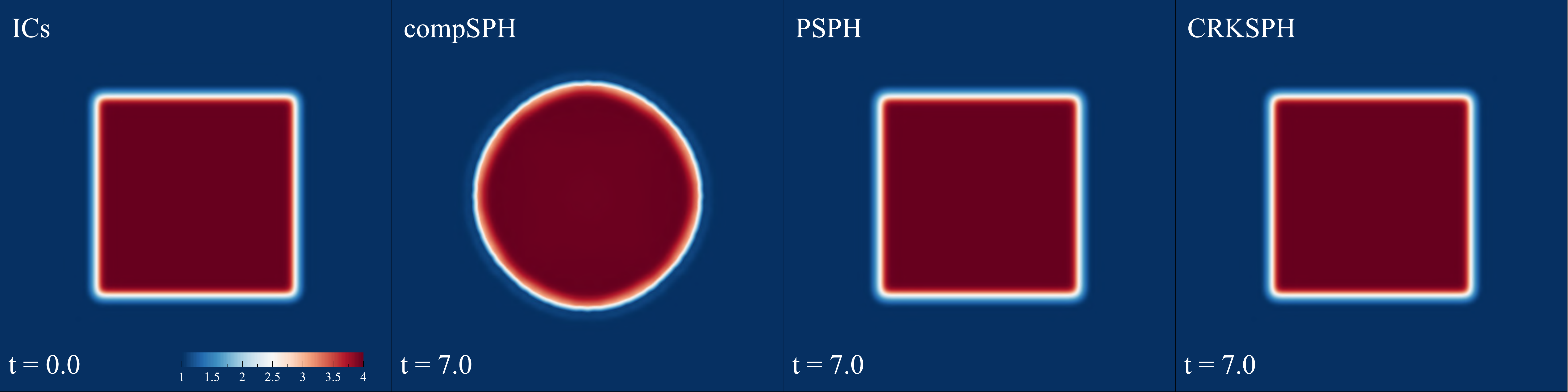}
\caption{The initial and final conditions (t=0, t=7) of the hydrostatic box test at resolution $N=100^2,$ for compSPH, PSPH, and CRKSPH, respectively. Artificial ``surface-tension'' errors at the contact discontinuity, incurred from assuming continuity in the material density,  cause classical SPH to deform the square into a circle. Both PSPH and CRKSPH maintain the equilibrium. PSPH avoids the tension error by discretizing in pressure (a smooth quantity in this problem), whereas CRKSPH uses corrected kernels to accurately interpolate the interface. }
\label{fig:box}
\end{figure}
\Cref{fig:box} compares the evolved hydrostatic box using $100^2$ points evenly seeded on an initial lattice. The left panel shows the initial condition in density, while the three succeeding panels show the final state at $t=7$, obtained by compSPH, PSPH, and CRKSPH.
The well-known erroneous result using ordinary SPH (represented by compSPH here) derives from a spurious numerical ``surface-tension'' like force due to the discontinuous nature of the pair-wise weighting of the SPH points across density separation boundaries.  Ordinary SPH discretizes volume based on the mass density, and, thereby, inherently assumes material continuity on the length scale of the sampling volume \cite{Hopkins2012,Saitoh2013}.  Contact discontinuities violate this underlying density smoothness assumption, leading to spurious forces that mimic an artificial ``surface-tension.''  Here, the resulting deformation minimizes surface area, where the high-density region transforms into a circle -- as seen in the compSPH panel of \cref{fig:box}. PSPH skirts this issue by weighting the point-to-point interactions by a function of the pressure, which is uniform in this case.  CRKSPH also avoids the problem because the corrected kernel is able to interpolate accurately across the surface. 

Traditional mesh-based hydrodynamic methods also trivially pass this test, however, 
it should be noted that such algorithms have truncation errors that are not Galilean invariant. Thus, unresolved mesh-based simulations can become corrupted if the entire system is given a constant translational motion (as demonstrated in \cite{Hopkins2015} among other studies). Each of the methods considered here (compSPH, PSPH, and CRKSPH) are Galilean invariant, and, therefore, these results are maintained regardless of any arbitrary boost given the frame of the problem.

Finally, it is important to recognize that the initial particle distribution for the hydrostatic box is equally spaced where our density evaluation (\cref{eq:CRKrho}) exactly evaluates the correct constant densities -- confirmed by the CRKSPH results in \cref{fig:box}. The traditional SPH sum density definition found in \cref{eq:SPHSum} does not have this property, and, if used, averages the density across the boundary, leaking error into the pressure evaluations. Nevertheless, \cref{eq:CRKrho} is not a perfect solution for surfaces with arbitrary particle geometries, as demonstrated in \cref{sec:consist} using the box test initialized with equal particle mass, where both the density inaccuracy and the inconsistency error described at the end of \cref{sec:crkmomderiv} degrade the equilibrium (though the results remain substantially superior to SPH). This deficiency highlights the need for a more sophisticated density (and volume treatment), to properly handle the surfaces of multi-material phenomena. We defer such multi-material related issues for future work.

\subsection{Shock Phenomena}
In this section, we consider a number of standard hydrodynamic test cases dominated by shocks.  The presence of shocks violates the inviscid assumption of the ordinary discretized fluid equations (i.e., \cref{eq:momeq2,eq:engeq2}) requiring the addition of artificial viscosity (the $Q_i, Q_j$ terms in \cref{eq:CRKDvDt}) to adequately model this phenomena.  The goals of this section are twofold: (1) we seek to examine how well CRKSPH handles strong shock phenomena in general, and (2) to illustrate the benefits of our simple limited artificial viscosity described in \cref{sec:limitedq}, particularly when compared with more complex artificial viscosity prescriptions, such as the method of Cullen \& Dehnen \cite{Cullen2010,Hopkins2015}. To that end, we consider three classic shock test cases: the Sod shock tube \cite{Sod1978} in \cref{sec:Sod}, the Sedov-Taylor blastwave \cite{Sedov1959,Taylor1950a} in \cref{sec:Sedov}, and the Noh implosion \cite{Noh1987} in \cref{sec:Noh}. Each of these problems have analytic solutions, extendable to two and three-dimensions in the cases of Noh and Sedov.

\subsubsection{Sod Problem}
\label{sec:Sod}
\begin{figure}[ht]
  \centering
  \includegraphics[width=0.55\textwidth]{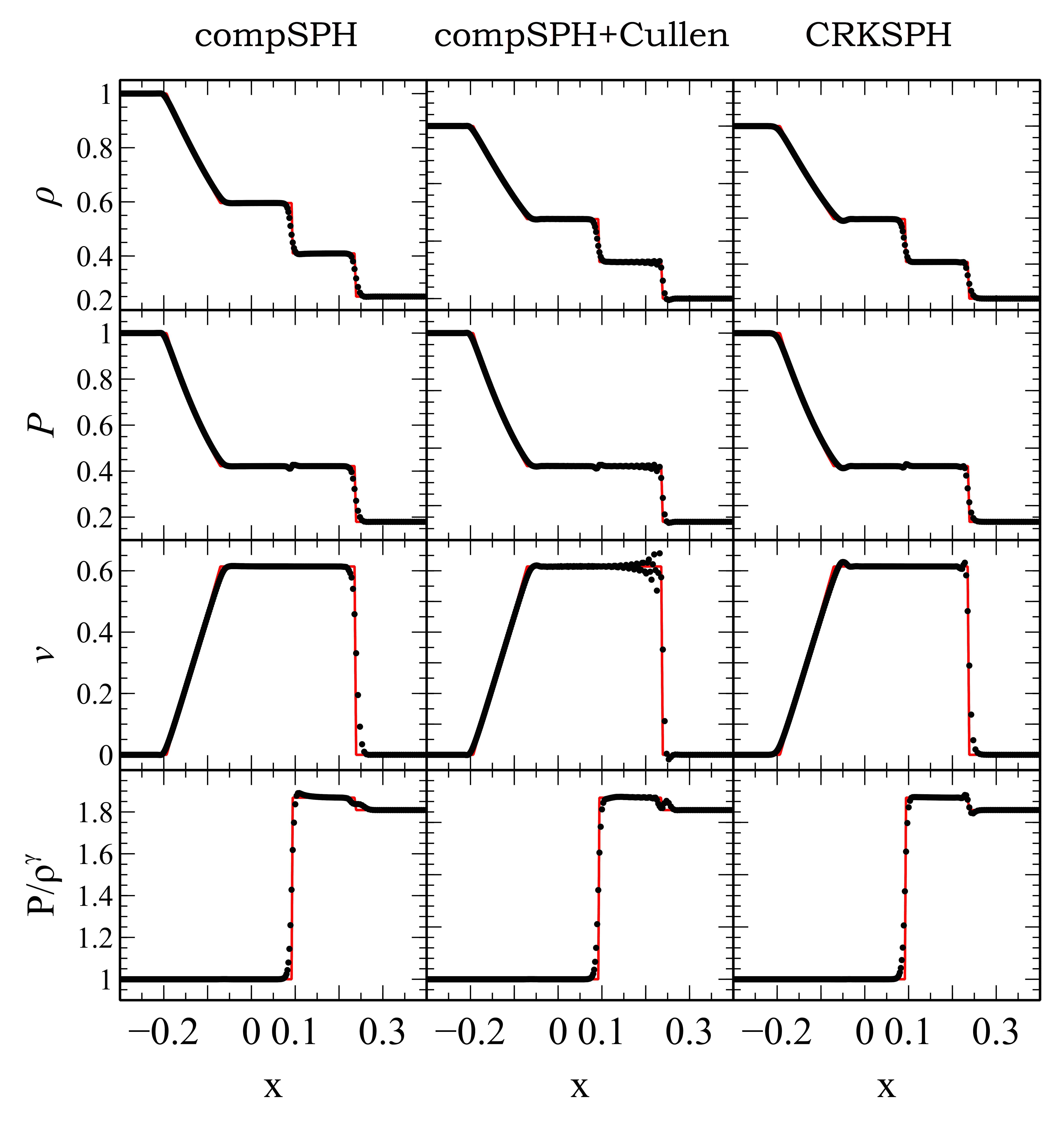}
  \caption{A comparison of the results of the Sod shock tube test run in 1D for compSPH (left column), compSPH with the Cullen-Dehnen viscosity (middle column), and CRKSPH (right column), at $t=0.15$. The simulations were run with equal mass particles, namely ($N_{\text{high}},N_{\text{low}}$) = (400,100), for the high and low pressure regions.  The analytical expectations are plotted (red) with lines. Top-to-bottom, we plot the density $\rho$, pressure $P$, velocity $v$ and entropy $P/\rho^\gamma$ for each solver.  Due to the non-limited viscosity model, compSPH demonstrates the most diffusion. That being said, all three solvers perform quite well on this simple problem. In general, CRKSPH resolves the shock transition region to the same fidelity as the Cullen-Dehnen algorithm, while suffering fewer ringing artifacts.}
  \label{fig:sod-planar}
\end{figure}
The 1D Sod problem \cite{Sod1978} is a shock tube test, in which two initially uniform gaseous regions with different initial pressures are brought into contact. A shock propagates into the initially lower-pressure region, while a rarefaction wave travels into the high-pressure gas.  A commonly tested instance of this problem used to benchmark SPH codes \cite[\dots]{Hernquist1989, Rasio1991, Springel2005, Springel2010, Hopkins2015} consists of the high pressure region on the left side ($x_{\text{high}} \in [-0.5,0]$) of the domain with $(\rho_{\text{high}}, P_{\text{high}}) = (1.0, 1.0)$, and a low pressure gas on the right ($x_{\text{low}} \in [0,0.5]$) region with initial conditions $(\rho_{\text{low}}, P_{\text{low}}) = (0.25, 0.1795)$. The velocity is initialized to be zero in both regions.  In this experiment we use a $\gamma=5/3$ ideal gas, with equal mass particles numbering ($N_{\text{high}},N_{\text{low}}$) = (400,100) in the two domains, where we impose reflective boundary conditions.  In order to create continuous initial conditions across the interaction boundary, we initialize the density and pressure profiles according to
\begin{align}
  \label{eq:Sod_initial}
  \rho(x) &= \rho_{\text{high}} + (\rho_{\text{high}} - \rho_{\text{low}}) \left[1 + e^{-x/\Delta_{\max}}\right]^{-1} \\
  P(x) &= P_{\text{high}} + (P_{\text{high}} - P_{\text{low}}) \left[1 + e^{-x/\Delta_{\max}}\right]^{-1}
\end{align}
where $\Delta_{\max} = \max(\Delta x_{\text{high}}, \Delta x_{\text{low}})/2$ and $(\Delta x_{\text{high}}, \Delta x_{\text{low}})$ is the unperturbed spacing on the left and right of the initial discontinuity. We maintain constant mass points and perturb their spacing in order to reproduce the profile of \cref{eq:Sod_initial}. We note that although the smoothed conditions used here were intended to be consistent with the SPH continuity assumptions of the state variables, we have found virtually identical results running with discontinuous interfaces for this problem as well, and the reader is encouraged to compare the results with publishings that have utilized the step-like initial conditions (e.g. \cite{Hernquist1989, Springel2005,Hopkins2015}).

\Cref{fig:sod-planar} demonstrates the numerical results of the Sod test performed in 1D using compSPH, compSPH+Cullen, and CRKSPH compared to the theoretical solution at $t=0.15$. The analytic Riemann solution consists of three regions: (1) a smooth rarefaction roughly in the domain $x \in [-0.2,-0.1]$), (2) a contact discontinuity at $x\approx0.1$, and (3) a shock at $x\approx0.25$. As is evident from \cref{fig:sod-planar}, CRKSPH fares similarly to compSPH+Cullen, capturing the shock in fewer points compared to the non-limited viscosity used in compSPH. However, CRKSPH demonstrates less post-shock ringing in the velocity when compared to compSPH+Cullen, as well as a more accurate solution for the shocked region in the entropy ($P/\rho^\gamma$).  CRKSPH does show a slightly more of an overshoot in velocity at the trailing edge of the rarefaction ($x\approx-0.1$), but the differences are small.  Overall CRKSPH solves this fairly mild shock problem well, capturing the shock transition region to the same fidelity as the Cullen-Dehnen algorithm, while suffering fewer ringing artifacts and superior entropy evolution.  

\begin{figure}[ht]
  \centering
  \begin{tabular}{cc}
    \textbf{Profiles} & \textbf{Positions} \\
    \includegraphics[width=0.4\textwidth,trim=70 70 0 0]{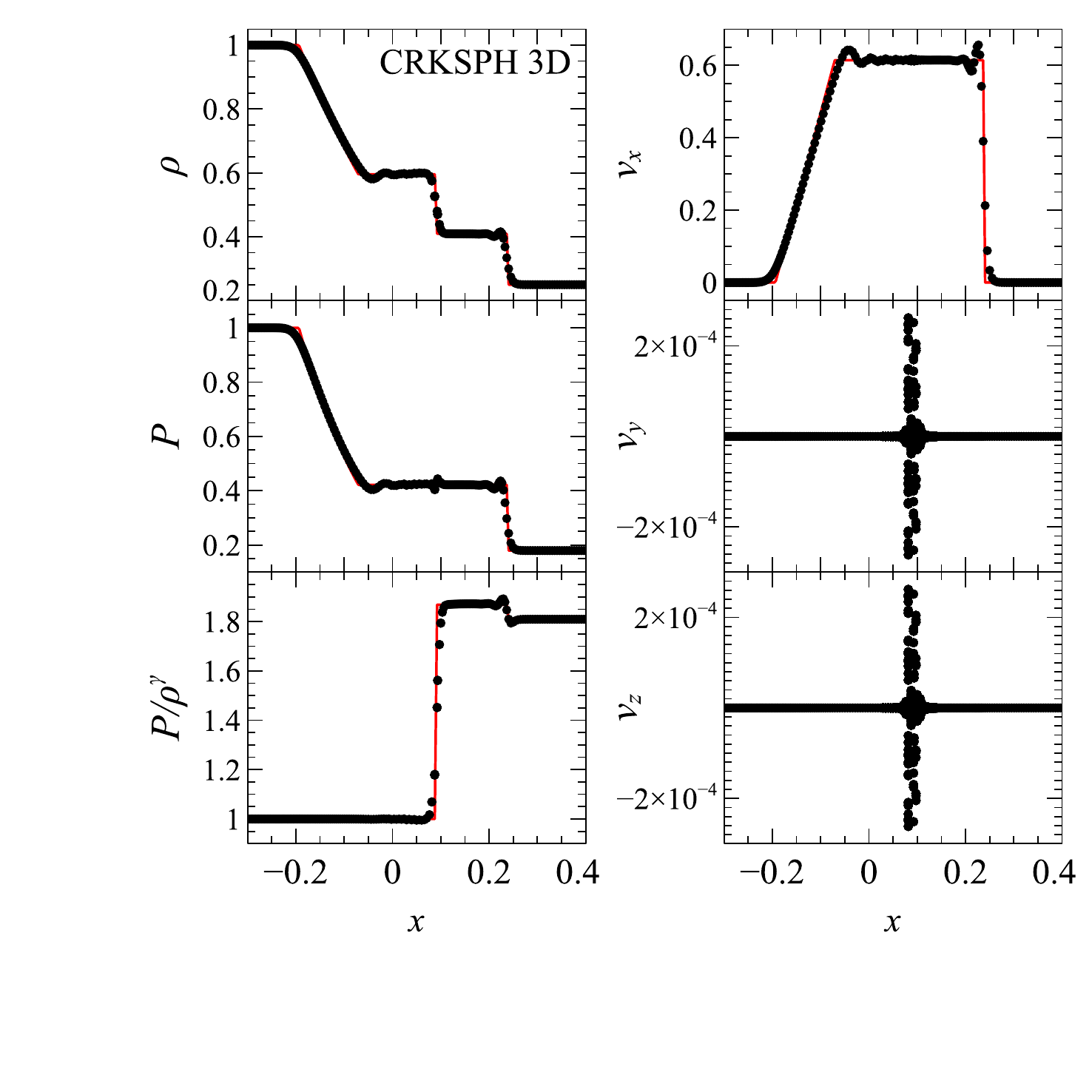} &
    \raisebox{0.5\height}{
      \begin{tabular}[b]{rl}
        \raisebox{0.5\height}{\rotatebox{90}{$t=0$}} & \includegraphics[width=0.52\textwidth]{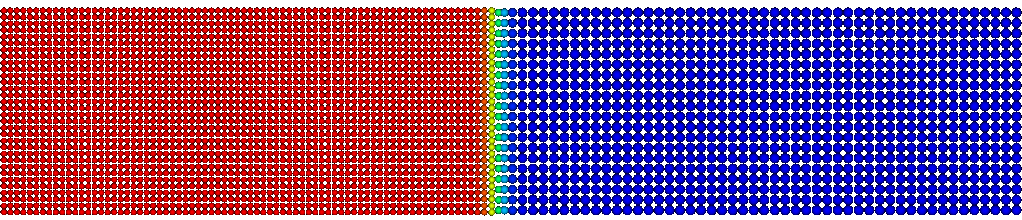} \\
        \raisebox{0.1\height}{\rotatebox{90}{$t=0.15$}} & \includegraphics[width=0.52\textwidth]{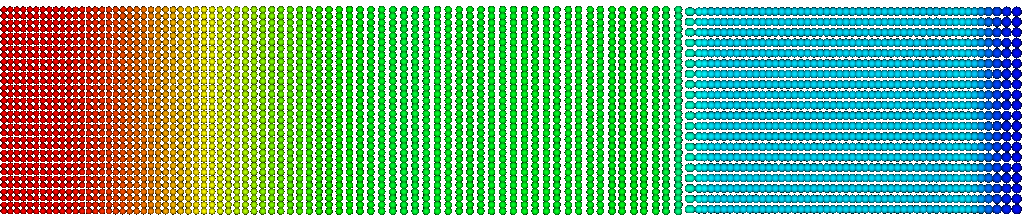}
      \end{tabular}
    }
  \end{tabular}
  \caption{The left panel displays profiles (vs.~$x$ coordinate) of the planar Sod problem run in 3D, with resolution $(N_x, N_y, N_z)_{\text{high}}= (160, 32, 32)$ and $(N_x, N_y, N_z)_{\text{low}} = (100, 20, 20)$ for the high and low pressure regions, respectively.  Note we plot all points here, so the lack of scatter in these profiles is an indication of how well the 1D solution (\cref{fig:sod-planar}) is maintained in 3D. The small deviation from null in the $v_y$ and $v_z$ velocity components also demonstrates excellent 1D symmetry.  The right panel plots the positions of all points (colored by the mass density) as seen from the positive $z$ direction for the initial conditions at $t=0$ and final state at $t=0.15$, again illustrating that planar symmetry is maintained.}
\label{fig:sod-planar-3d}
\end{figure}
In order to demonstrate how well CRKSPH maintains the 1D Sod solution in three dimensions, we repeat this experiment using CRKSPH in a 3D volume $(x,y,z) \in ([-0.5,0.5], [0, 0.1], [0, 0.1])$, employing reflecting boundaries in $x$ and periodic boundaries in $(y,z)$.  In order to maintain mass matching of the points in each region we initialize the points on lattices of dimension $(N_x, N_y, N_z)_{\text{high}} = (160, 32, 32)$ in the high density region and $(N_x, N_y, N_z)_{\text{low}} = (100, 20, 20)$ in the low density region, but adjust the $x$ spacing of the points to reproduce the smoothed profiles of \cref{eq:Sod_initial}.  The right panels in \cref{fig:sod-planar-3d} plot the initial ($t=0$) and final ($t=0.15$) positions of the points in this calculation, demonstrating that the planar symmetry of the problem is well maintained despite using simple lattice initial conditions to seed the points.  The profiles in the left panels of \cref{fig:sod-planar-3d} show the 3D CRKSPH solution does a fine job reproducing the 1D Sod results -- note we plot all the points vs.~$x$ in these profiles, again demonstrating how little scatter there is away from the 1D solution.  We also plot the $y$ and $z$ velocity components, showing that the maximum velocities away from the planar $x$ evolution occur near the discontinuity in the initial two lattices (as expected), but even here this error in the velocity is down more than three orders of magnitude compared with the $x$ velocity component.

\subsubsection{Sedov-Taylor Blastwave}
\label{sec:Sedov}
\begin{figure}[ht]
  \centering
  \begin{tabular}{cc}
    \textbf{\;\;\;\;\;\;Planar} & \textbf{\;\;\;\;\;\;Cylindrical} \\
    \includegraphics[width=0.45\textwidth]{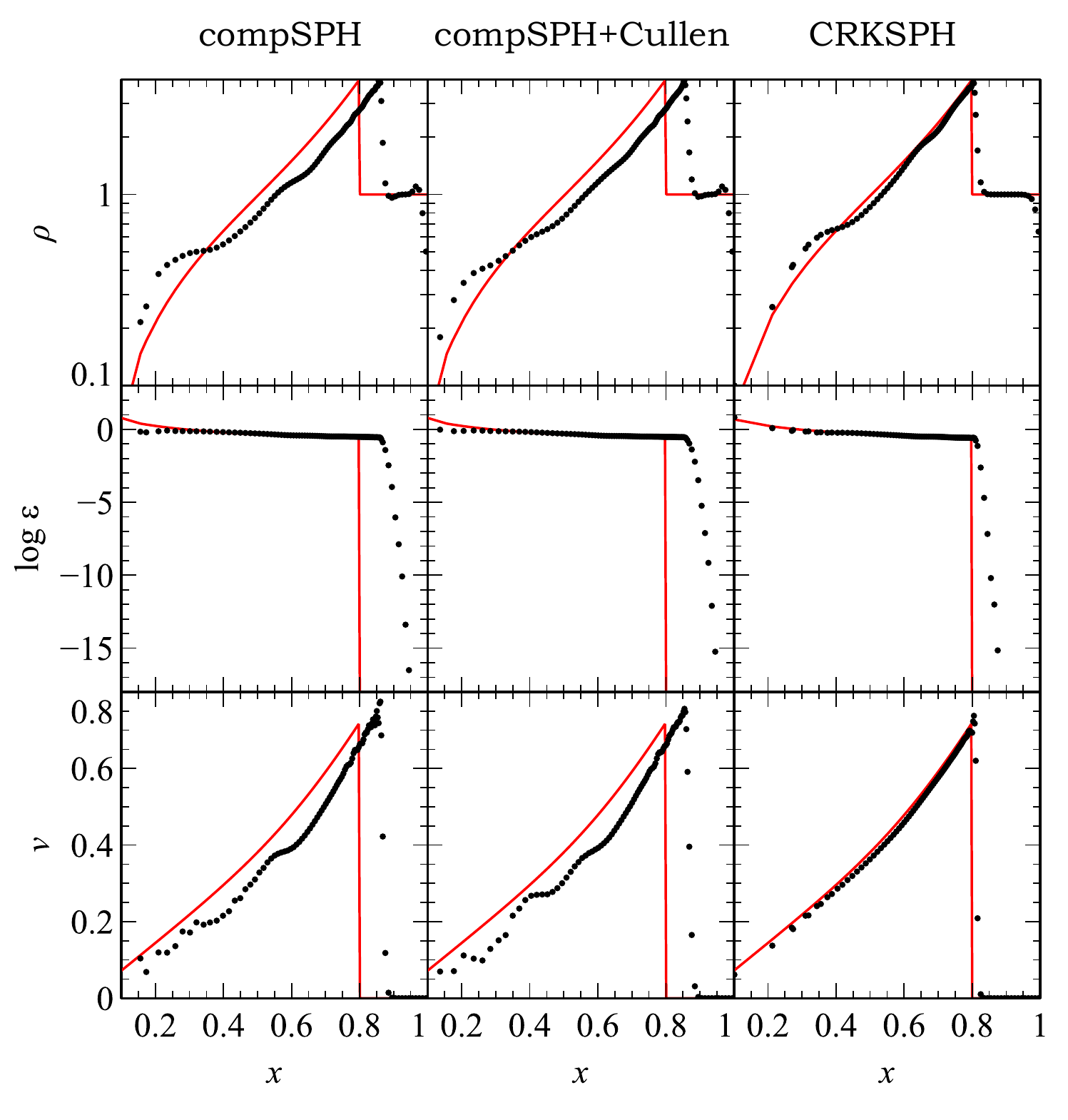} &
    \includegraphics[width=0.45\textwidth]{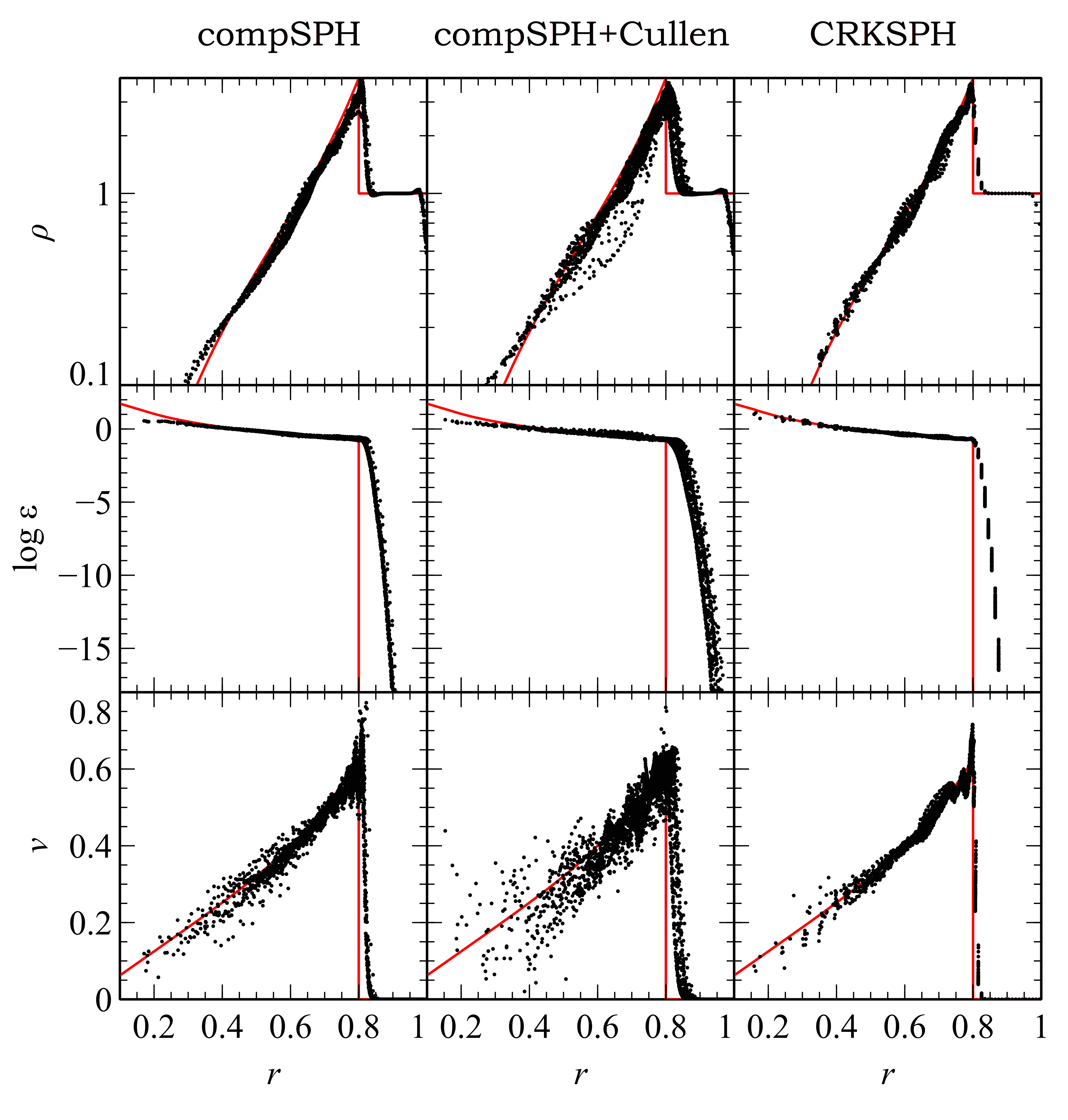}
  \end{tabular}
  \caption{Profiles of density, internal energy, and velocity of the Sedov-Taylor blastwave when the analytic shock position is predicted to be $r_s=0.8$. The left and right panels show the planar $(\nu=1)$ and cylindrical $(\nu=2)$ cases, respectively. The points plot the simulation results (at resolution $N=100^\nu$ particles), as a function of radius, while the red lines show the analytic solution. Note, we plot all simulation points in these profiles, so the scatter gives an indication of the symmetry. Owing to the numerically difficult spike initial conditions used, the 1D -- and to a lesser extent -- 2D solutions of both variants of compSPH overshoot the shock-positions. CRKSPH accurately resolves the shock-position, while generally demonstrating less scatter when compared to the other cases. }
  \label{fig:sedov-planar}
\end{figure}
\begin{figure}[ht]
\centering
\includegraphics[width=0.45\textwidth]{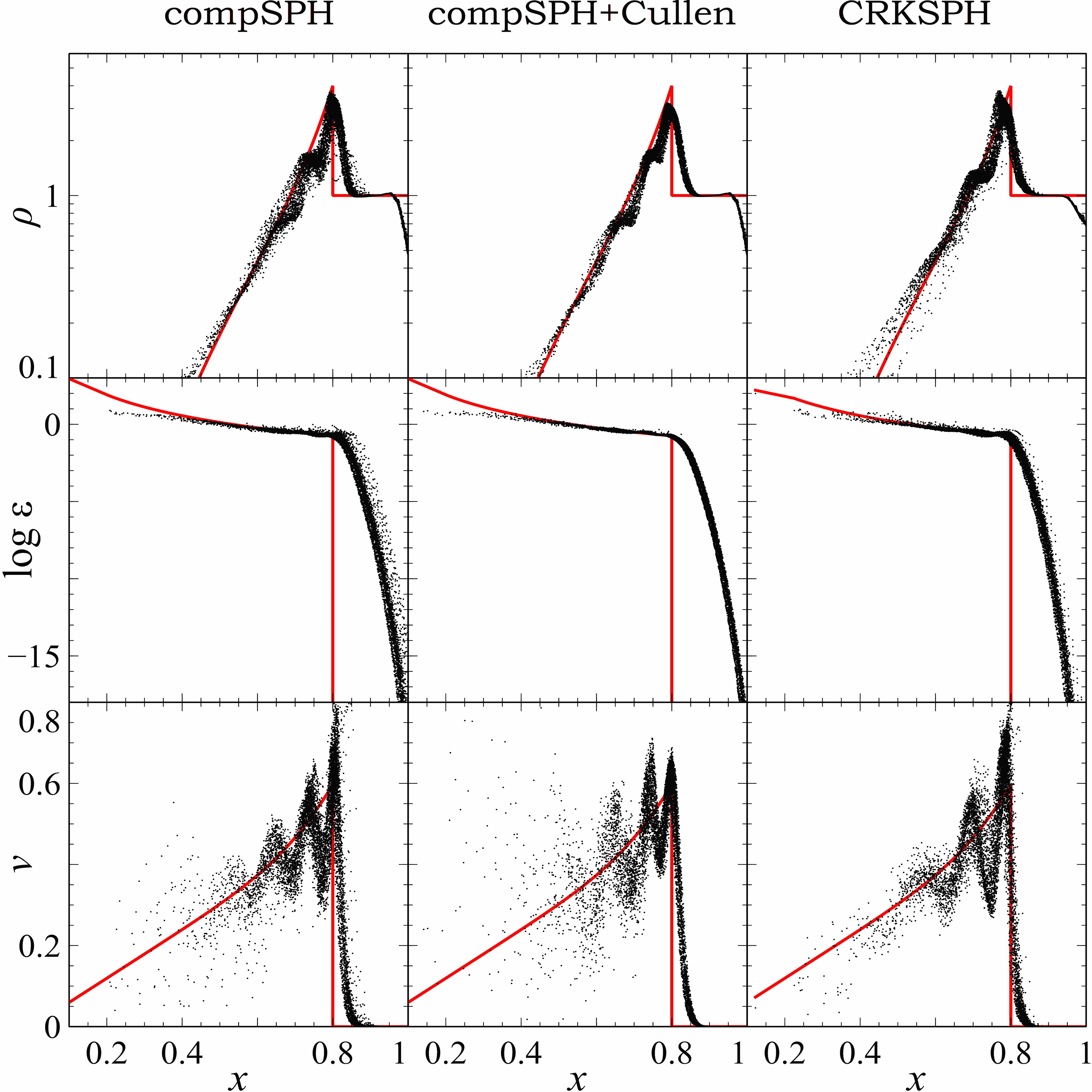}
\includegraphics[width=0.45\textwidth]{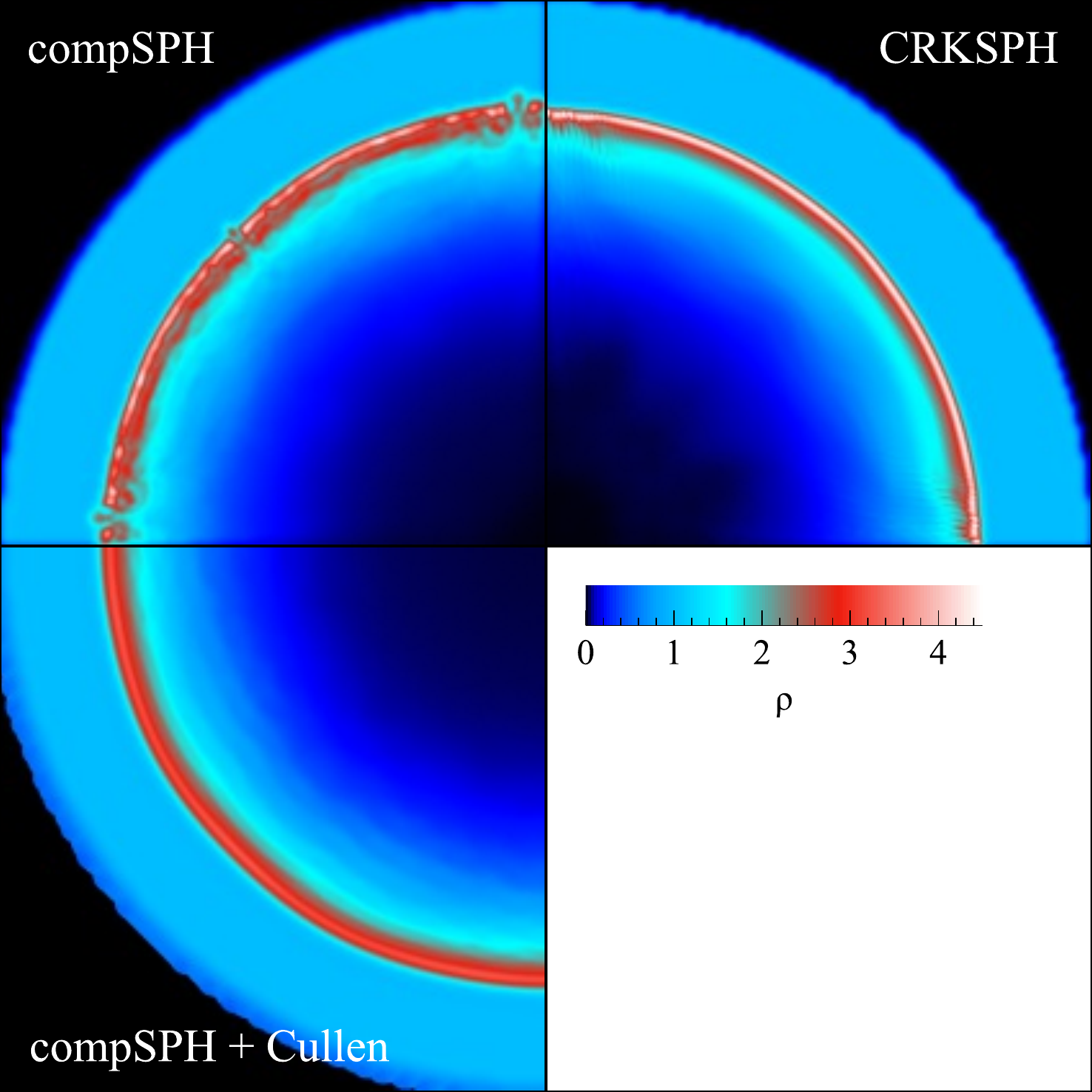}
\caption{The spherical $(\nu=3)$ Sedov-Taylor blastwave results when the predicted shock position is $r_s=0.8$, using $N=100^3$ particles.  The left figure shows the radial profiles (as was shown for the planar and cylindrical results of \cref{fig:sedov-planar}), while the right figure shows images of a slice through the mass density in each calculation along the $z=0$ plane. The Cullen viscosity prescription demonstrates an improved density solution when compared to \cref{fig:sedov-planar}, while, once again, CRKSPH demonstrates the least scatter in the velocity field. The density slices illustrate how CRKSPH resolves the sharpest shock-front when compared to the other two methods, where compSPH exhibits the most lattice imprinting of the three solvers.  }
\label{fig:sedov-spherical}
\end{figure}

The Sedov-Taylor blastwave test \cite{Sedov1959,Taylor1950a,Taylor1950b} consists of an initially homogenous pressureless static fluid of density $\rho_0=1$, into which is introduced an explosive point source of energy $E_0$ at the origin. This results in an isotropic blastwave with a shock-front traveling at radius $r(t)=\beta (E_0 t^2/\rho_0)^{1/(2 + \nu)}$ (in $\nu$ dimensions) and velocity $v(t)=dr(t)/dt$, where $\beta$ is a constant determined by solving the equations of motion. The constant $\beta(\gamma,\nu)$ can be solved to an arbitrary precision based on the relations in \cite{Sedov1959}; for the cases considered here $\beta(\gamma,\nu) \approx 1.11, 1.12$, or $1.15$ for $\nu = (1, 2, 3)$, respectively, and $\gamma=5/3$. The density at the expanding shock-front is a constant $\rho_{s}=\rho_0(\gamma +1)/(\gamma-1)$, while the velocity and pressure are decaying functions of time, i.e.  $v_{s}=2v/(\gamma+1) \propto t^{-\nu/(2+\nu)}$ and $p_{s}=2\rho v^2/(\gamma+1) \propto t^{-2\nu/(2+\nu)}$, as the wave travels away from the origin. Outside the shock ($r > v_s t$) the density remains at the initial constant $\rho_0$, while in the interior post-shock region ($r < v_s t$) density rapidly decays, vanishing at the origin; the solution is demonstrated in \crefrange{fig:sedov-planar}{fig:sedov-spherical} at a time when the analytic solution predicts the shock front has overtaken 80\% of the domain.

\Cref{fig:sedov-planar} plots the radial profiles for the planar ($\nu=1$) and cylindrical ($\nu=2$) cases, while \cref{fig:sedov-spherical} shows the spherical ($\nu=3$) results.  In each case, we present models using compSPH, compSPH+Cullen, and CRKSPH vs.~the analytic solution.  Note, the radial profiles plot all the points in the simulation, providing a good measure of how symmetric the results are in the 2D and 3D cases.  We have deliberately chosen to initialize these problems in the most difficult manner for the methods to cope with: the points are created on a lattice of radial dimension 100 (so in 2D we have a $100^2$ initial point lattice, while 3D is $100^3$), with all the initial energy placed on a single particle at the origin.  In 1D we model half the domain $x \in [0,1]$, 2D a quadrant $(x,y) \in ([0,1], [0,1])$, and 3D an octant $(x,y,z) \in ([0,1], [0,1], [0,1])$, in each case with reflecting boundaries. We model an initial spike energy $E_0=1$, where the reflective symmetries imply our single particle gets an energy of $E_0/2$ in 1D, $E_0/4$ in 2D, and $E_0/8$ in 3D.  

It is helpful to reflect on the ramifications of the selected initial conditions. Placing the initial energy spike on a single particle is sympathetic with the delta function nature of the initial conditions from which the analytical solution was derived. However, the numerical scheme is left with the difficult task of relaxing the sub-resolution energy peak to a resolved solution, and, for many methods, this results in unphysical oscillations and numerical instabilities.  It is also worth emphasizing that this method of sourcing the energy is analogous to how many astrophysicists couple energy sources from other physics, such as supernova feedback in galaxy formation models or nuclear energy release in models of supernova burning \cite{Raskin2010,2012MNRAS.423.1726S}.  Additionally, by placing our initial points on a lattice in 2D and 3D, we are testing a problem with rigorous spherical symmetry on a point distribution that does not reflect that symmetry.  For many techniques (particularly low-order meshed methods) this results in various levels of distortion or imprinting in the spherical symmetry of the shock. Lastly, running the simulation to a shock position encompassing 80\% the domain ($r_s=0.8$ in our geometry) is unusually long, and thus exposes the accumulated evolution error of the methods analyzed.  

We begin by considering the planar Sedov results in \cref{fig:sedov-planar}, where we note that both compSPH and compSPH+Cullen overshoot the analytic shock position.  This was noted in \cite{Owen2014}, and is a result of depositing the initial energy on a single point.  If, instead, the energy is deposited smoothly according to the local interpolation kernel values (as was done in \cite{Owen2014}), this overshoot is ameliorated and the problem shows good numerical convergence with increasing resolution.  Impressively, CRKSPH does not require initial smoothing of the energy deposition, and accurately predicts the shock position and solution with a single particle energy source.  Turning to the 2D cylindrical Sedov result on the right of \cref{fig:sedov-planar}, we see the overshoot in shock position for the SPH methods is reduced (though in fact it is still present), but the CRKSPH solution is still clearly preferable, with less scatter around the analytic solution, particularly in the density and velocity profiles.  In this case, the Cullen-Dehnen viscosity prescription is actually hurting the solution relative to the ordinary viscosity in the compSPH example, while our simple viscosity limiter in CRKSPH does not show similar problems.

Finally, the 3D Sedov solutions in \cref{fig:sedov-spherical} present a more complicated picture.  The radial profiles show all three methods do a reasonable job representing the solution, and, while the Cullen-Dehnen viscosity shows more scatter in the velocity profiles, it demonstrates improved density behavior compared with the previous 2D example.  Though the CRKSPH density in the evacuating post-shock region shows more scatter than our previous examples, this is not a large effect.  On the right of \cref{fig:sedov-spherical}, we show a pseudocolor slice of the density for each simulation.  We can see compSPH shows the most imprinting on the shock-front density due to our initial lattice of points; the Cullen-Dehnen viscosity modification helps clear up these shock-front imprinting artifacts.  CRKSPH demonstrates relatively clean spherical symmetry of the shock.  All three models show good symmetry, though CRKSPH captures a sharper shock front than either compSPH or compSPH+Cullen.

\subsubsection{Noh Problem}
\label{sec:Noh}
\begin{figure}[ht]
  \centering
  \begin{tabular}{cc}
    \textbf{\;\;\;\;\;\;Planar} & \textbf{\;\;\;\;\;\;Cylindrical} \\
    \includegraphics[width=0.45\textwidth]{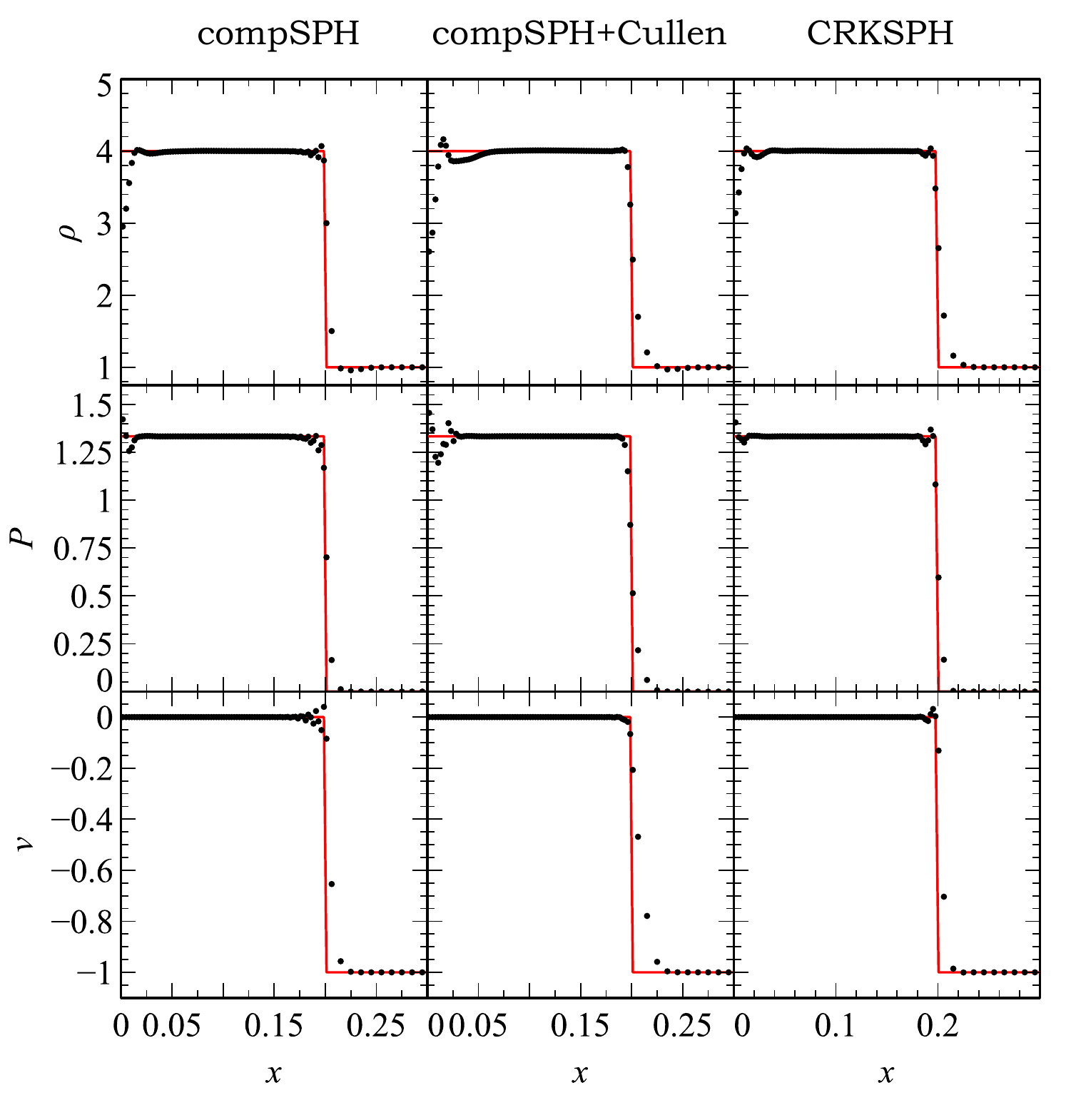} &
    \includegraphics[width=0.45\textwidth]{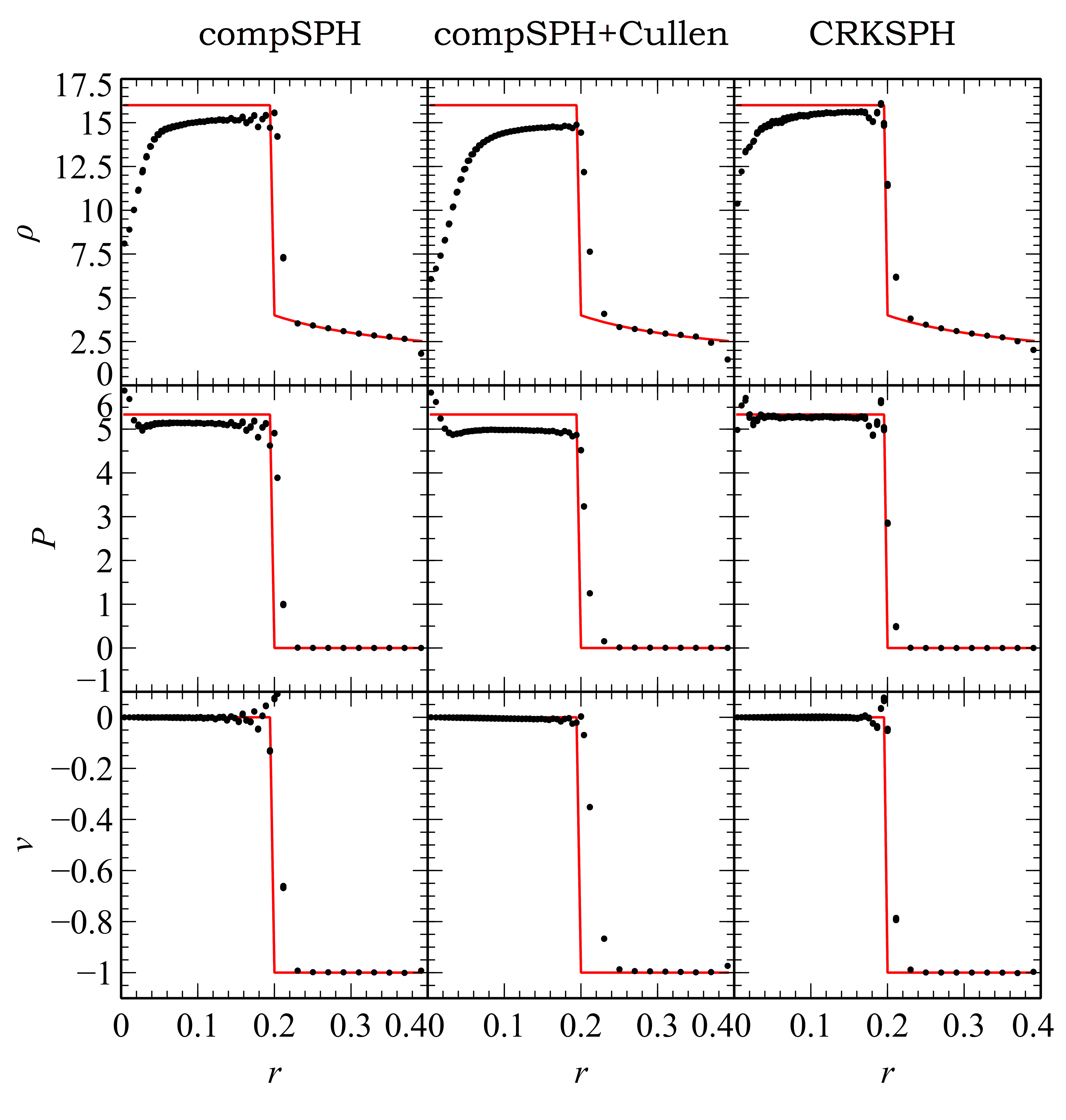}
  \end{tabular}
  \caption{Profiles of the density, pressure and velocity for the Noh implosion test vs.~the analytic solution at $t=0.6$, when the shock is predicted to be at radius $r_s=0.2$.  The left figure shows the planar $(\nu=1)$ results, and the right the cylindrical $(\nu=2)$ case. The simulations were run with $100^\nu$ particles.  As in the previous Sedov-Taylor radial profiles, we plot all points in these radial profiles, and draw (in red) lines indicating the analytic solution. All three solvers accurately resolve the 1D case, owing to the exact energy conservation of the methods. In 2D, both compSPH methods demonstrate unphysical pre-shock heating due to their viscosity treatments, causing the post shock density to be underestimated. CRKSPH more accurately captures the shock position and post-shock density, though there is some slight post-shock ringing. All methods demonstrate classic ``wall-heating'' at the origin, though CRKSPH is the least susceptible}
  \label{fig:noh-planar}
\end{figure}
We next consider the challenging Noh implosion test case in $\nu=$ 1, 2 and 3 dimensions \cite{Noh1987}.  In this problem, a pressureless $\gamma=5/3$ gas is initialized with uniform inward motion toward the origin: $(\rho_o, v_0^\alpha, P_0) = (1, -\hat{r_i}^\alpha, 0)$ where $\hat{r_i}^\alpha$ is the unit-vector of node $i$'s position.  For $\nu=1$, this corresponds to two streams of a fluid impacting along a plane; $\nu=2$ implies cylindrical convergence to a line; and $\nu=3$ represents spherical inflow to a point.  These conditions result in a self-similar solution of a shock moving away from the origin with velocity $v_s=1/3$.  In the post-shock region $r \in [0, v_s t]$ the fluid stagnates with $\rho_s = \rho_0 \left((\gamma + 1)/(\gamma - 1)\right)^\nu$, $P_s=v_s \rho_s$.  Ahead of the shock ($r > v_s t$), the fluid undergoes adiabatic shockless compression according to $\rho(r,t)=\rho_i(1 - t/r)^{\nu-1}$, while $P=0$, $v_i^\alpha = -\hat{r_i}^\alpha$.

The numerical challenges of the Noh problem are two-fold.  First, the implosion is singular in that the initial velocity field has a singularity at the origin.  This, initially unresolved, point of convergence yields the classic problem of ``wall-heating'' in Lagrangian methods, characterized by the thermal energy overshooting the analytic solution near the origin, and being compensated for by an undershoot in the mass density, such that the proper post-shock pressure is maintained.  This wall-heating effect was one of the original motivations for the development of artificial heat conduction \cite{Noh1987}, which has more recently been suggested as an approach to help deal with SPH's shortcomings in dealing with fluid mixing near density discontinuities \cite{Price2008}.  The second major difficultly in modeling the Noh problem for $\nu=$ 2 or 3, is the pre-shock adiabatic compression, which offers a severe test of the artificial viscosity formalism.  The simple SPH pair-wise viscosity of \crefrange{eq:visc}{eq:unlimmu} is active for any compression, and, therefore, will unphysically heat the pre-shock inflowing gas, making it less compressible,  thereby artificially driving the shock too quickly and underestimating the post-shock mass density.  Note, even using the full Von Neumann-Richtmyer viscosity will suffer this error as the velocity divergence is $\partial_\alpha v^\alpha < 0$ in the pre-shock region.  Therefore, the two and three-dimensional variants of the Noh problem provide an excellent test of the artificial viscosity models.

As in the previous Sedov-Taylor tests, we model a unit volume of the problem (i.e., $x \in [0,1]$ for the 1D planar problem, $(x,y) \in ([0,1], [0,1])$ in 2D, and $(x,y,z) \in ([0,1], [0,1], [0,1])$ in 3D) and employ reflecting boundary conditions to complete the geometry.  We assume convergence at the origin, and initialize $100^\nu$ points on a lattice for our initial conditions.   \Cref{fig:noh-planar}, shows the radial profiles of our Noh test solutions using compSPH, compSPH+Cullen, and CRKSPH for the $\nu=1$ (left) \& $\nu=2$ (right) cases at $t=0.6$.  We note that all three methods do an excellent job on the planar ($\nu=1$) symmetry.  This is typical for Lagrangian methods that are exactly energy conserving.  The 2D cylindrical results on the right of \cref{fig:noh-planar} are more interesting.  Both compSPH and compSPH+Cullen tend to undershoot the post-shock density (curiously the Cullen extension is a bit worse in this metric), while CRKSPH does a better job of capturing the proper shock position, as well as resolving the post-shock density value and discontinuity.  These results indicate that both compSPH forms are suffering from higher unphysical heating in the pre-shock inflow regime, almost certainly due to the artificial viscosity.  All three methods demonstrate the wall heating error at the origin (evidenced by the undershoot in density near $r=0$) though CRKSPH also fares better in this metric compared to the other methods.  CRKSPH does show some evidence of post-shock ringing in the profiles, though this is damped after only one or two oscillations behind the shock.  This could be due to either over-suppression of the viscosity or excitation of unphysical high-frequency error modes in the point distribution (i.e., unresolved modes below the numerical resolution of the method).  Either way these post-shock oscillations are not large, and are in fact comparable to the post-shock oscillations in compSPH, even though compSPH has an unlimited artificial viscosity.

\begin{figure}[ht]
\centering
\includegraphics[width=0.45\textwidth]{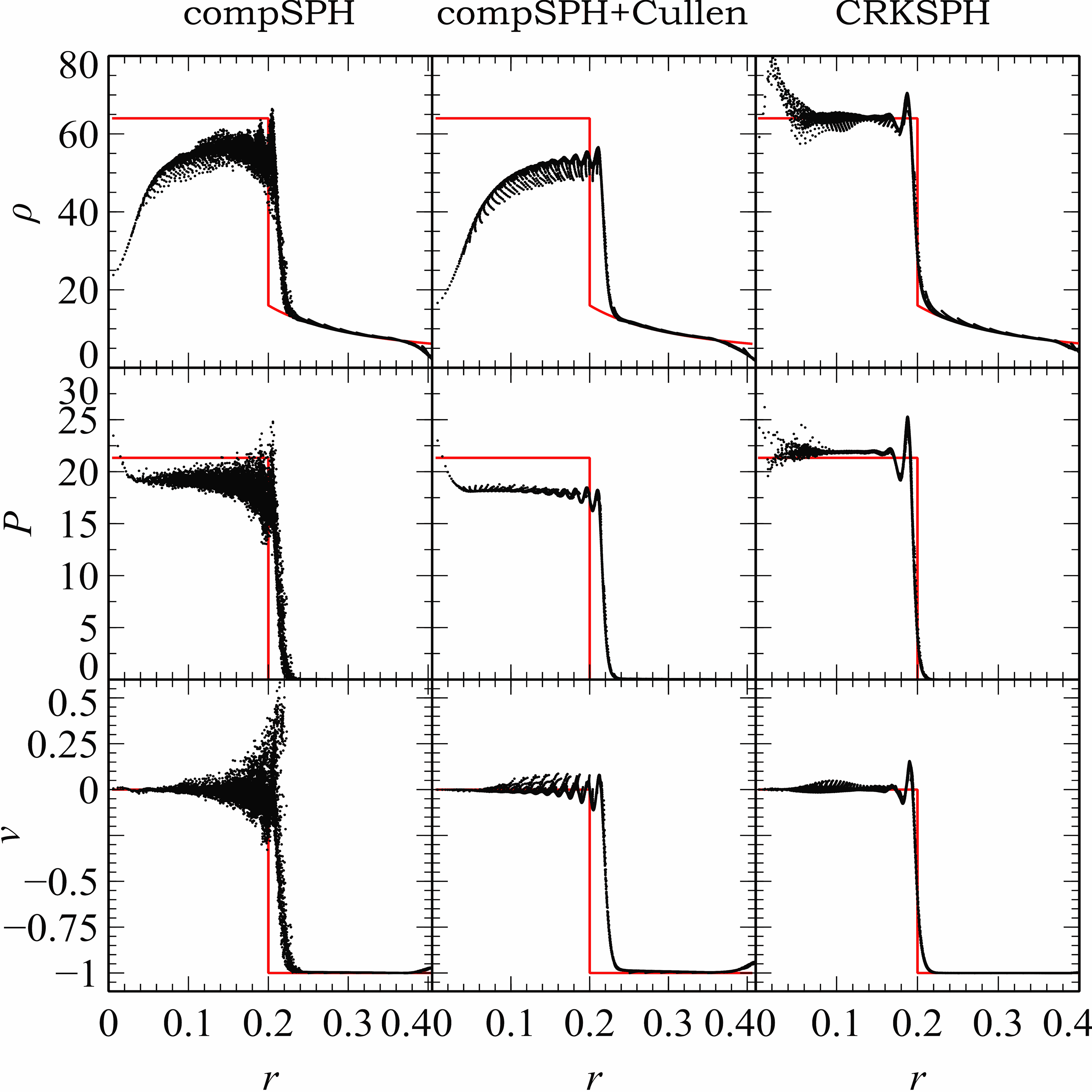}
\includegraphics[width=0.45\textwidth]{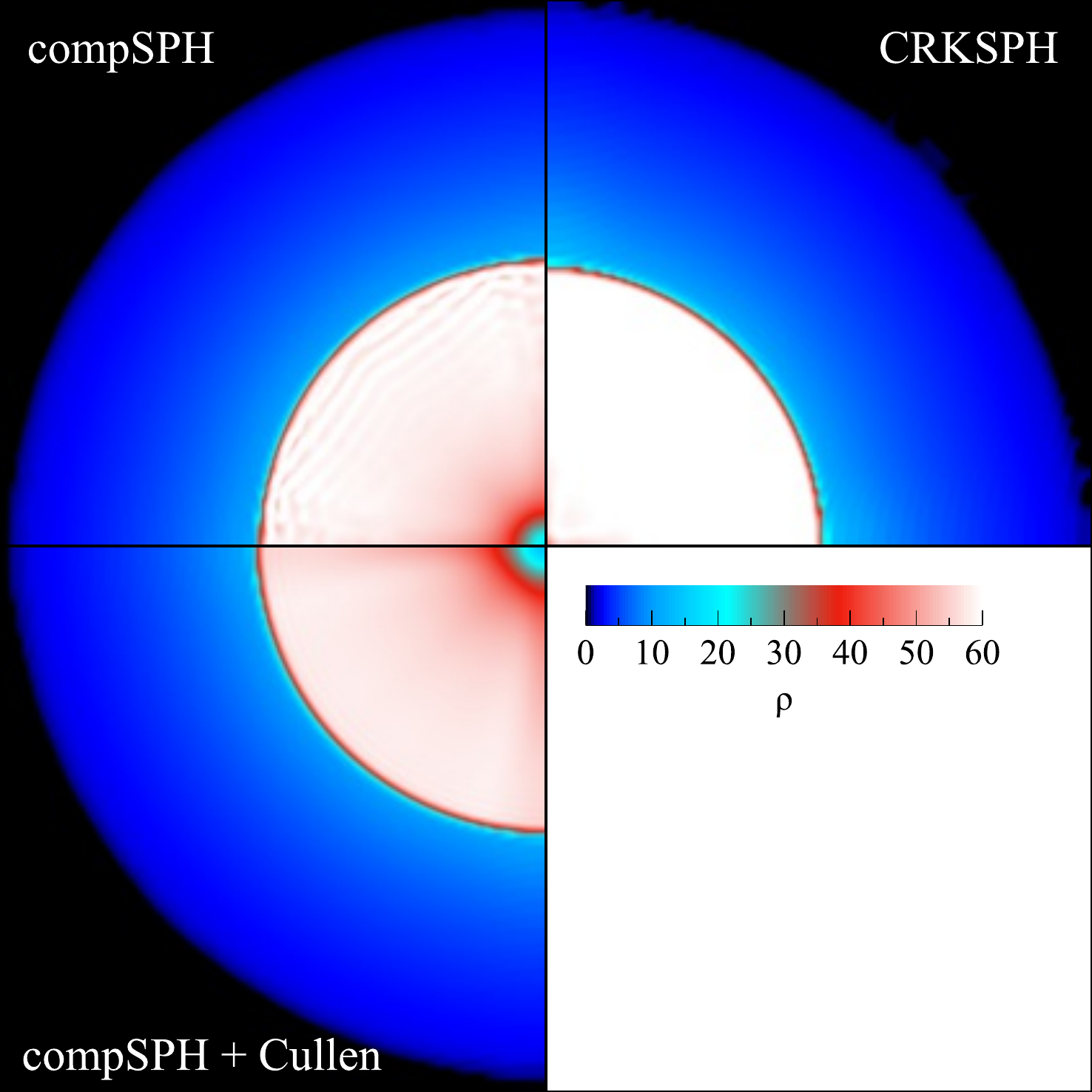}
\caption{The spherical $(\nu=3)$ Noh implosion results at $t=0.6$ and $N=100^3$ particles.  On the left we plot the radial profile scatter plots as was shown for the planar and cylindrical geometries in \cref{fig:noh-planar}, while the right-hand pseudocolor plots show a slice through the mass density along the $z=0$ plane. When compared to \cref{fig:noh-planar}, all three methods demonstrate significantly more scatter, with CRKSPH affected the least. The ``wall-heating'' effect at the origin is drastically reduced in the CRKSPH model, when compared to the other two solutions. CRKSPH is also the only method to correctly resolve the post-shock density, illustrating the effectiveness of the viscosity limiter. The density slices show how all three methods preserve spherical symmetry, while CRKSPH resolves the sharpest shock-front, as held true in the Sedov experiment in \cref{fig:sedov-spherical}. }
\label{fig:noh-spherical}
\end{figure}
Finally, \cref{fig:noh-spherical} demonstrates the 3D spherical Noh implosion results.  We see more scatter in the 3D profiles than we saw in the 1D or 2D results, but we note that CRKSPH exhibits the least scatter in the radial profiles of the three methods tested.  Additionally, CRKSPH is the only method to achieve the correct post-shock density $\rho_s=64$, and demonstrates the least wall-heating near the origin.  The post-shock oscillations in CRKSPH are somewhat more pronounced than we noted in the 2D results of \cref{fig:noh-planar}, but, again, these oscillations are damped very quickly in the post-shock region and are no worse than the post-shock scatter seen in compSPH.  The symmetry of all three methods is excellent (as evidenced by both the radial profiles and mass density slice images), though CRKSPH has the best overall symmetry and sharpest shock transition.

These results, particularly the $\nu=2$ and $\nu=3$ cases, demonstrate the utility of our viscosity limiter in \cref{eq:limmu}.  We find the unphysical pre-shock heating, endemic to the traditional SPH viscosity, is almost entirely removed, allowing us to accurately capture the shock position and post-shock density in these problems.  Once the inflowing material transitions through the shock, the viscosity correctly turns on and allows CRKSPH to resolve the shock and damp any post-shock oscillations at least as effectively as the unlimited viscosity used in the compSPH examples.  Additionally, CRKSPH demonstrates excellent symmetry preserving properties, avoiding imprinting from the initial lattice seeding of the points or the so-called ``carbuncle'' instability \cite{Peery1988}, wherein the shock preferentially propagates along preferred directions in the discretization (such as preferred mesh directions or point alignments).

\subsubsection{Double Interacting Blastwaves}
\begin{figure}[ht]
\centering
\includegraphics[width=0.45\textwidth]{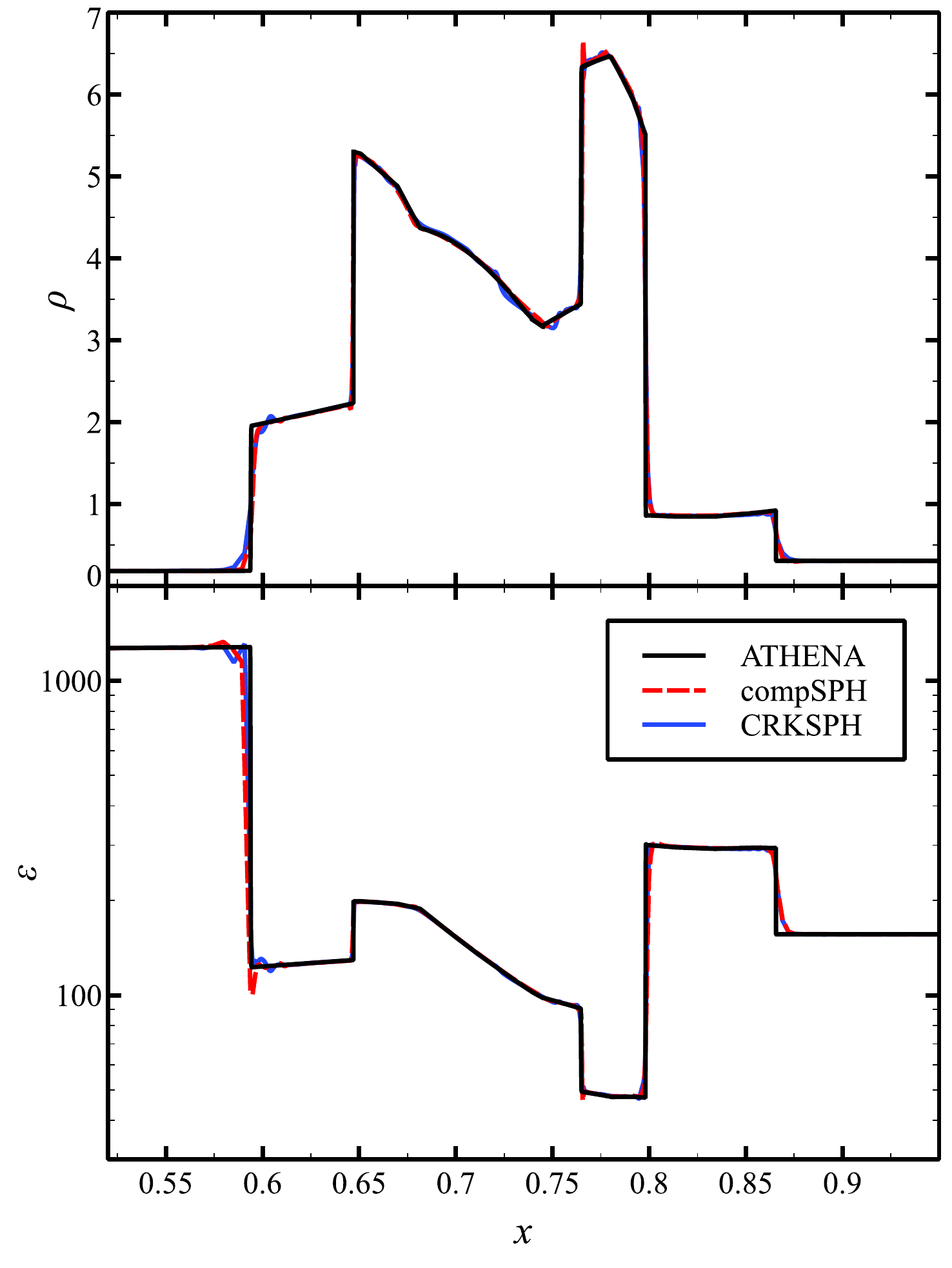}
\caption{Profiles of the density (top) and specific thermal energy (bottom) for the Woodward-Colella double blastwave problem at $t=0.38$.  We compare $N=1000$ node compSPH (dashed red lines) and CRKSPH (solid blue) models with a highly resolved $N=10,000$ Godunov Eulerian solution produced by ATHENA \cite{Stone2008}. Both meshfree methods demonstrate excellent solutions, where CRKSPH exhibits a somewhat improved behavior with respect to monotonicity, whereas compSPH slightly suffers from trajectory offshoots at a few of the transition points. }
\label{fig:dbw}
\end{figure}
Our final 1D shock test is the Woodward-Colella double blastwave \cite{Woodward1984}.  In this problem, a gas of adiabatic index $\gamma=1.4$, $v=0$, and unit density is initialized in three distinct pressure regions of a unit box: $P_0=1000$ for $x \in [0,0.1]$; $P_0=0.1$ for $x \in [0.1,0.9]$; $P_0=100$ for $x \in [0.9,1]$.  The boundaries of the box at $x=0$ and $x=1$ are reflecting.  The evolution involves two blastwaves launching from the two high-pressure regions into the initially low-pressure domain in between, eventually undergoing multiple shock and rarefaction interactions.  The resulting composite double peak density solution is demonstrated in  \cref{fig:dbw} at time $t=0.038$.  Unfortunately, although the double blastwave test offers an elaborate 1D shock probe, there is no analytical solution for comparison.  We, therefore, adopt as our reference a numerical solution from the high-order Godunov Eulerian grid code, ATHENA \cite{Stone2008}, using a high resolution of $N=10,000$ zones and a rather conservative courant number of 0.1.

\Cref{fig:dbw} compares two medium resolution calculations ($N=1000$) using compSPH and CRKSPH against the $N=10,000$ ATHENA reference.  We deliberately use these medium resolutions for compSPH and CRKSPH in order to highlight the differences, which become tiny at the extreme resolutions such as those used for the ATHENA reference here.  Both compSPH and CRKSPH perform well in this shock-dominated problem, mirroring the complex solution gradients and resolving the resulting double shock peaks well, when compared to the ATHENA reference.  The differences between compSPH and CRKSPH are minor, though compSPH does demonstrate some over/under-shoots near the transition points of $x \approx 0.6$ (in the specific thermal energy) and $x \approx 0.77$ (in the density).  Both meshfree methods appear to resolve the shock and rarefaction transitions to roughly the same level.  This test demonstrates that CRKSPH handles these sorts of complex strong shock interactions as well as compSPH, though with somewhat improved monotonicity for equivalent resolution.

\subsubsection{Convergence}
\begin{figure}[ht]
\centering
\includegraphics[width=0.45\textwidth]{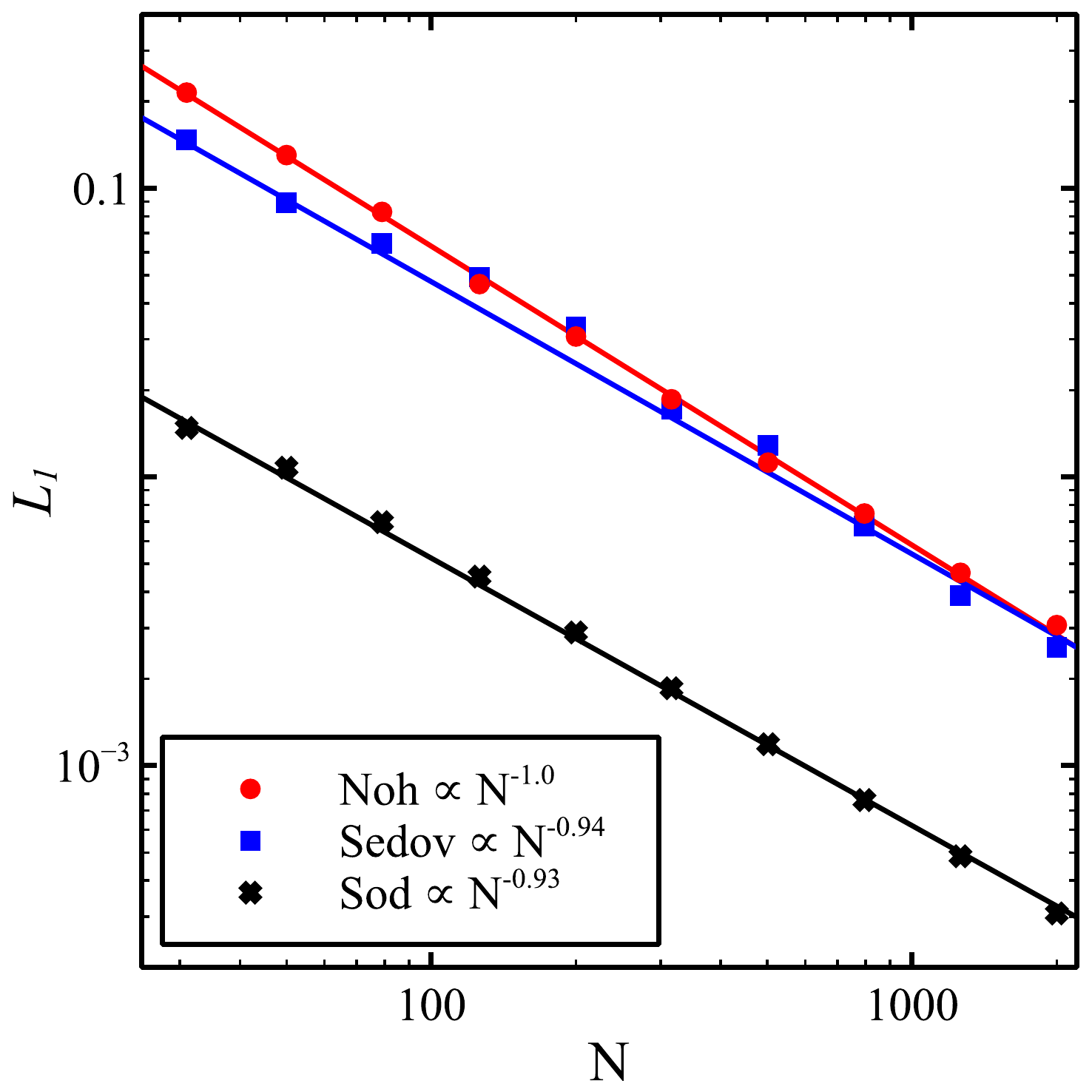}
\caption{$L_1$ density norms for the CRKSPH solutions of the planar shock tests (Sod, Sedov, and Noh) as a function of resolution $N$, i.e., number of CRKSPH points.  Also plotted are the fitted convergence rates for each case, which theoretically should be first-order ($\propto N^{-1}$) for shock-dominated problems such as these.}
\label{fig:1dconverge}
\end{figure}
We conclude our examination of shock-dominated problems with a measurement of the convergence rate for CRKSPH. \Cref{fig:1dconverge} measures the $L_1$ norm of the density for the 1D (planar) shock problems that are accompanied by analytic solutions: Sod, Sedov, and Noh. We expect, at best, first-order convergence, as these problems have discontinuities; reassuringly, all three cases demonstrate linear convergence rates.  Of this set, the Sedov-Taylor test is notoriously difficult, owing to the complication of representing a point-like injection of energy in the initial condition.  The challenge of demonstrating convergent behavior for the Sedov problem with the entirety of the initial energy on a single point has been noted previously in models of compSPH \cite{Owen2014}, as well as other studies that argue these singular initial conditions may preclude convergent behavior entirely \cite{Cook2013}.  Here, however, we find that even with a point-like energy source, our Sedov-Taylor convergence rate is near unity for CRKSPH.

Lastly, we note that Eulerian solvers also demonstrate linear convergence of the Sedov-Taylor blastwave as they utilize implicit viscous smoothing.  This inherent diffusion in Eulerian methods is also the reason that such methods avoid the ``wall heating'' often noted in Lagrangian models of the Noh problem, i.e., thermal energy overshoot/mass density undershoot at the convergent point of the problem (evident in the compSPH models of the Noh problem in \cref{sec:Noh}).  This has at times been put forth as an advantage of Eulerian solvers; however, it is worth noting that this implicit numerical diffusion implies an inescapable unphysical transport of entropy.  The artificial heat conduction introduced in \cite{Noh1987} is motivated by an effort to deliberately introduce similar entropy diffusion into Lagrangian methods, yet, in general, many Lagrangian implementations forgo such terms in preference for maintaining the strict entropy preserving nature of such schemes.  Indeed, this strict lack of unphysical entropy transport is viewed as a strength of Lagrangian methods for many problems (particularly those where avoiding unphysical transport of entropy is important) compared with Eulerian discretizations.

\subsection{Angular Momentum Preservation, Vorticity, and Hydrodynamically Unstable Interfaces}
In this section, we focus on two-dimensional problems in the absence of shock hydrodynamics.  The Gresho (\cref{sec:Gresho}) and Yee isentropic (\cref{sec:Yee}) vortices examine how well CRKSPH preserves local angular momentum.  We also look at two classic hydrodynamic instability tests: the Kelvin-Helmholtz shear driven instability in \cref{sec:KelvinHelmholtz}, and the gravitationally driven Rayleigh-Taylor growth in \cref{sec:RayleighTaylor}.  These are all phenomena where classic SPH has been demonstrated to have problems in the past, and so we wish to examine how well CRKSPH handles these problems.  In these examples, we include comparisons with ``pressure based'' or ``density independent'' SPH, a.k.a.~PSPH \cite{Saitoh2013,Hopkins2012,Hopkins2015}, the details of which are summarized in \cref{sec:PSPH}.  The PSPH modification of SPH was developed to help remedy some of the problems found in classical SPH implementations modeling complex mixing flows, such as those we examine in this section (particularly near density discontinuities), and, therefore, it is relevant to compare CRKSPH with PSPH in these examples.

\subsubsection{Gresho Vortex}
\label{sec:Gresho}
\begin{figure}[ht]
\centering
\includegraphics[width=0.98\textwidth]{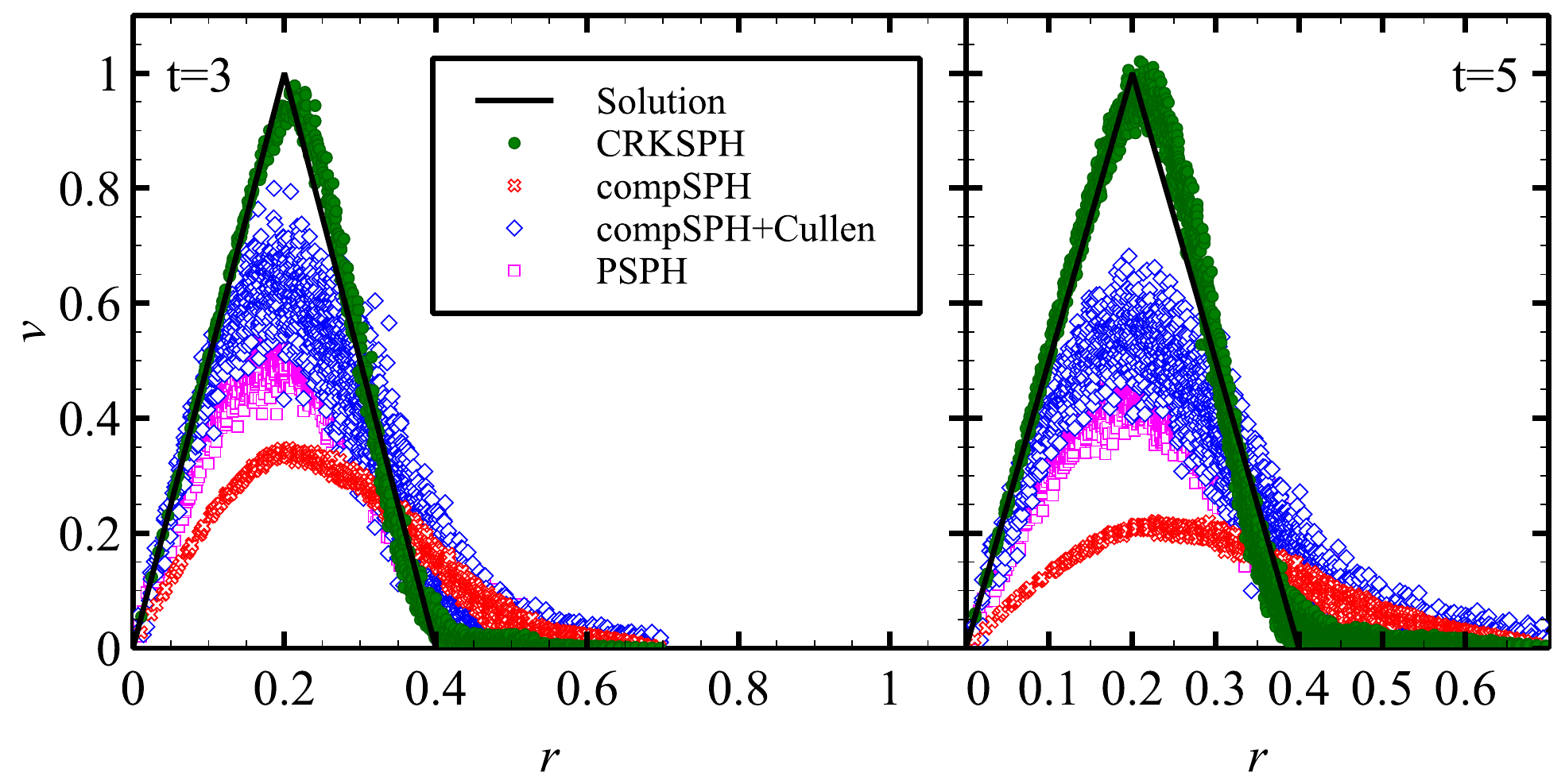}
\caption{Azimuthal velocity of the Gresho Vortex test using $64^2$ particles for compSPH with both the Monaghan and Gingold (red crosses) and Cullen (blue diamonds) viscosity prescriptions, PSPH (magenta squares), and CRKSPH (green circles) at time $t=3$ (on the left) and $t=5$ (right panel). The unlimited viscosity of compSPH significantly damps the evolution of the vortex. The Cullen viscosity treatment in PSPH and compSPH noticeably improves the diffusion at the sacrifice of significant scatter. CRKSPH obtains the best solution, exhibiting less scatter while accurately maintaining the theoretical peak velocity, preserving a robust solution as far out as t = 5.0. }
\label{fig:Gresho}
\end{figure}

The Gresho vortex \cite{Gresho1990} is a 2D triangular (in rotational velocity) vortex in steady-state equilibrium. A $\gamma=5/3$ ideal gas is initialized in a periodic box of unit length centered at the origin, with uniform unit density. The vortex is defined by an azimuthal velocity profile complemented with a radial pressure given as
\begin{align}
  \label{eq:Gresho}
  P(r), \; v_{\phi}(r) = \left\{\begin{array}{l@{\quad}l@{\quad}l}
    12.5r^2+5,              & 5r,    & 0 \le r < 0.2 \\
    12.5r^2-20r+4\ln(5r)+9, & 2-5r,  & 0.2 \le r < 0.4 \\
    3+4\ln(2),              & 0,     & 0.4 \le r.
  \end{array}\right. 
\end{align}
The solid line in \cref{fig:Gresho} shows this triangular velocity profile.  The pressure gradient is constructed to balance the centrifugal force of the vortex, which in the absence of viscosity should be stable and rotate indefinitely.  Measuring deviations from this initial profile is a sensitive test of how well a numerical hydrodynamic method can maintain such dissipationless flow; in particular, the shearing velocity can cause the SPH artificial viscosity to activate, leading to unphysical transport of angular momentum and the degradation of the vortex. It is also notable that the presence of cusps in these initial conditions (at $r=0.2$ and $r=0.4$) will cause each method to deviate from the ideal solution as these sharp edges are rounded out to some degree.

\Cref{fig:Gresho} demonstrates the results at times $t=3$ (left) and $t=5$ (right) of running PSPH and CRKSPH, as well as compSPH with both Monaghan and Cullen viscosity prescriptions. In each case, we initialize the problem on a lattice consisting of $64^2$ particles.  This is conformal with our unit box shape for the initial conditions, but antagonistic to the physical symmetry of the problem.  Thus, there is an adjustment period early in the evolution, as the points settle into a more natural configuration for the physical geometry of the vortex, stressing how well each method handles such perturbations.  
Running the problem to late times (such as $t=5$) demonstrates extreme degradation of the vortical flow by the standard SPH techniques.  Proceeding from worst to best, we see that the ordinary unlimited Monaghan-Gingold viscosity (\crefrange{eq:visc}{eq:unlimmu}) in compSPH (red crosses) has almost completely halted the rotational flow.  A combination of both the E0-error and the overly diffusive Monaghan-Gingold viscosity introduces substantial viscous errors into the solution, culminating in the near destruction of the vortical motion.  A significant improvement is achieved with the addition of the Cullen viscosity treatment to compSPH, represented by the blue diamonds in \cref{fig:Gresho}.  This case gives the second-best results in terms of maintaining the strength of the rotational flow, albeit with a great deal of noise evidenced by the scatter in this curve.  The PSPH solution (magenta squares) performs similarly to compSPH+Cullen as it also utilizes the Cullen viscosity treatment.  The CRKSPH case (green circles) yields the best solution, maintaining a near theoretical peak rotational velocity even as late as $t=5$ (a time well past what is usually shown for the Gresho test).  This is due to a combination of the improved interpolation afforded by RK theory, as well as our limiting modification of the CRKSPH viscosity in \cref{eq:limmu}.  Of the two, the limited viscosity is the dominant effect in maintaining the peak rotational speed, as it nearly eliminates the unphysical activation of the viscosity, thereby reducing unphysical angular momentum transport.


\subsubsection{Isentropic Vortex}
\label{sec:Yee}
\begin{figure}[ht]
  \centering
  \includegraphics[width=0.45\textwidth]{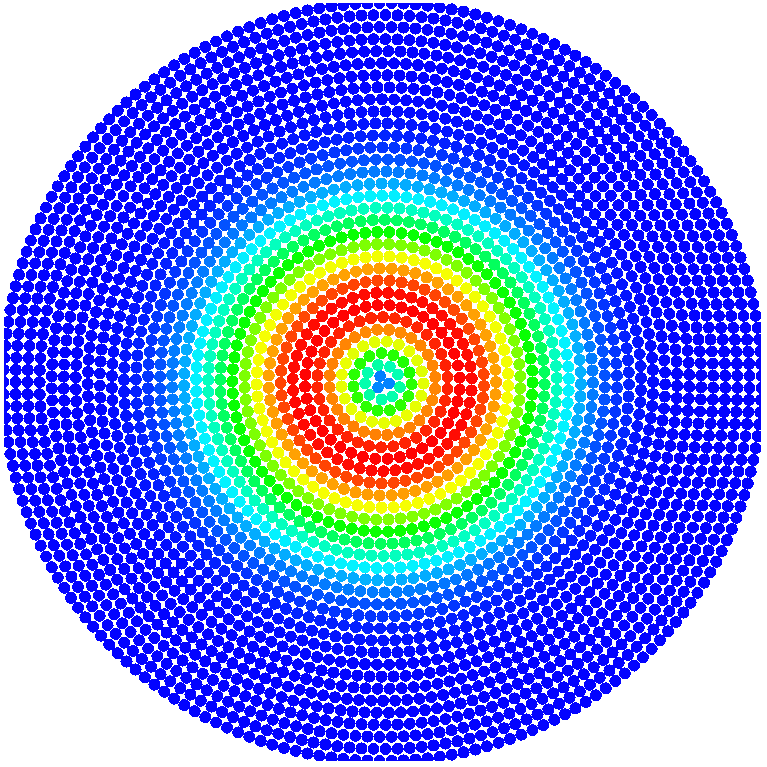}
  \caption{Plot of point positions colored by velocity for the $N_r=32$ simulation of the Yee vortex at $t=8$.}
  \label{fig:YeePoints}
\end{figure}
The Gresho test described in \cref{sec:Gresho} is a well known vortical flow problem, yet it has one important drawback for examining hydrodynamic solver performance: although the problem is shock free, it contains discontinuities in the initial conditions at the cusp points of $r=0.2$ and $r=0.4$ (see \cref{eq:Gresho}).  These discontinuities complicate using the Gresho vortex as a test for the convergence rate of a given hydrodynamical method, as we cannot expect to achieve the nominal convergence rate in a discontinuous problem.  In order to examine our convergence properties for a more complicated multi-dimensional test than the one-dimensional acoustic-wave examined in \cref{sec:acoustic}, we would like to have a problem similar to the Gresho test with smooth properties.  The isentropic Yee vortex \cite{Yee2000} is a 2D steady-state equilibrium vortex test in a free-stream flow, yet is smooth and continuous everywhere, and, therefore, more amenable to measuring convergence. Additionally, since the Yee vortex is an inherently 2D scenario, it is useful for demonstrating higher-order convergence for more than a trivially one-dimensional flow (as opposed to the acoustic wave test).

The initial conditions for the Yee vortex can be expressed as perturbations about a central point $(x_c, y_c)$, viz. 
\begin{align}
  \label{eq:Yee}
  \begin{pmatrix} \delta v_x \\ \delta v_y \end{pmatrix} &= \frac{\beta}{2 \pi}e^{(1-r^2)/2} \begin{pmatrix} -(y-y_c) \\ x-x_c \end{pmatrix} \\
  \delta T &= -\frac{(\gamma - 1)\beta^2}{8\gamma \pi^2} e^{1-r^2} \nonumber
\end{align}
where $r^2 = (x-x_c)^2 + (y-y_c)^2$, and the constant $\beta$ controls the vortex strength. The density and pressure are given by $\rho = T^{1/(\gamma-1)}$ and $P = \rho T$,  where $T  = T^\infty + \delta T$.  Note that unlike the Gresho problem, the Yee vortex formally extends to infinity, though the perturbations decrease rapidly with increasing $r$.  For our example, we take the free-stream parameters to be $v_x^\infty=v_y^\infty=0$, $P^\infty = \rho^\infty = 1$, and $T^\infty = P^\infty/\rho^\infty$, for an ideal gas of $\gamma=1.4$.  We also assume the vortex is centered at the origin, $(x_c, y_c) = (0,0)$, and choose a vortex strength parameter of $\beta = 5$.  We initialize the Yee vortex differently than our previous box-like examples due to the formally infinite extent of the initial conditions; rather than creating our points in a box we create a circularly symmetric distribution of points out to a maximum radius $r_{\max}=5$, as depicted in \cref{fig:YeePoints}.  We parameterize the points by the number of radial rings used $N_r$; the number of points seeded in each ring is chosen to most closely match the fixed radial spacing $\Delta r = r_{\max}/N_r$.  The mass of each ring is set based on the desired density profile at that radius, and the masses of the points set appropriately.  In order to represent the infinite extent of this problem, we create 10 rings of non-dynamical or ``ghost'' points for $r > r_{\max}$ on which we impose the fixed initial conditions from \cref{eq:Yee}.  Note, this is similar to how we impose the external boundary conditions for the Kidder isentropic implosion in \cref{sec:KidderTest}.

\begin{figure}[ht]
  \centering
  \includegraphics[width=0.45\textwidth]{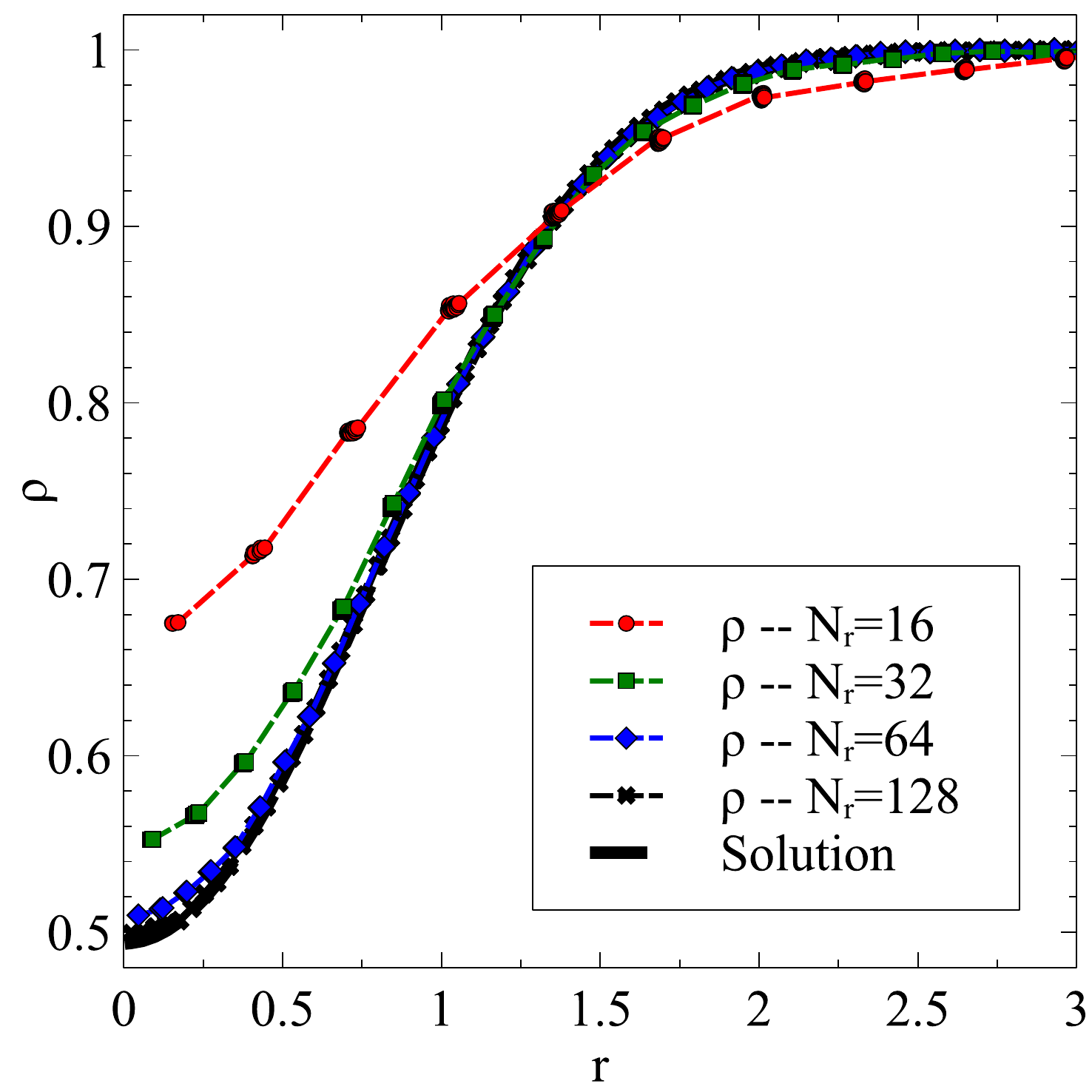}
  \includegraphics[width=0.45\textwidth]{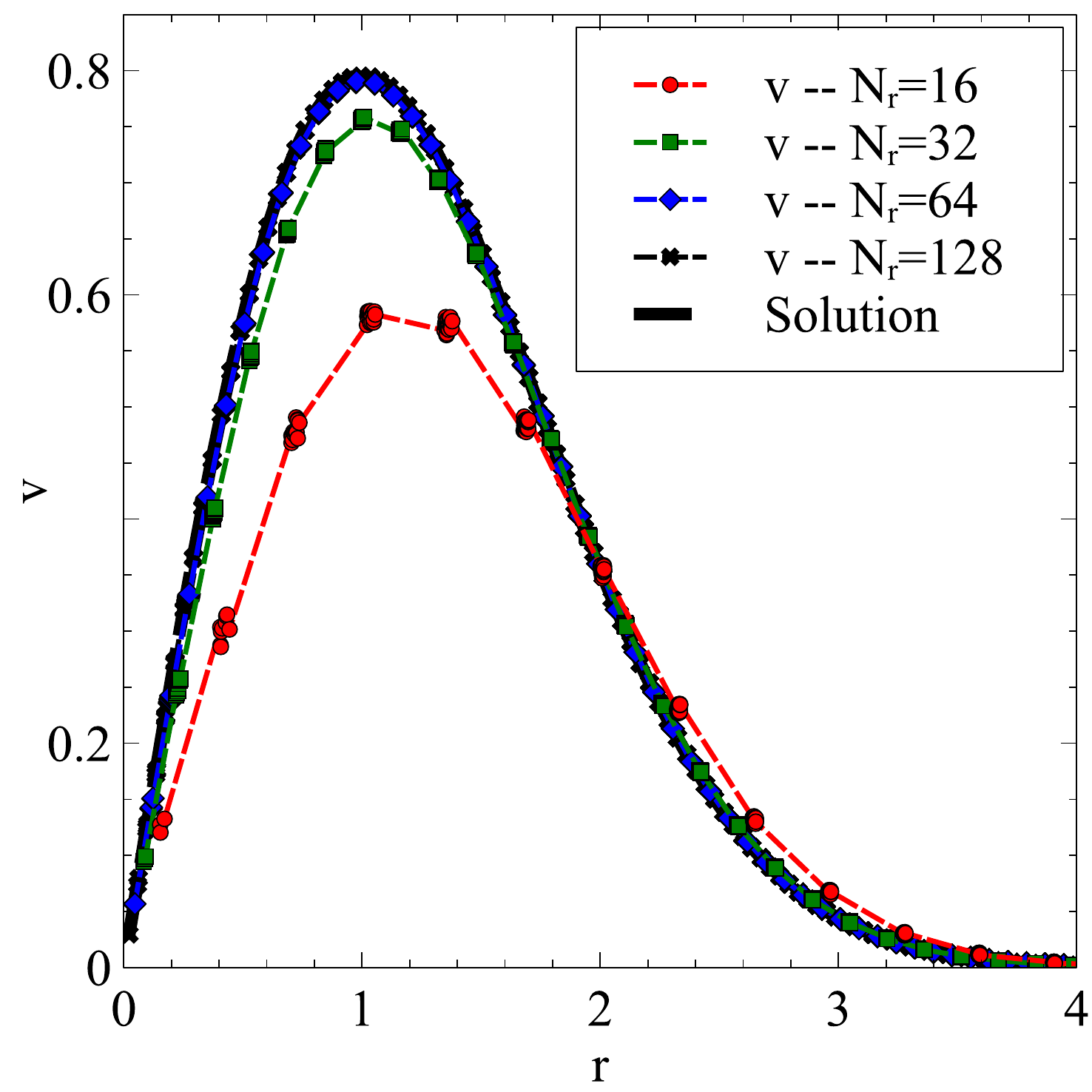}
  \caption{Radial profiles of $\rho(r)$ (left) and $v(r)$ (right) for the Yee vortex at $t=8$.  Results are shown for simulations using $N_r \in (16, 32, 64, 128)$ radial shells. There is no significant scatter in points of the same radial bin, indicating accurate symmetry preservation. }
  \label{fig:YeeProfiles}
\end{figure}
\begin{figure}[ht]
  \centering
  \includegraphics[width=0.45\textwidth]{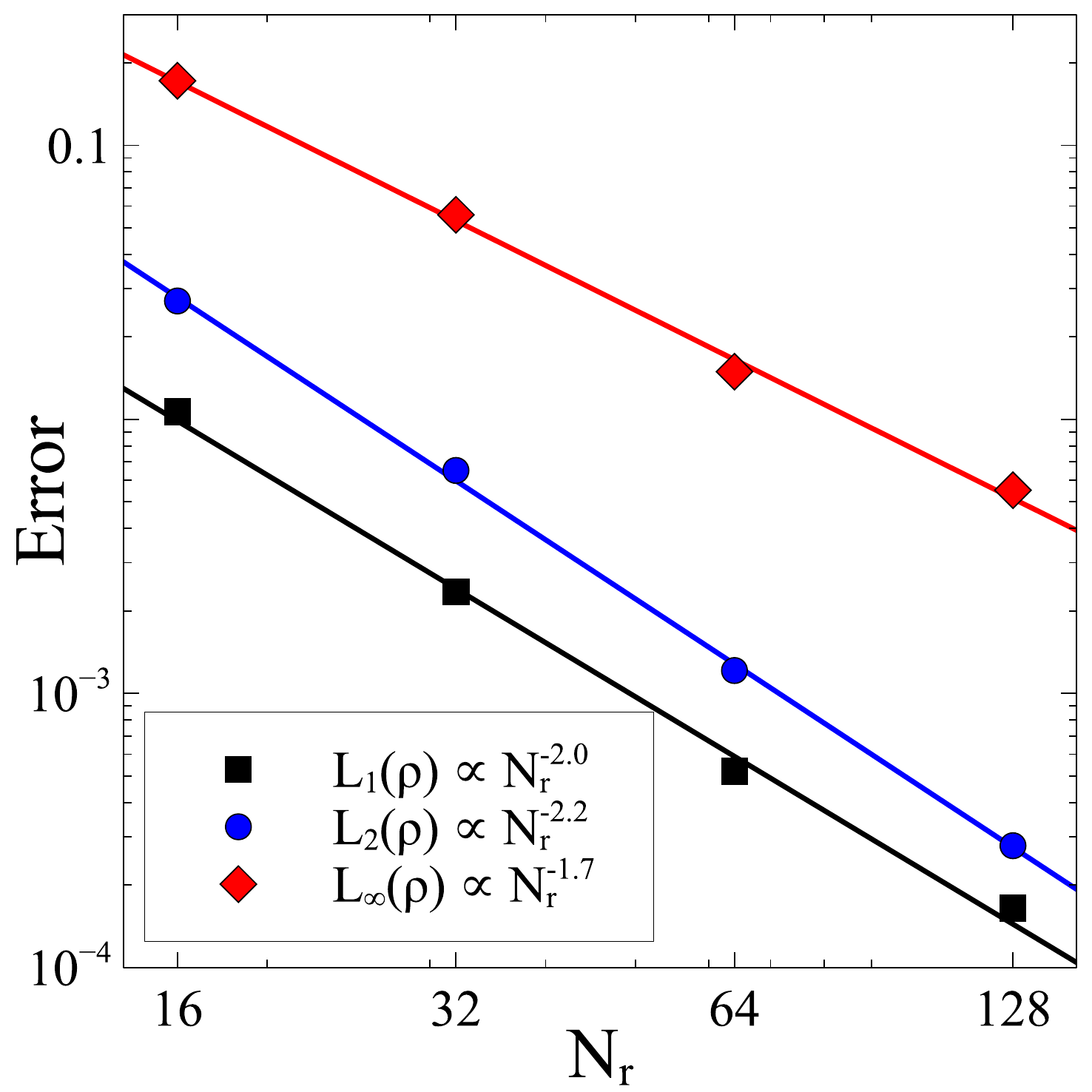}
  \includegraphics[width=0.45\textwidth]{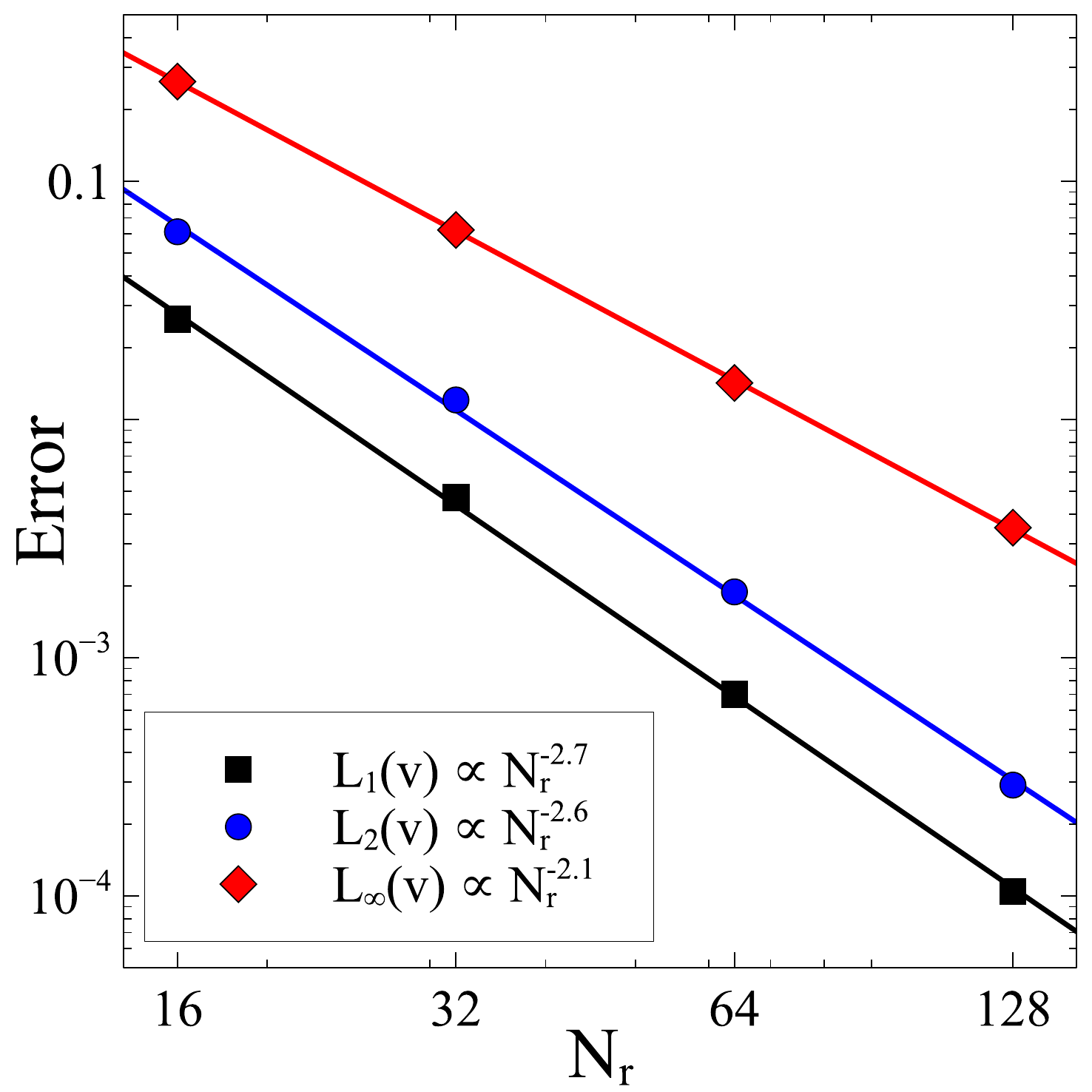}
  \caption{$L_1$, $L_2$, and $L_\infty$ errors and measured convergence rates for CRKSPH models of the Yee isentropic vortex as a function of the radial number of shells of points $N_r \in (16, 32, 64, 128)$.  Convergence rates are shown for the mass density (left) and velocity (right), measured at simulation time $t=8$. CRKSPH demonstrates second-order convergence, as expected for a smooth problem given our utilization of linear-reproducing kernels in the solver.}
  \label{fig:YeeConverge}
\end{figure}
\Cref{fig:YeeProfiles}, shows radial profiles of the mass density and velocity for a series of CRKSPH simulations ($N_r \in (16, 32, 64, 128)$) of the Yee vortex at $t=8$.  Once again we plot all points in these calculations, so at each radius for each calculation we can see the scatter (or lack thereof) in each radial ring of points.  Clearly, these simulations maintain the radial symmetry of each ring well, and it is evident, with increasing resolution, how the models rapidly approach the analytical expectation (shown as the solid lines).  \Cref{fig:YeeConverge} plots the $L_1$, $L_2$, and $L_\infty$ measures of the errors in each simulation for the mass density and velocity vs.~the radial resolution.  We draw the fitted convergence for each norm and quote the resulting order of convergence.  Since we are using linear-order reproducing kernels we expect second-order convergence for a smooth problem, and, indeed, we find the order of convergence in each norm is right around 2.  The velocity shows somewhat higher rates of convergence ($m \in [-2.1, -2.7]$), while the mass density converges at almost exactly the expected value of 2.

\subsubsection{Kelvin-Helmholtz Shearing}
\label{sec:KelvinHelmholtz}
\begin{figure*}[ht]
\centering
\includegraphics[width=0.95\textwidth]{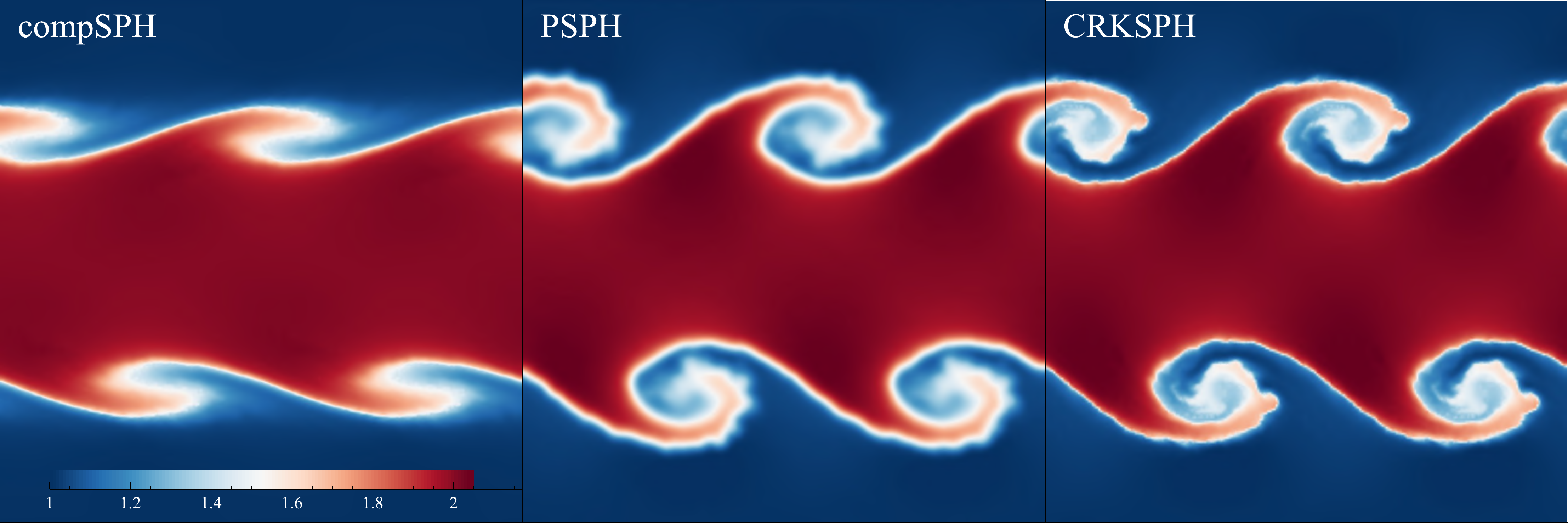}
\includegraphics[width=0.63\textwidth]{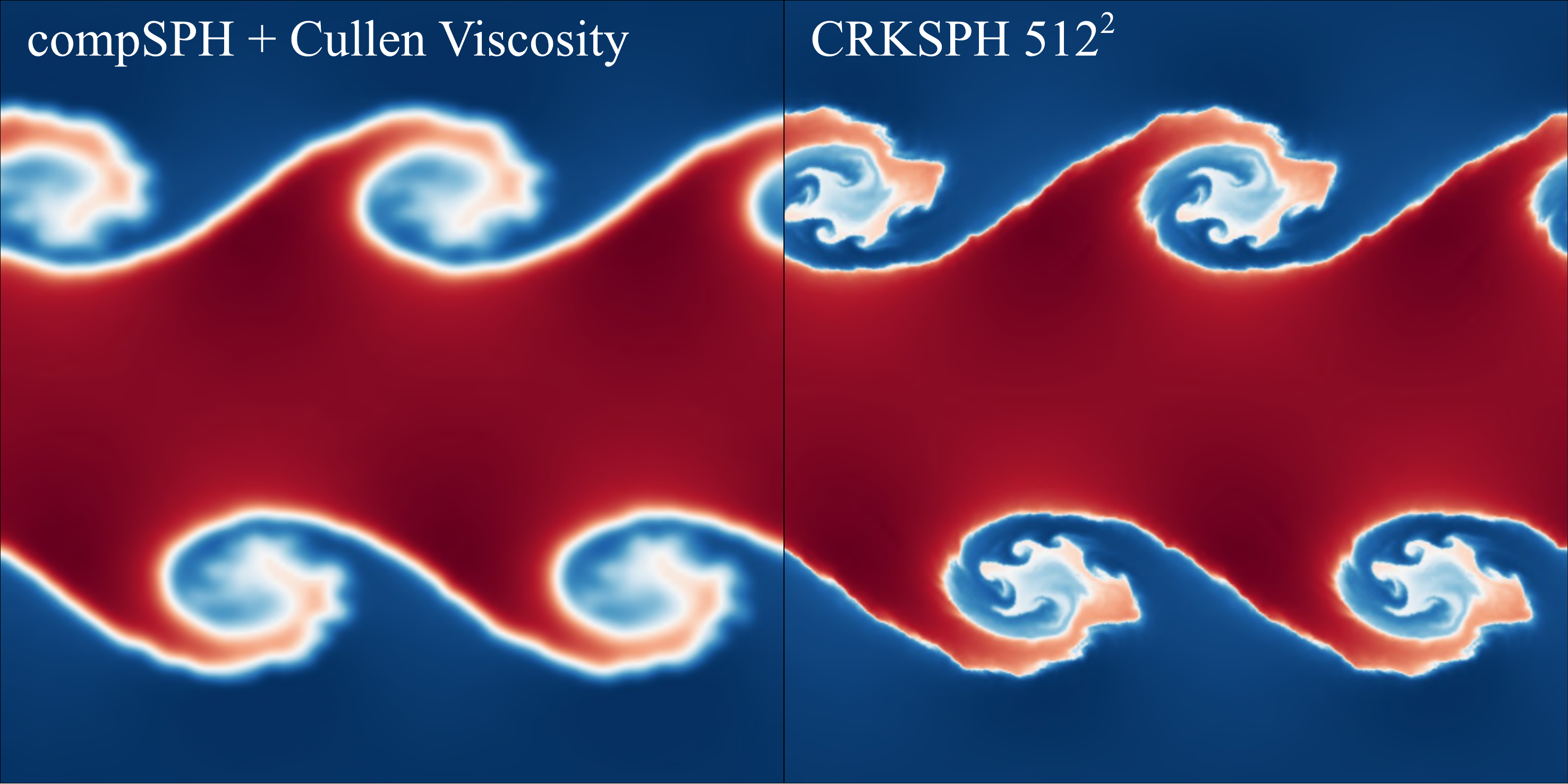}
\caption{Top: pseudocolor snapshots of the mass density at $t=2.0$ ($\approx 2\tauKH$) of the growth of a Kelvin-Helmholtz instability for compSPH, PSPH, and CRKSPH, all with $256^2$ particles. Bottom: same as above for compSPH with the Cullen form of viscosity (left) and a higher resolution CRKSPH result at $512^2$ particles (right). The over-diffusion of the compSPH viscosity treatment is apparent in the first panel. All other treatments visually demonstrate appropriate qualitative mixing behavior with slight variations of diffusivity in the vortex when compared to the higher resolution run. These results indicate that the dominant compSPH error can be attributed to the overactivity of the standard viscosity model. }
\label{fig:KHCompare}
\end{figure*}
The classical Kelvin-Helmholtz (KH) fluid mixing instability arises when a velocity shear occurs over a perturbed interface.  The shearing motion is transformed into growing vortical motion by the KH instability, a process which can be important for a variety of physical phenomena in astrophysics, and, in general hydrodynamics \cite{McNally2012}.  Traditional SPH has previously demonstrated unphysical suppression of such fluid instabilities, particularly in the presence of density inhomogeneities.  This issue was the focus of a study by \cite{Agertz2007}, demonstrating that for relevant astrophysical scenarios, unaltered SPH seriously underrepresents the effects of this type of fluid instability. In a succeeding study comparing SPH and grid codes on an idealized KH model, \cite{McNally2012} quantitatively confirmed problems in standard SPH modeling of such phenomena.   The dominant sources of error for this problem in SPH appear to be two-fold: the E0 error (such as described in the box tension test in \cref{sec:HydroBox}) and over-activity of the artificial viscosity.  It is clear that an error term that behaves like an unphysical surface tension, e.g.  the E0 error, can retard the growth of an unstable interface. 
As was found in \cite{Agertz2007}, if one removes the density discontinuity from such unstable interfaces the inaccuracies of the E0 error are alleviated, leaving the artificial viscosity the dominant difficulty -- in such scenarios so long as the viscosity is adequately suppressed SPH is capable of following the resulting fluid instabilities. 
Indeed, when considering phenomena involving shearing and/or vortical flows, the damping introduced by a non-limited artificial viscosity -- as noted in the discussion of the Gresho vortex (\cref{sec:Gresho}) -- is an important source of diffusion that can damp the growth of instabilities.
Over the years, a number of investigators have suggested remedies for SPH models of the KH instability, such as introducing artificial heat conduction \cite{Price2008}, increasingly sophisticated filters applied to the artificial viscosity \cite{Balsara1995,Cullen2010}, efforts to apply such viscous limiters more broadly in the SPH equations \cite{Read2010,Read2012}, replacing the weighting of ordinary SPH with functions of pressure to form PSPH \cite{Saitoh2013,Hopkins2012}, and Godunov based hybrid methods such as MFM and MFV \cite{Hopkins2015}.  In this section, we examine how  CRKSPH fares on this test case.  For comparison, we test compSPH (both using the simple Monaghan-Gingold viscosity and the full sophisticated treatment of Cullen-Dehnen) and PSPH.

In our example, we use the smooth 2D Kelvin-Helmholtz initialization described by \cite{McNally2012}: a shear flow in an ideal gas of adiabatic index $\gamma = 5/3$ is initialized in a periodic box of unit length and uniform pressure $P=2.5$. The smooth interface transitions are characterized by the density and $x$-velocities in regions, namely
\begin{align}
  \label{eq:KH}
  \rho(y), \; v_x(y) = \left\{\begin{array}{l@{\quad}l@{\quad}l}
    \rho_1 - \rho_m e^{(y-1/4)/\Delta}, & v_1 - v_m e^{(y-1/4)/\Delta}, & y \in [0, 1/4] \\
    \rho_2 + \rho_m e^{(1/4-y)/\Delta}, & v_2 + v_m e^{(1/4-y)/\Delta}, & y \in [1/4, 1/2] \\
    \rho_2 + \rho_m e^{(y-3/4)/\Delta}, & v_2 + v_m e^{(y-3/4)/\Delta}, & y \in [1/2, 3/4] \\ 
    \rho_1 - \rho_m e^{(3/4-y)/\Delta}, & v_1 - v_m e^{(3/4-y)/\Delta}, & y \in [3/4, 1] \\
  \end{array}\right.
\end{align}
with regional densities $\rho_1=1.0$, $\rho_2=2.0$, $\rho_m=(\rho_1-\rho_2)/2$, velocities $v_1=0.5$, $v_2=-0.5$, $v_m=(v_1-v_2)/2$ and smoothing parameter $\Delta = 0.025$. The $y$-velocity perturbation is initialized to be $v_y(x)=\delta_y \text{sin}(2\pi x/\lambda)$, with mode amplitude $\delta_y=0.01$ and wavelength $\lambda = 1/2$. The classical growth-rate \cite{Chandra1961} expected for a sharp interface layer is 
\begin{align}
  \label{eq:tauKH}
  \tauKH = \frac{(\rho_1 + \rho_2)\lambda}{\sqrt{\rho_1\rho_2} |v_1 - v_2|}.
\end{align}
For our parameters, we expect $\tauKH \approx 1.06$.  Note, however, we do not have a sharp transition region by design, so this estimate is approximate.

\begin{figure}[ht]
\centering
\includegraphics[width=0.45\textwidth]{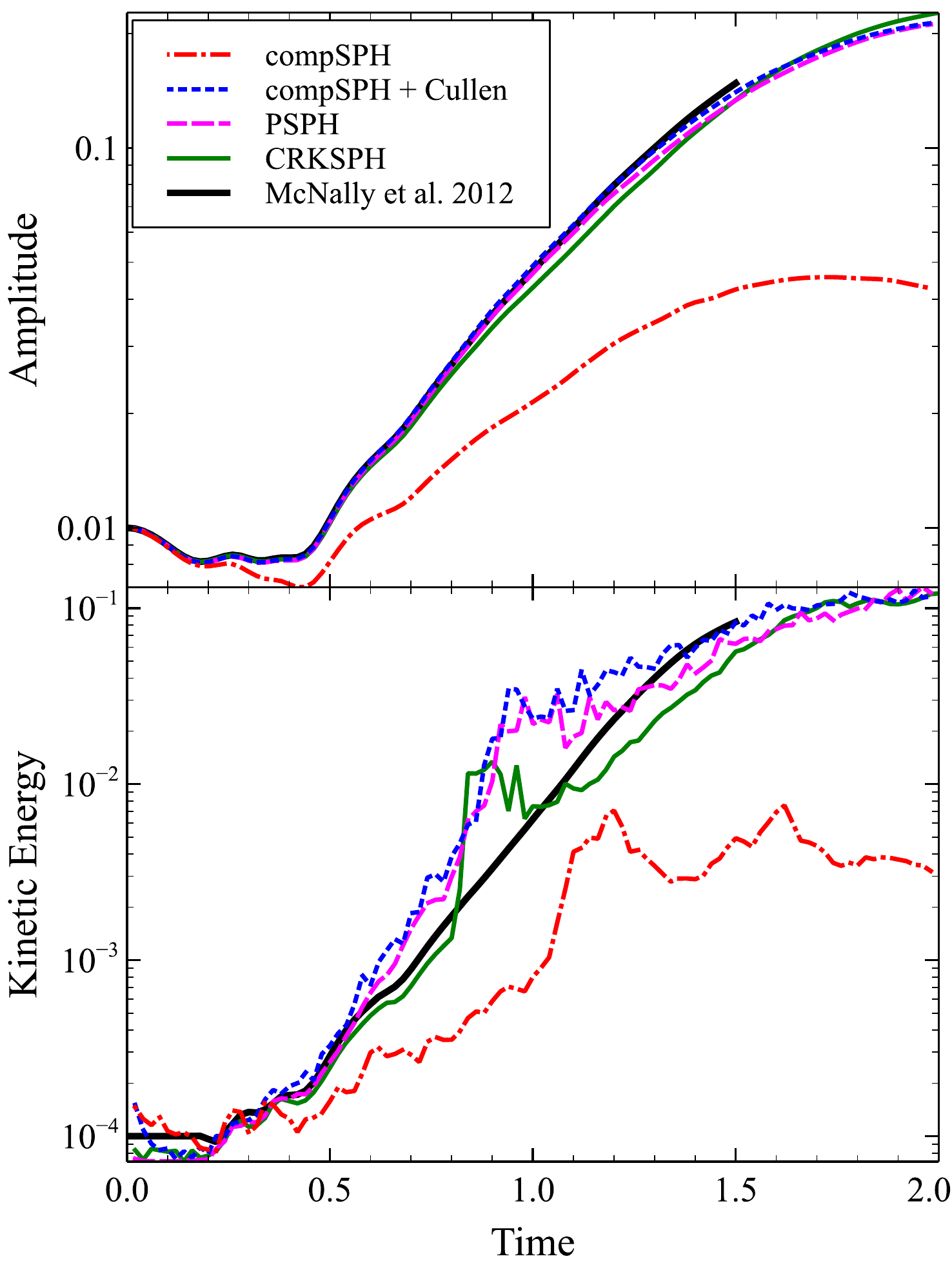}
\caption{Top: the $y$-velocity mode (defined in \cite{McNally2012}) of the mixing region as a function of time for the five $256^2$ models of the Kelvin-Helmholtz instability, compared with the high-resolution reference of \cite{McNally2012}.  Bottom: time evolution of the maximum $y$ direction kinetic energy in the same calculations. Once again, the overly active viscosity in compSPH diffuses the solution. All three remaining methods do a reasonable job of matching the reference of \cite{McNally2012}, where CRKSPH is slightly damped towards later times. As for the kinetic energy, CRKSPH reproduces the reference solution more accurately than the other methods, before slowing down for $t > 1.2$. The late-time slowing of the CRKSPH mixing growth is primarily due to the build-up of small-scale noise, leading to activation of the artificial viscosity.}
\label{fig:KHMixing}
\end{figure}
\begin{figure}[ht]
\centering
\includegraphics[width=0.45\textwidth]{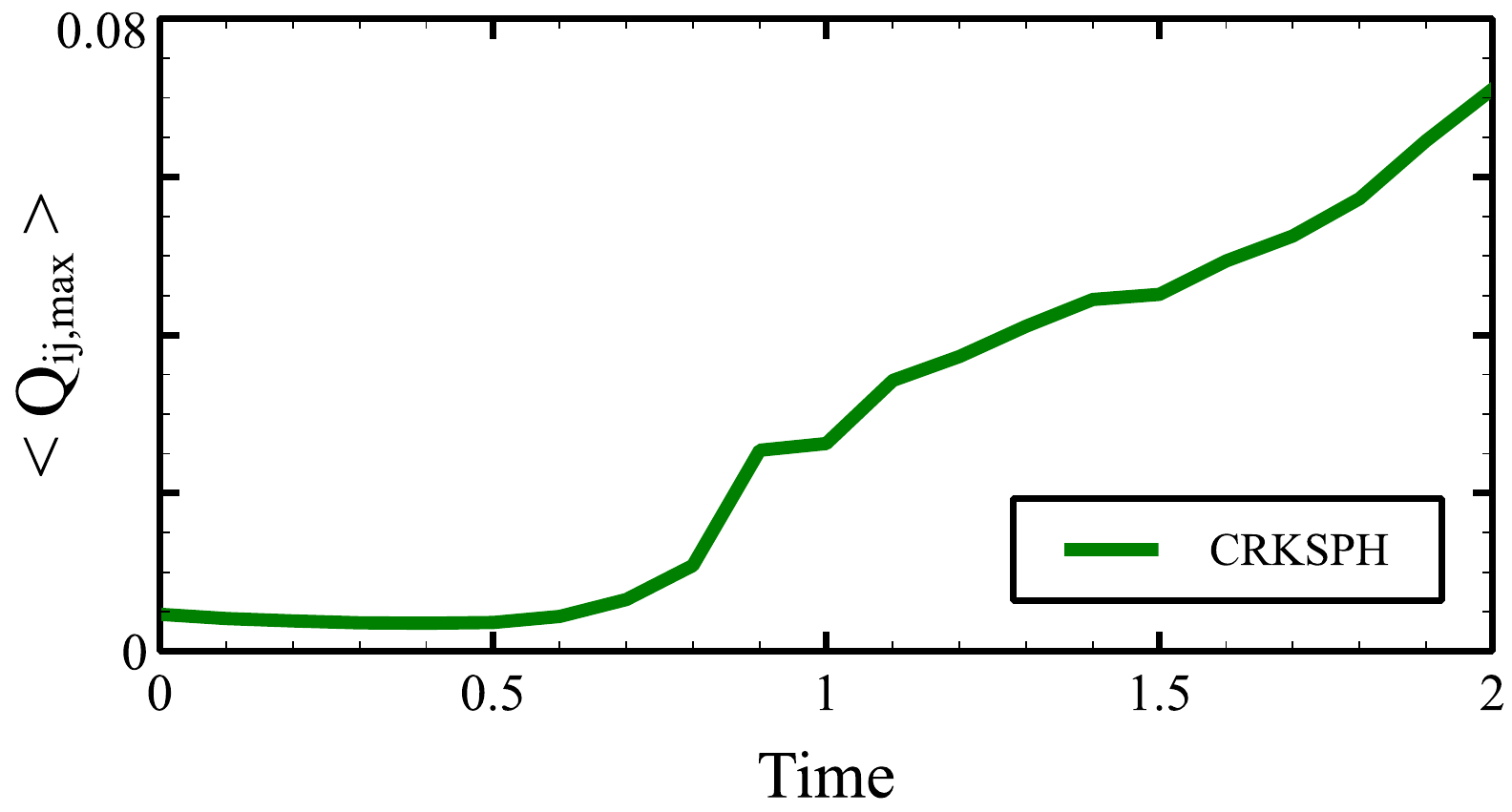}
\caption{The simulation-wide average of the maximum pair-wise viscous pressure over time for CRKSPH on the Kelvin-Helmholtz test. Note, the sharper increase around $t\approx1$, seeded by small error-modes, leads to the slight suppression of the mixing growth at late times seen in \cref{fig:KHMixing}, likely pointing the way to future improvements for CRKSPH.}
\label{fig:viscosityHistory}
\end{figure}
\Cref{fig:KHCompare} shows pseudocolor images of the density field at $t=2$ (roughly $2 \tauKH$) for four different models of the KH problem: compSPH, compSPH+Cullen, PSPH, and CRKSPH, all using $256^2$ points initially placed on a lattice.  For comparison, we also show a $512^2$ CRKSPH model.  Visually we can see that all calculations produce reasonable looking vortical mixing regions, with the exception of compSPH using the standard viscosity.  
Assuringly, compSPH+Cullen yields a very good result in this visual metric, suggesting that the artificial viscous damping is the dominant error term for compSPH, in agreement with \cite{Agertz2007} as we have a smoothed density transition where the E0 effects should be reduced. 
There is also some evidence of noise at different levels in each panel of \Cref{fig:KHCompare}, resulting is some feathery structures or secondary instabilities inside the vortices.

For a more quantitative comparison, \cref{fig:KHMixing} plots the growth of the $y$-velocity mode of the mixing region and the time evolution of the maximum $y$-direction kinetic energy ($\max\limits_i\frac{1}{2} m v^2_{y,i}$).  In order to facilitate quantitative comparisons with our results, the authors of \cite{McNally2012} kindly provided us with the highly resolved reference solution used in their paper, also plotted in \cref{fig:KHMixing}.  The $y$-velocity mixing mode, shown in the upper panel of \cref{fig:KHMixing}, is computed using the method described in \cite{McNally2012} (Eqs.~10 to 13 of that reference).  We find that all methods, with the exception of compSPH, do well on this measure.  The evolution of the maximum $y$-direction kinetic energy in the bottom panel of \cref{fig:KHMixing} is even more interesting.  We note that both PSPH and compSPH+Cullen show very similar evolution, while compSPH with the standard viscosity is markedly slowed relative to the reference solution.  This again suggests that the viscosity is the dominant source of error in the compSPH example, and if we reran the PSPH case with the standard SPH viscosity of \crefrange{eq:visc}{eq:unlimmu} it would  badly suppress the KH evolution.  The CRKSPH kinetic energy is remarkably close to the reference solution (closer than any other method) until $t > 1$, after which time it follows a similar slope just below the reference solution.  It is also worth noting all the results shown here are closer to the reference in this figure than the SPH comparisons in \cite{McNally2012}, e.g., the bottom three panels of Figure.~8 of \cite{McNally2012}.

Overall, CRKSPH performs well for the KH test, but not without some caveats.  It matches the reference solution at least as well as the other meshfree methods in the mixing amplitude, and outperforms the others in the kinetic energy comparison to the reference.  However, there is evidence of some noise arising in these calculations -- this is the source of the deviations from perfect rollup in the vortices in \cref{fig:KHCompare}, and perhaps some slowing of the major mixing scale at late time.  These effects are related; as small-scale noise in the calculation begins to grow from small-scales upward, it eventually begins to trigger the artificial viscosity to damp variations in the pair-wise velocity.  This is illustrated in \cref{fig:viscosityHistory} where we plot the average of the maximum pair-wise viscous pressure over time. What is most evident from the figure is the rapid growth of this average at $t\approx1$. Any triggering of the artificial viscosity will lead to spurious slowing of the mixing region growth, such as is hinted at at late times in \cref{fig:KHMixing}.  This is a key difference between the simple high-order limiting of the viscosity used in CRKSPH vs.~the physics based reasoning that goes into the Cullen-Dehnen viscous algorithm.  The CRKSPH viscosity will activate for any non-linear pair-wise particle motions, regardless of the type of flow the points are embedded in.  We view this as a desirable trait of the viscous method, as it errs toward a robust shock-capturing approach and does not involve complex analysis of overall flow and tuning of shear vs.~compressional terms.  That being said, the KH results potentially provide a clue for future improvements to the CRKSPH approach: likely CRKSPH would benefit from methods to prevent the appearance and growth of such small-scale noise.

\subsubsection{Rayleigh-Taylor}
\label{sec:RayleighTaylor}
\begin{figure*}[ht]
\centering
\includegraphics[width=0.95\textwidth]{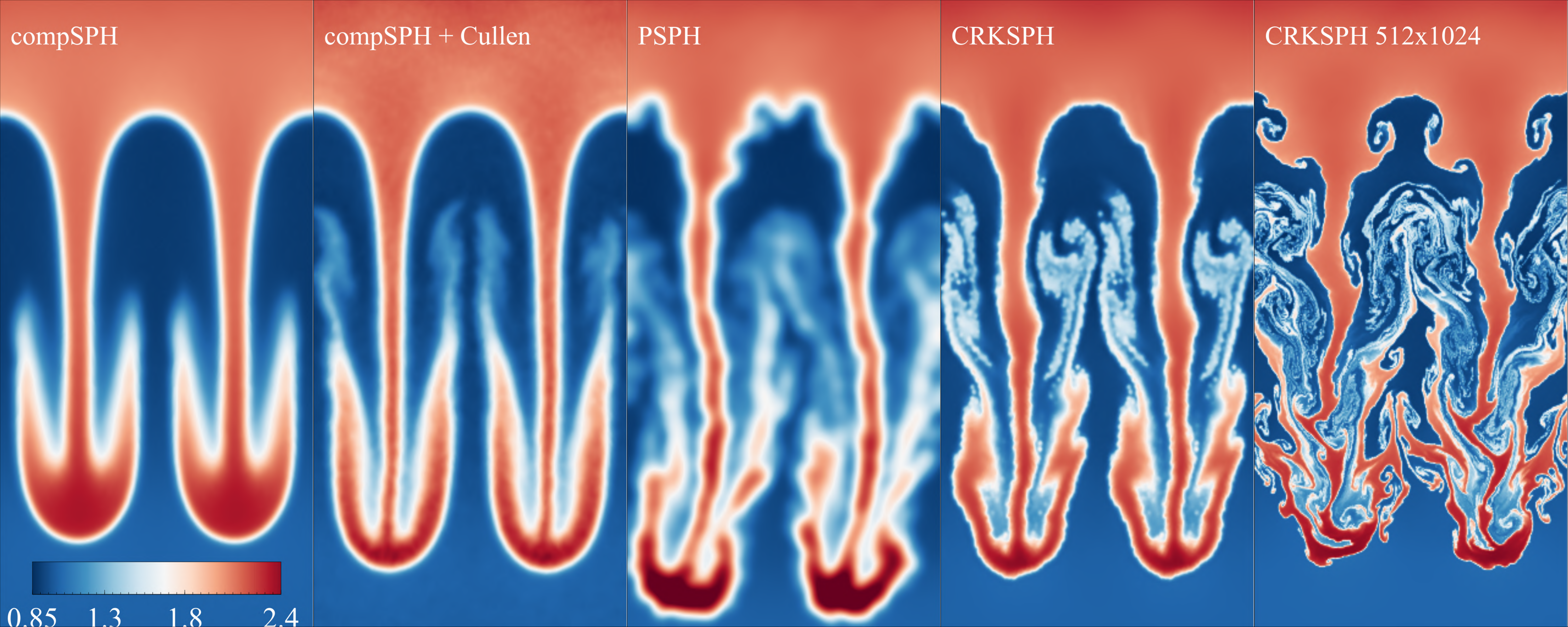}
\caption{Pseudocolor plots of the density in the Rayleigh-Taylor instability models.  From left-to-right is shown compSPH with Monaghan-Gingold viscosity, compSPH with the Cullen viscosity prescription, PSPH, and CRKSPH, all at $128\times256$ particle resolution at a time $t=4$. For reference we include a high-resolution run of 512x1024 using CRKSPH (far right). Both compSPH methods perform reasonably, primarily due to our resolution choice of 4 radial neighbors, as well as a quintic kernel to avoid pairing instabilities.  CompSPH shows more diffusion than the Cullen variant, as has been found in the other tests. PSPH shows the greatest plunge depth and the more perturbations growing along the rising bubble interfaces. CRKSPH maintains the sharpest interfaces of all the methods and is able to resolve secondary instabilities in the trailing plumes, all qualitative indications of less unwanted viscosity activation and noise control.  The high-resoultion CRKSPH calculation on the far right clearly shows the most growth of secondardy instabilities, but agrees with the lower-res calculations overall mixing layer size.}
\label{fig:RT}
\end{figure*}
Rayleigh-Taylor (RT) instabilities \cite{Rayleigh1883,Taylor1950c,Chandra1961} occur in a variety of astrophysical phenomena (e.g. \cite{Kane2000}), and have become a standard probe for the growth of subsonic perturbations \cite{Stone2008,Hopkins2015,Springel2010}.  The RT instability is another example where ordinary SPH has been demonstrated to substantially retard, or entirely suppress, the growth of the mixing layer \cite{Abel2011,Saitoh2013} -- yielding another useful test of the applicability of CRKSPH.  In their simplest form, RT instabilities grow from an interface between two fluids of differing density in a constant acceleration field, with the heavier fluid on top of the lighter.  In this example, we adopt the smoothed 2D RT problem setup described in \cite{Abel2011,Hopkins2015}: we assume a computational volume $(x,y) \in ([0,1/2], [0,1])$, wherein a dense fluid ($\rho_T=2.0$) is initialized for $y > 1/2$, resting atop a low density ($\rho_B=1.0$) fluid in the domain $y < 1/2$.  The system is subject to a constant gravitational acceleration $g = 1/2$ in the negative $y$ direction.  At the interface the density rolls smoothly between $\rho_T$ and $\rho_B$ according to
\begin{align}
  \rho(y) = \rho_B + (\rho_T-\rho_B)[1+e^{-(y-0.5)/\Delta}]^{-1}
\end{align}
with smoothing parameter $\Delta=0.025$. The interface is seeded with a smooth velocity perturbation in the $y$-direction according to
\begin{align}
  v_y(x,y)=\delta_y[1+\text{cos}(8\pi(x+0.25))][1+\text{cos}(5.0\pi(y-0.5))], y \in [0.3,0.7]
\end{align}
with amplitude $\delta_y=0.025$, and zero velocity otherwise.  We assume a single material $\gamma=1.4$ ideal gas in pressure equilibrium with the gravitational acceleration, giving us a pressure profile of
\begin{align}
  P(y) = P_0 - g \rho(y) (y - 1/2)
\end{align}
where $P_0 = \rho_T/\gamma$ such that the sound speed is near unity around the interface.  The computational volume is bounded by periodic boundaries in the $x$-direction ($x = 0$ and $x = 1/2$). In order to maintain pressure equilibrium along the vertical ($y$) direction, in the presence of the constant gravitational acceleration, we establish 20 extra rows of points above and below the problem (i.e., $y > 1$ and $y < 0$) and enforce the constant equilibrium conditions on these external ghost nodes.  This is similar to how we establish boundary conditions in the Kidder isentropic implosion (\cref{sec:KidderTest}) and Yee vortex (\cref{sec:Yee}).

\Cref{fig:RT} shows snapshots of the growth of the RT instability by $t=4$ for our four solvers. One thing of immediate note is that both compSPH models yield a credible result for this problem.  The ordinary compSPH calculation appears to suffer a bit more numerical diffusion vs.~the compSPH+Cullen case, as we might expect since the shearing flow along the interpenetrating material will spuriously activate the ordinary Monaghan-Gingold viscosity.  PSPH shows the greatest plunge depth, predicting a larger mixing layer than the other methods, and thicker transition regions along the interface between the heavy and light fluids.  Gratifyingly, CRKSPH demonstrates reasonable and consistent growth of the instability (comparing low and high-res CRK in the depth of the mixing layer), and the sharpest transition between the heavy and light fluids.  CRKSPH also most clearly captures the secondary instabilities that cause roll-up and distortion of the trailing plumes along the spikes and bubbles of the primary perturbation for equivalent resolution vs.~compSPH or PSPH.  The high-resolution ($512 \times 1024$) CRKSPH model demonstrates much more growth in the Kelvin-Helmholtz driven secondary instabilities along the bubbles and spikes, and even captures the roll-up along the sides of the rising bubbles (i.e., the blue material rising into the red).  Due to the lack of a physical viscosity to damp such growth we should expect these structures to occur as we go to higher and higher resolution even though they are numerically seeded, but the consistent scale of the overall mixing layer going from low to high-resolution is encouraging.

Overall, it appears CRKSPH handles this problem well: it shows reasonable growth of the mixing layer, the least evidence for numerical diffusion that would wash out details (e.g. the rollups along the plunging plumes), and the sharpest transition layer between the heavy and light regions. It is also worth noting that both our compSPH examples perform reasonably on this problem (compare for example with the SPH results in Figures 4 \& 5 of \cite{Abel2011}); this is also true, to a lesser extent, for the KH results in \cref{sec:KelvinHelmholtz}. We have found the difference can be attributed to our resolution and kernel choice. It is well known that increasing the number of neighbors per point improves SPH results on problems such as this RT example.  As mentioned previously, we chose our number of neighbors in order to give the best results for all the methods examined (settling on 4 radial neighbors for all of the presented tests), preferring resolved solutions over the computational savings of undersampling per point. Combining the increased neighbor count with a quintic or higher-order kernel -- importantly used to mediate pairing instabilities -- results in a significantly improved solution over the traditional SPH implementations that use cubic splines with low neighbor counts. That all being said, CRKSPH goes beyond these improvements, and demonstrates enhanced mixing for fluid instabilities, as shown in the above examples. 

\subsubsection{Combining Shears \& Shocks -- the Shearing Noh Test}
For our last idealized test case, we turn our attention to a problem combining both strong shocks and shearing flows.  The shearing Noh problem was first introduced in \cite{Owen2004}, and is designed as an extreme test of shock hydrodynamics codes in the presence of an arbitrarily strong shear.  This is a 2D problem, based on the planar Noh test case (\cref{sec:Noh}).  The idea is to establish a planar Noh-like shock, propagating orthogonally across a shearing flow.  We create a 2D domain $(x,y) \in ([0,0.2], [0,1])$ filled with a $\gamma=5/3$ ideal gas of initially unit density and zero pressure.  The velocity field is initialized to be
\begin{align}
  \begin{pmatrix} v_x \\ v_y \end{pmatrix} =
  \begin{pmatrix}
    v_s \cos(2 \pi y_i) \\
    -1
  \end{pmatrix}.
\end{align}
Periodic boundaries are enforced at the $x$-boundaries ($x=0$ and $x=0.2$), while a reflecting condition is created at $y=0$.  Note that for $v_s=0$ this is simply the ordinary planar Noh problem running in the $y$-direction.  However, for $v_s \ne 0,$ there is a continuously shearing component added to the velocity in the $x$-direction.  Since the problem is periodic in $x,$ we expect the points to endlessly cycle around the computational volume in the $x$-direction, while the planar Noh problem progresses up the $y$-direction.  As we turn up the shearing component, this becomes an extremely challenging problem for any numerical hydrodynamic scheme.  It is critical that the method be able to separate the convergent and shearing components of the velocity field, a test most solvers struggle with resulting in greater and greater departures away from the analytic solution as more shear is applied.  Additionally, because the fluid is shearing, it will be subject to the Kelvin-Helmholtz instability.  We are not inserting any perturbation into the initial conditions, however, if the solver introduces any perturbative numerical errors, the Kelvin-Helmholtz instability will cause such perturbations to grow,  resulting in an even larger departure from our analytic expectation.  In \cite{Owen2004}, this problem was designed as a torture test for artificial viscosity methods; here, we are interested in how well our modified CRKSPH viscosity handles this situation, while also examining how well symmetry is maintained, or, rather, not lost to Kelvin-Helmholtz amplified numerical noise.

\begin{figure}[ht]
  \centering
  \begin{tabular}{cc}
   $\boldsymbol{\;\;\;\;\;\;v_s = 1}$ & $\boldsymbol{\;\;\;\;\;\;v_s = 5}$ \\
    \includegraphics[width=0.45\textwidth]{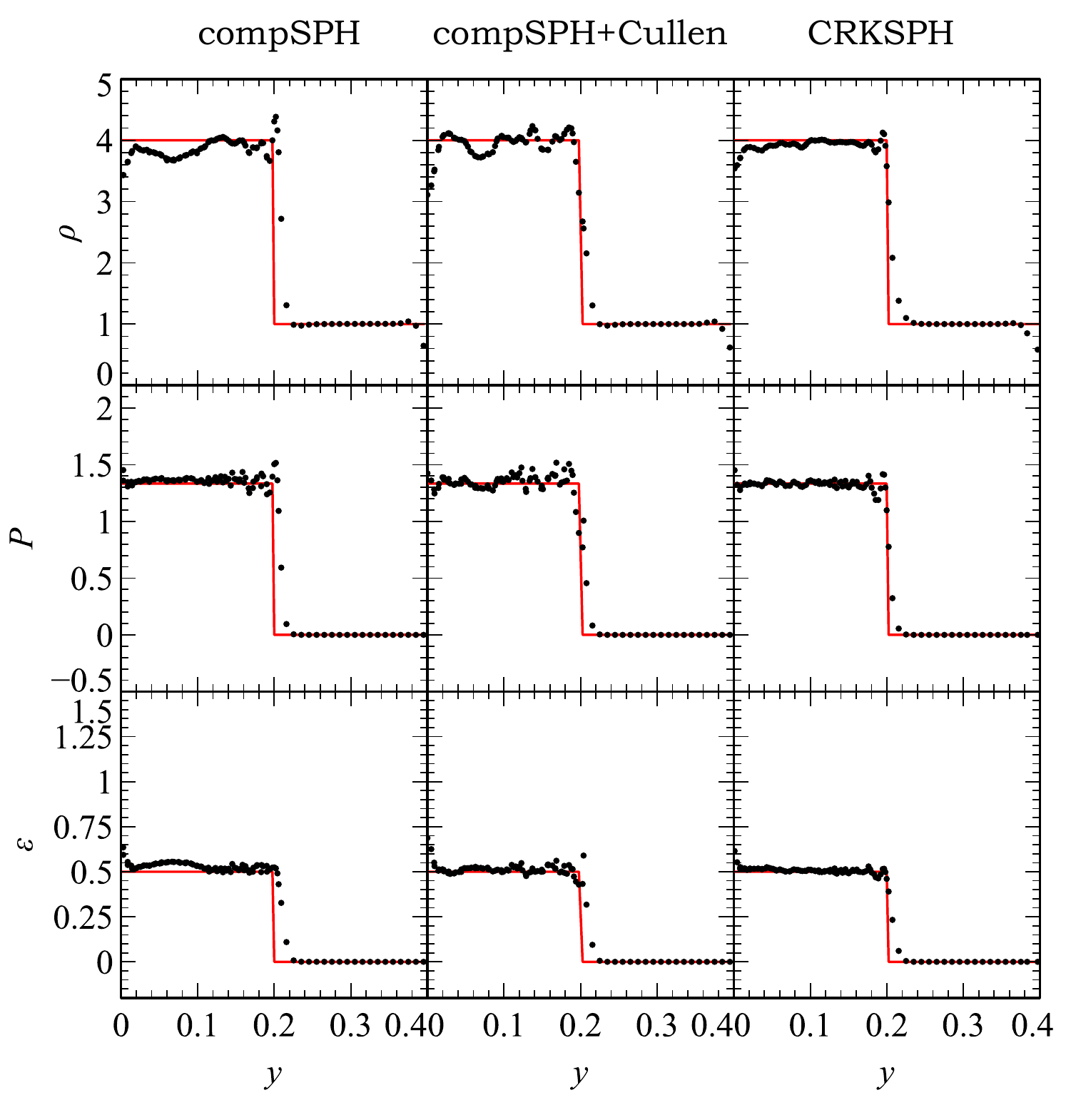} &
    \includegraphics[width=0.45\textwidth]{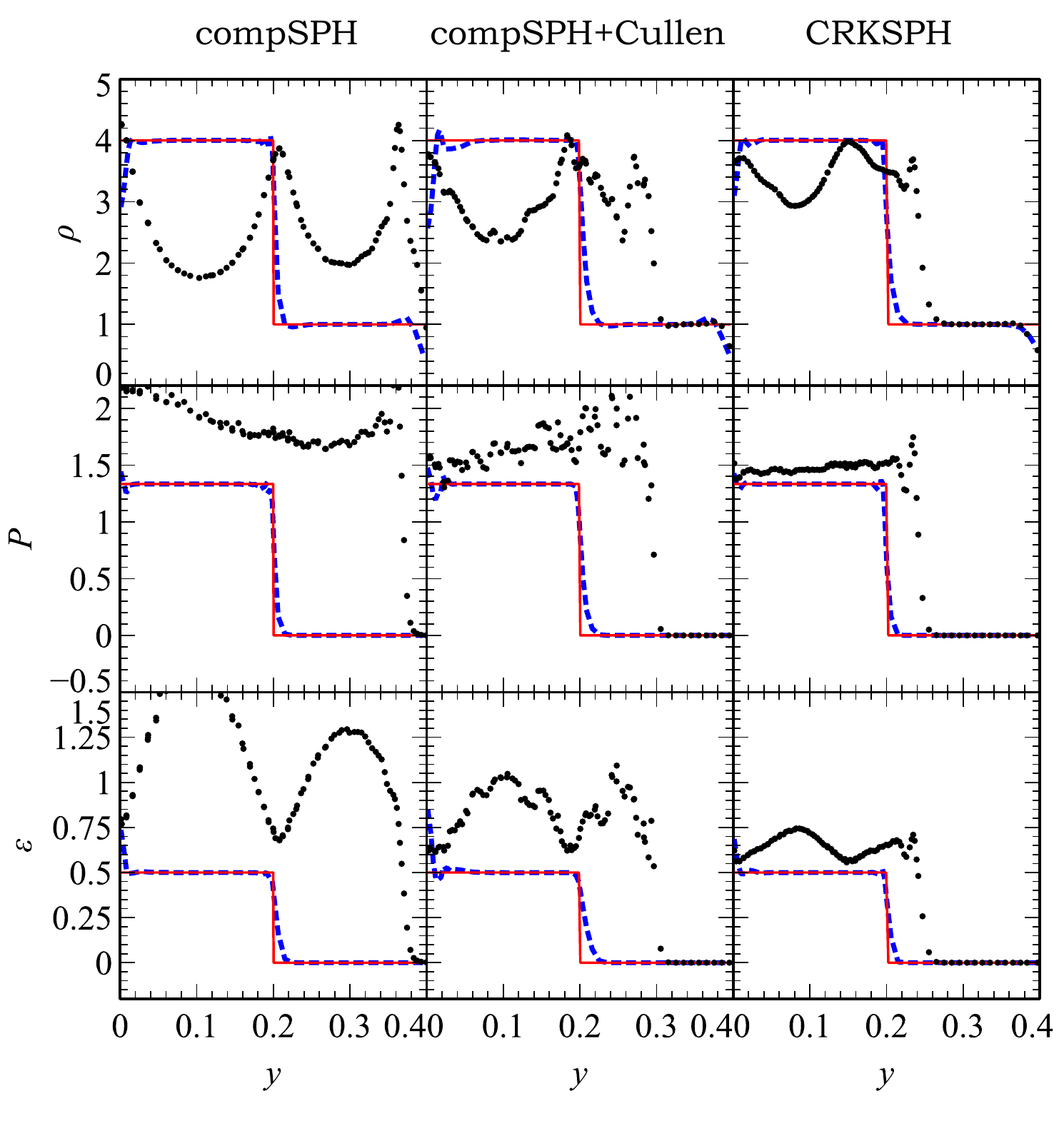}
  \end{tabular}
  \caption{Results of the Noh shearing test for $v_s=1.0$ (left) and $v_s=5.0$ (right) at resolution $N_y=100$ in the shock direction (as was true in \cref{fig:noh-planar}), and $N_x=20$ in the shearing component. The analytic solution is drawn red. For reference, we include the numerical results of the planar Noh test from \cref{sec:Noh} in blue (on the right). All of the methods maintain reasonable solutions in the low-shear case, although clearly not as accurately as the shear-free planar Noh reference results. CRKSPH demonstrates the best match to the analytical solution. In the high-shear case, the CRKSPH improvement is more notable: it resolves the shock front position more accurately, and possesses the least noise, albeit still producing a solution quite far from the shear-free case, a testament to the difficulty of the problem setup.}
  \label{fig:noh-shear}
\end{figure}
We examine the compSPH, compSPH+Cullen, and CRKSPH solutions of this problem for two different shear components: $v_s=1$ and $v_s=5$.  In each case we use $(N_x, N_y) = (20 \times 100)$ points initially seeded on a lattice in the domain $(x,y) \in ([0,0.2], [0,1])$, and run to time $t=0.6$, where the shock is predicted to be at $y=0.2$.  \Cref{fig:noh-shear}, plots the final profiles as a function of $y$ vs.~the analytically expected (planar) Noh solution for the density, pressure, and specific thermal energy.  Note, in each of the panels we have plotted the results for all points in these simulations; thus, if the 20 points in the $x$ direction (per row) from the initial conditions maintain the expected planar symmetry, we should see no scatter in these profiles.  All three methods demonstrate excellent maintenance of this symmetry, showing little scatter for the most part.  For the moderate shear case ($v_s=1$, left panel of \cref{fig:noh-shear}) we see that the models match the analytic prediction reasonably well, though certainly not as well as we see in the generic planar Noh problem in \cref{fig:noh-planar} (redrawn in blue on the right panel for reference).  The degradation of compSPH is expected, as the Monaghan-Gingold viscosity is unable to distinguish the shearing component of the velocity field from compression, forcing compSPH to deviate from the solution due to unphysical heating contributed by this shearing component.  Interestingly, the addition of the Cullen viscosity modifier for compSPH+Cullen does not yield a better solution. If anything, there is the most evidence of scatter in the profiles, as well as the same or more deviation from the analytic expectation, when comparing compSPH+Cullen vs.~compSPH alone.  The CRKSPH model by contrast does the best on this problem, showing very little scatter and a good match to the analytic solution for the $v_s=1$ case.

Turning our attention to the extreme shearing case of $v_s=5$ (right-side of \cref{fig:noh-shear}), we see that all three methods struggle.  The unphysical heating of the ordinary Monaghan-Gingold viscosity causes the ordinary compSPH model to miss the shock position by nearly a factor of two, greatly underpredicting the post-shock density while over-predicting the post-shock thermal energy and pressure.  The addition of the Cullen viscosity in the center column improves the shock position somewhat (though it is still off by 50\% in $y$). Unfortunately, the refined shock position is at the cost of increasing noise/scatter in the post-shock profiles, which is most evident in the pressure.  The CRKSPH model suffers the least unphysical viscous heating due to the shear, though the error is still far from negligible with the shock position off by nearly 25\% in $y$.  CRKSPH shows the best match to the post-shock solution and demonstrates less scatter/symmetry compared with the other methods.

Based on these results CRKSPH holds up to this extreme test well.  In order to do a better job, we will likely need to extend the CRKSPH formalism with a tensor viscosity in order to properly account for directionality in the shock physics (such as was the subject of \cite{Owen2004} for which this problem was designed). Moreover, particle regularization treatments can reduce the perturbation errors, seeded by particle scatter, which were unnecessarily amplified. Nonetheless, this problem is illuminating, and demonstrates the utility of the CRKSPH artificial viscosity limiter for complex multi-dimensional problems such as this.  It is also worth noting that the effects this problem is testing are not esoteric; in many astrophysical scenarios, modeling shocks over shearing flows is a potentially important process, such as gas falling into a galaxy, inflow to accretion disks, inflow to proto-planetary disks, etc.

\subsection{Practical Capability Demonstrations}
Thus far, we have focused on idealized test cases that are typically accompanied by analytic solutions or expectations. For our final tests, we examine two scenarios that are more akin to practical examples studied with numerical hydrodynamic techniques, albeit with simpler constructions; namely, we investigate the popular ``Blob" test of \cite{Agertz2007}, as well as the demanding multi-material ``Triple point shock" problem.  In both examples, not only are compressible strong shocks present -- critically requiring a conservative solver -- but they also include complex vortical or shearing flows. Our goal here is to examine how the CRKSPH dynamic equations, combined with our improved viscosity treatment, handle these problems relative to SPH. 
\subsubsection{``Blob'' Test}
\label{sec:blob}
In 2007, \cite{Agertz2007} presented a systematic comparison of SPH and Eulerian mesh-based methods applied to the problem of a dense gaseous blob embedded in a diffuse supersonic wind.  Their investigation found a stark difference between the examined methods: Eulerian models showed a disruption of the dense blob, due to Kelvin-Helmholtz (KH) and Rayleigh-Taylor (RT) instabilities interacting with the wind and complicated by the presence of a bow-shock in front of the blob; SPH models, however, tended to suppress background mixing of the dense blob, even over multiple Kelvin-Helmholtz times ($\tauKH$, \cref{eq:tauKH}).  This problem demonstrated conclusively SPH's deficiency at modeling mixing instabilities, and spurred a number of efforts to remedy this issue (e.g. \cite{Price2008,Read2010,Springel2010,Hess2010,Saitoh2013,Hopkins2012,Hopkins2015}).  In this section, we examine how CRKSPH fares on this test case, with comparisons to compSPH for context.

The test consists of an initially spherical cloud (the ``blob'') of density $\rhocloud=10,$ embedded in a background material of density $\rhoam=1$; both materials are $\gamma=5/3$ ideal gasses, initialized in pressure-equilibrium with value $P_0=1$.  The cloud begins at rest, whereas the light background material is born with a velocity valued at Mach number $\mathcal{M}=2.7$.  The formation of a bow-shock in front of the cloud precedes a turbulent evolution of ram-pressure stripping, shearing, and mixing of the blob into the background material. For a full analysis and discussion of the problem, we refer the reader to \cite{Agertz2007}. We cite the predicted Kelvin-Helmholtz growth rate defined over the ``crushing time'' of the cloud, viz. 
\begin{align}
  \label{eq:tauCR}
  \tauCR &= \frac{2 \rcloud \chi^{1/2}}{v} \\
  \label{eq:tauCloudKH}
  \tauKH &\approx 1.6 \tauCR
\end{align}
where $\rcloud$ is the radius of the cloud, $\chi=\rhocloud/\rhoam=10$ is the density contrast, and $v\equiv \mathcal{M}c_s$ is the velocity of the ambient background (with $c_s$ denoting the sound speed). $\tauKH$ is defined to be the approximate time when the largest (most disruptive) KH mode -- i.e. a wavelength $\propto$ cloud radius --  has fully grown, providing a reasonable time-scale for the cloud rupture. In our chosen scenario, these variables work out to be roughly $\tauCR \approx 1.81$ and $\tauKH \approx 2.9.$   
\begin{figure*}[ht]
  \centering
  \begin{tabular}{rcccc}
    & $1 \tauKH$ & $4 \tauKH$ & $8 \tauKH$ \\
    \raisebox{0.5\height}{\rotatebox[origin=l]{90}{compSPH}} &
    \includegraphics[height=3cm]{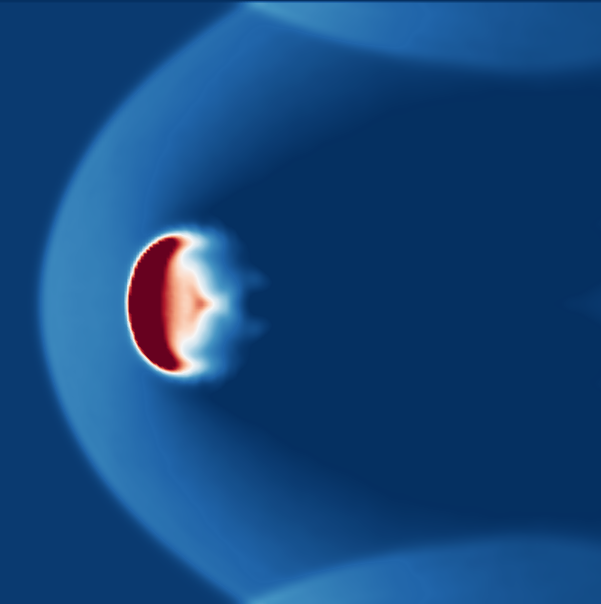} &
    \includegraphics[height=3cm]{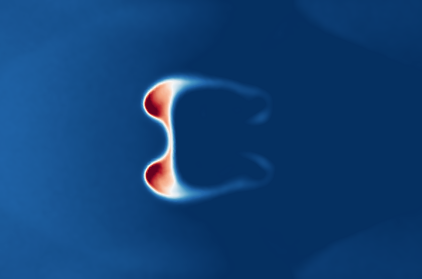} &
    \includegraphics[height=3cm]{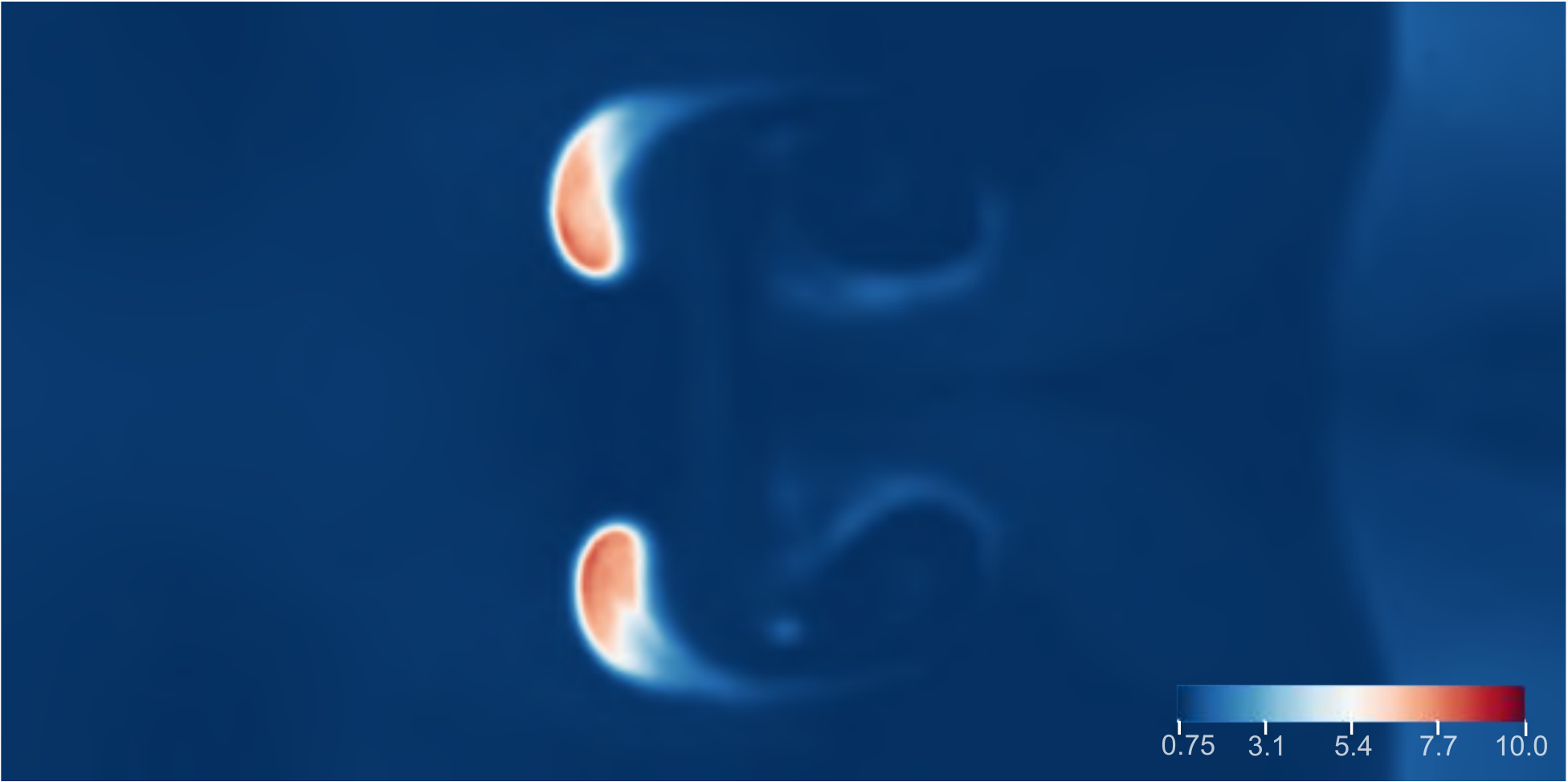} \\
    \rotatebox[origin=l]{90}{compSPH+Cullen} &
    \includegraphics[height=3cm]{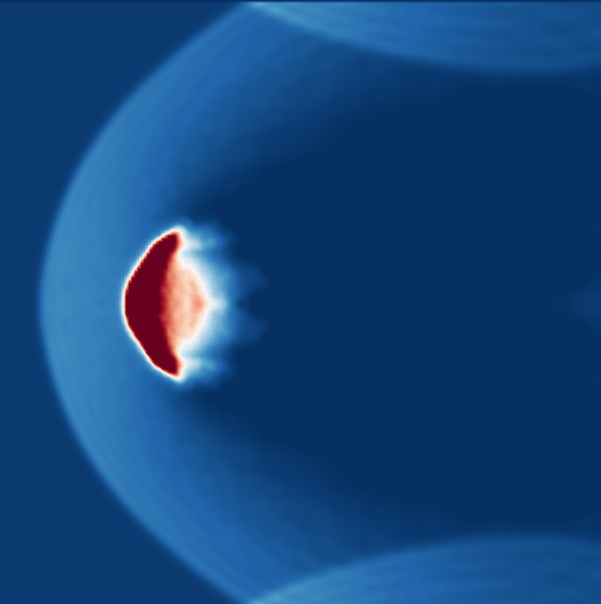} &
    \includegraphics[height=3cm]{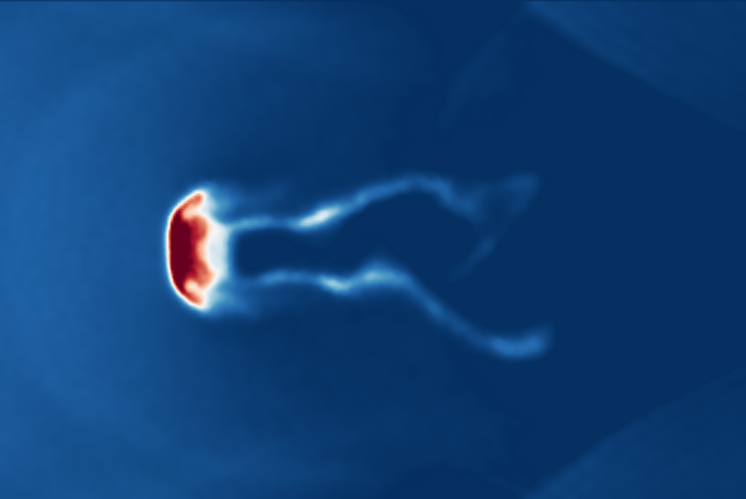} &
    \includegraphics[height=3cm]{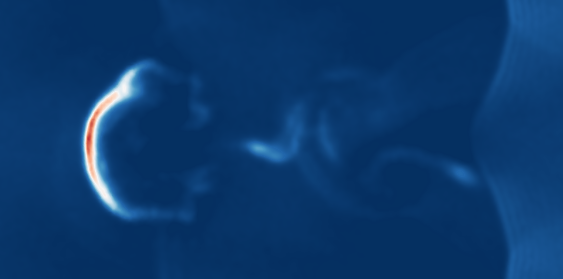} \\
    \raisebox{0.5\height}{\rotatebox[origin=l]{90}{CRKSPH}} &
    \includegraphics[height=3cm]{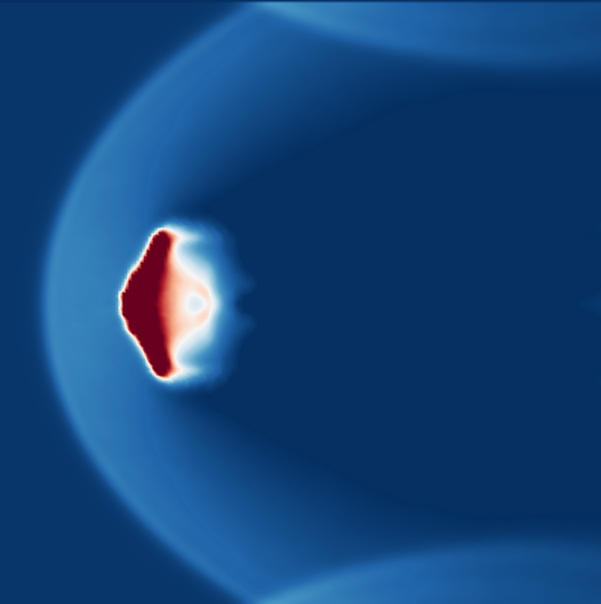} &
    \includegraphics[height=3cm]{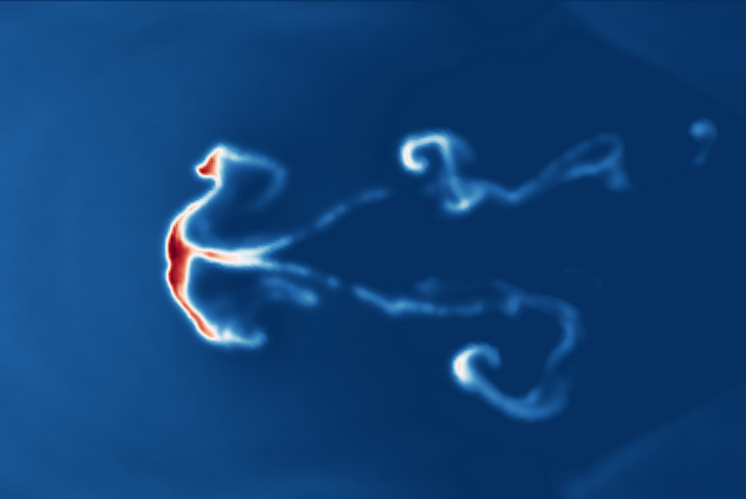} &
    \includegraphics[height=3cm]{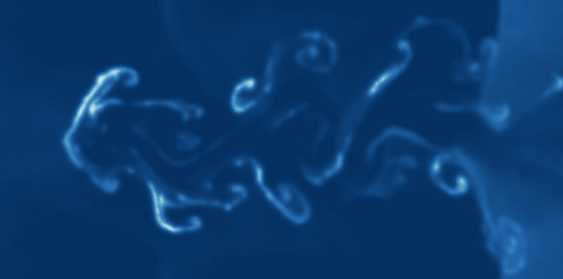}
  \end{tabular}
  \caption{Pseudocolor plots of the mass density in the 2D calculations of the blob test at times $t/\tauKH \in (1, 4, 8)$ for $512 \times 128$ particles in the ambient medium, and a cloud initially seeded with mass matched particles (ten times as dense).  All of the methods present similar solutions in the shock-dominated evolution shown in the first panel. The non-linear evolution in the remaining panels is another story, where the CRKSPH model evaporates the cloud by $t \approx 8 \tauKH$. On the other hand, the compSPH methods demonstrate the classic ``pancaking'' effect from incorrectly suppressing mixing due to E0 errors and overactive viscosity models, albeit improved with the Cullen prescription.}
  \label{fig:blob2d}
\end{figure*}

We begin with a 2D version of the problem (similar to the example shown in \cite{Cha2010}).  In this case we assume a rectangular volume $(x,y) \in ([0,40], [0,10])$ with periodic boundaries.  The cloud is initially centered at $(x_c, y_c) = (5,5)$ with radius $\rcloud = 1$, and the background wind material is moving in the positive $x$-direction.  We perform three simulations of this scenario -- compSPH, compSPH+Cullen, and CRKSPH -- using a resolution of $512 \times 128$ for the ambient medium seeded on a uniform lattice.  We excise a spherical region from this uniform background, and seed the cloud on a lattice with points mass matched to the background particles -- implying that the cloud points are 10 times more densely packed than the ambient material.  \Cref{fig:blob2d} shows a time-series of the mass density in the three calculations at times $t = $ 1, 4, and 8$\tauKH$. As expected, the methods agree on modeling the shock structure (shown in the first panel), but differ in the severity of cloud disruption: compSPH demonstrates the least mixing, resulting in a pancaking blob shape consistent with \cite{Agertz2007}; compSPH+Cullen evinces similar shock structures with more evolution of the blob; CRKSPH shows the most extreme distortion of the cloud due to instability growth, resulting in complete fragmentation.  Both compSPH results confirm the findings of prior studies; ordinary SPH appears to artificially retard the disruption of the cloud, likely due to an artificial surface tension seeded by E0 errors in combination with overactivity of the artificial viscosity.  The fact that adding the Cullen-Dehnen viscosity model improves the situation, somewhat demonstrates the attributable error to viscosity deficiencies. CRKSPH, on the other hand, shows the most realistic case, with further distortion and shedding of the cloud material.

\begin{figure*}[ht]
  \centering
  \begin{tabular}{rcccc}
    & $0.25 \tauKH$ & $1 \tauKH$ & $1.75 \tauKH$ & $4 \tauKH$ \\
    \raisebox{0.5\height}{\rotatebox[origin=l]{90}{compSPH}} &
    \includegraphics[height=3cm]{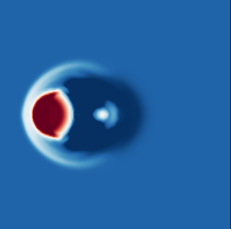} &
    \includegraphics[height=3cm]{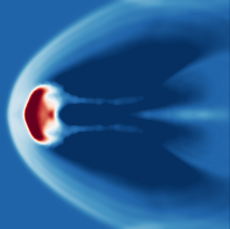} &
    \includegraphics[height=3cm]{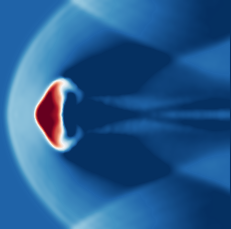} &
    \includegraphics[height=3cm]{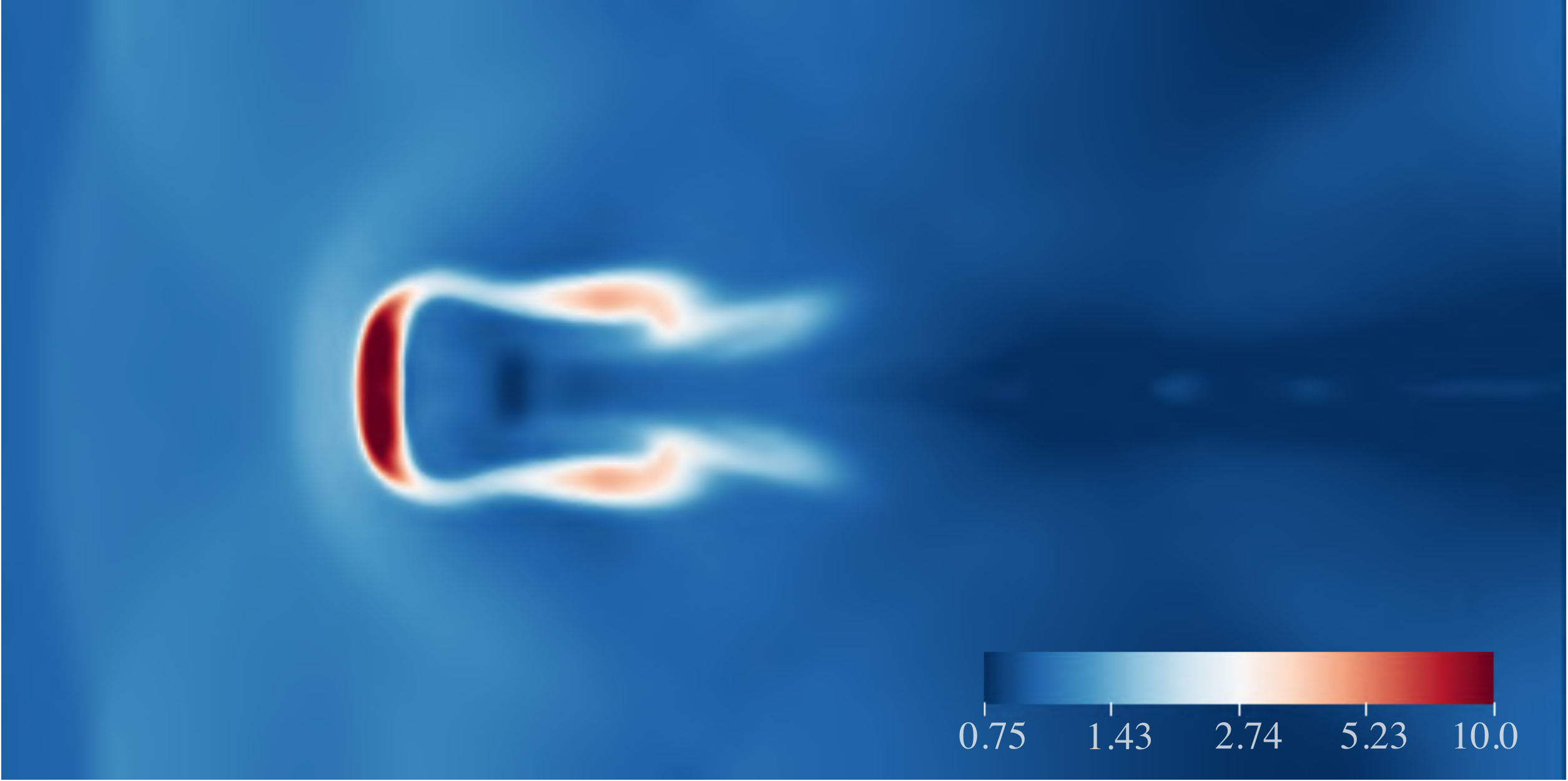} \\
    \rotatebox[origin=l]{90}{compSPH+Cullen} &
    \includegraphics[height=3cm]{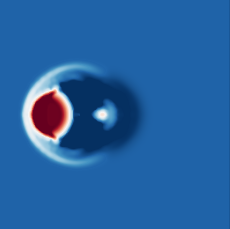} &
    \includegraphics[height=3cm]{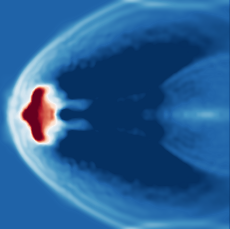} &
    \includegraphics[height=3cm]{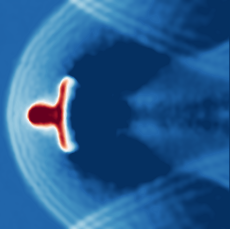} &
    \includegraphics[height=3cm]{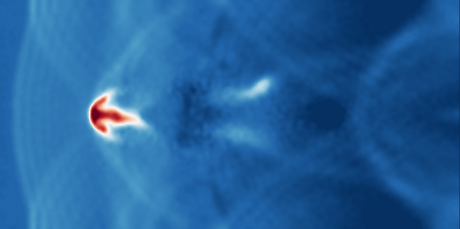} \\
    \raisebox{0.5\height}{\rotatebox[origin=l]{90}{CRKSPH}} &
    \includegraphics[height=3cm]{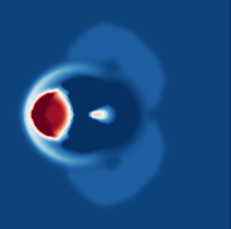} &
    \includegraphics[height=3cm]{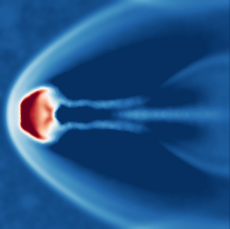} &
    \includegraphics[height=3cm]{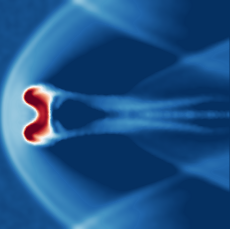} &
    \includegraphics[height=3cm]{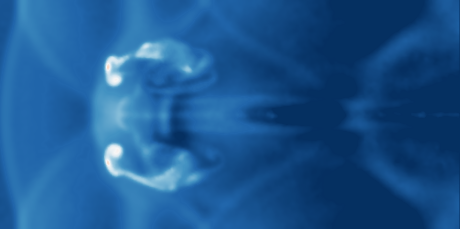}
  \end{tabular}
  \caption{Slices of the mass densities in the 3D calculations of the blob test along the $x=5$ plane at times $t/\tauKH \in (0.25, 1, 1.75, 4)$, for $128 \times 128 \times 512$ particles in the background medium and a mass matched cloud. As noted in \cref{fig:blob2d}, the methods agree in the early shock-dominated regime (here plotted at $t=0.25\tauKH$). However, as the evolution becomes increasingly dominated by non-linear fluid-instabilities,  the CRKSPH model effectively evaporates the cloud by $t \approx 3.5 \tauKH$, whereas the compSPH methods suppress full cloud disruption. }
  \label{fig:blob3d}
\end{figure*}
\Cref{fig:blob3d} shows the results for the full 3D blob test problem, presented at times $t = $ 0.25, 1, 1.75, and 4$\tauKH$.  In this case we model a periodic volume $(x,y,z) \in ([0,10], [0,10], [0,40])$, with the cloud centered at $(x_c, y_c, z_c) = (5,5,5)$, radius $\rcloud = 1$, and the background wind aligned in the positive $z$-direction.  We again seed the ambient medium points on an initial lattice (here of dimension $128 \times 128 \times 512$), with a sphere for the cloud excised and filled with mass matched points.  Similar to the 2D case, all schemes comparably model the shock-dominated dynamics; however, both compSPH models fail to entirely disrupt the cloud in the non-linear regime, whereas CRKSPH fully shreds the blob (as seen in the last panel).

\begin{figure}[ht]
  \centering
  \begin{tabular}{cc}
    2D & 3D \\
    \includegraphics[width=0.4\textwidth]{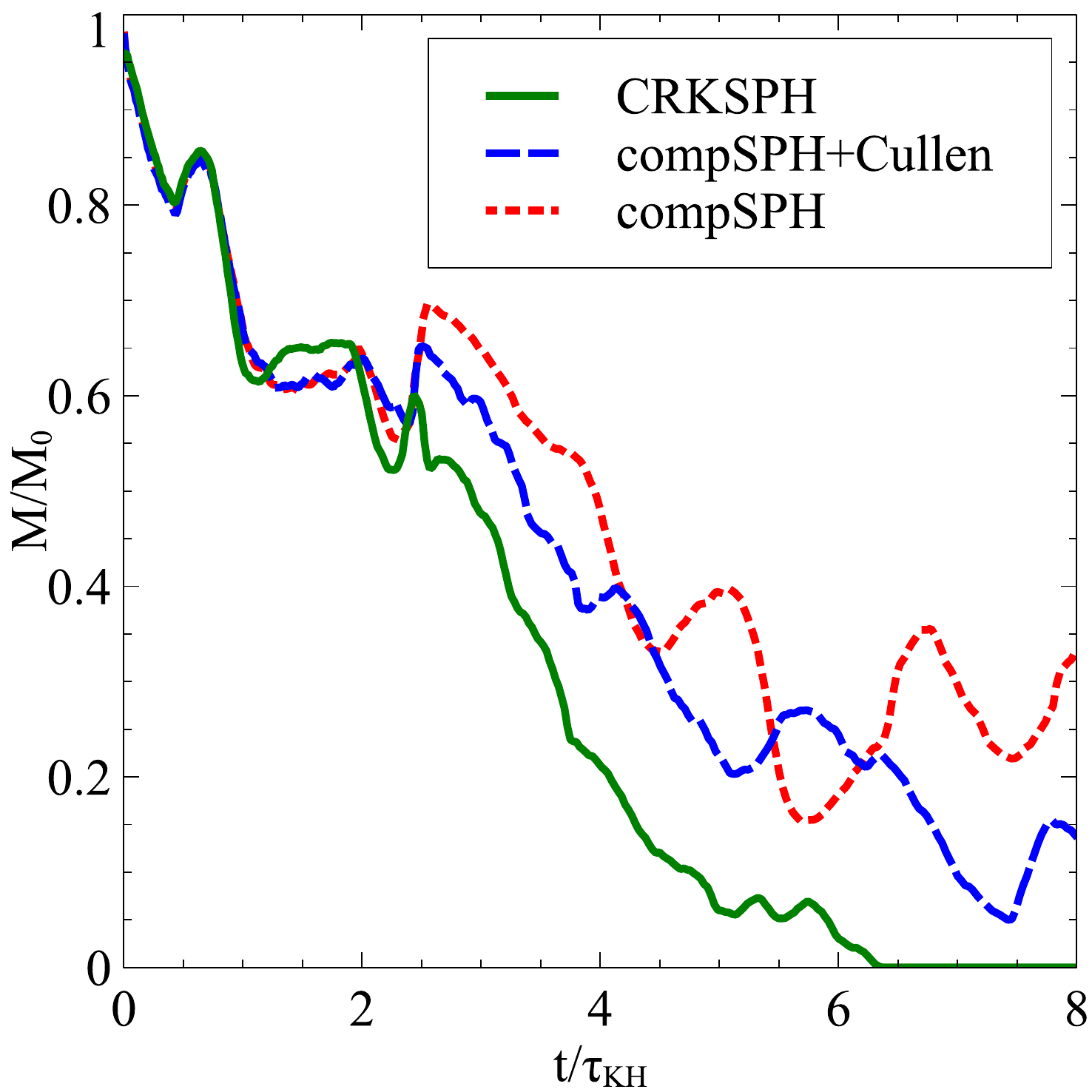} &
    \includegraphics[width=0.4\textwidth]{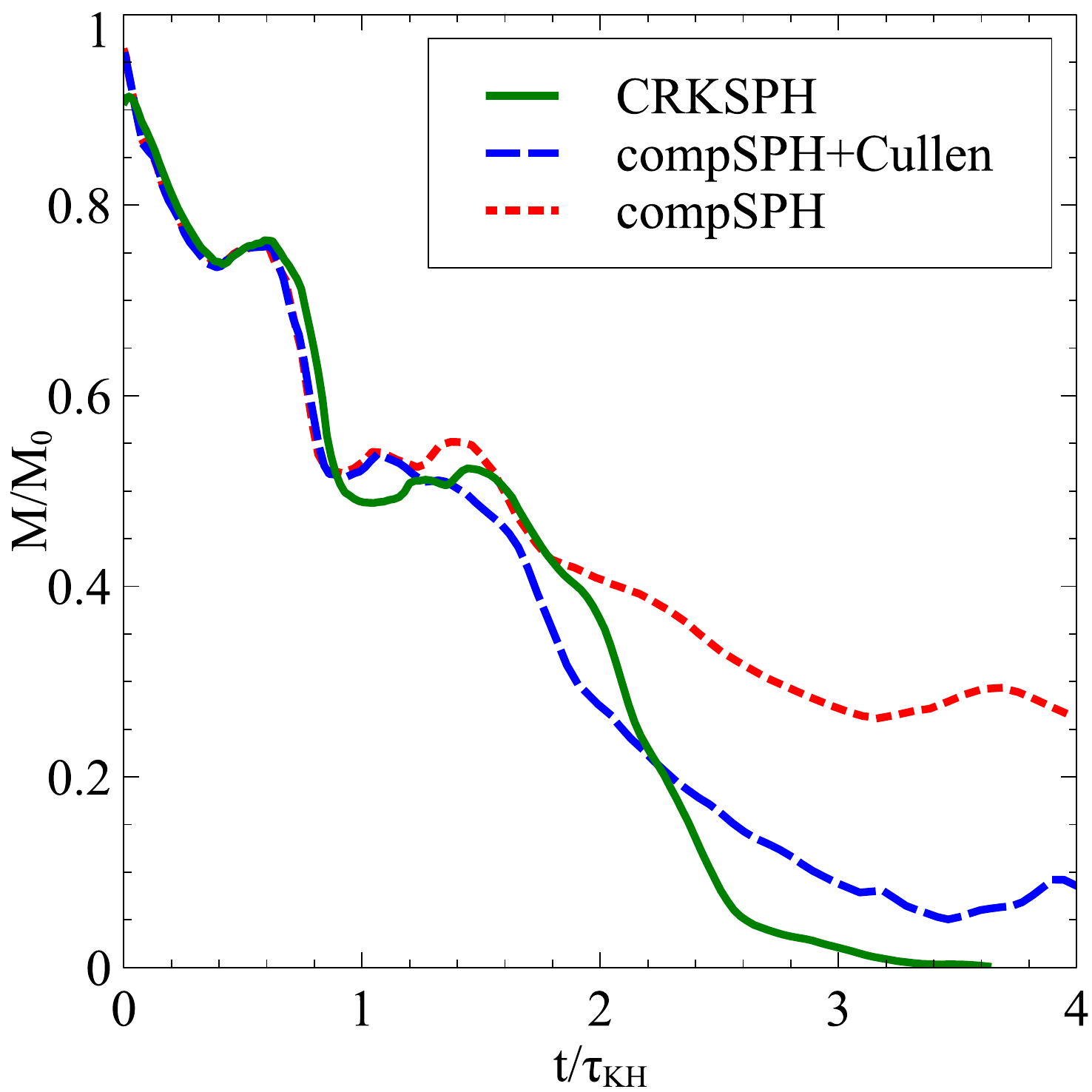}
  \end{tabular}
  \caption{Evolution of the cloud mass fraction as a function of time for the 2D (left) and 3D (right) blob simulations, where time is expressed in units of $\tau_{\text{KH}} \approx 2.903$. Evolution in the shock-dominated period ($\lessapprox 1 \tauKH$) agrees well between all three methods, but CRKSPH demonstrates significantly more mixing in the late-time instability driven regime.  These results confirm the qualitative interpretations of \cref{fig:blob2d,fig:blob3d}.}
  \label{fig:cloudfrac}
\end{figure}
To provide a more quantitative measurement of the cloud evaporation, we calculate the time dependent cloud mass fraction $M(t)/M_0$ (as defined in \cite{Agertz2007}): $M_0$ is the initial mass of the cloud at $t=0$, and any gas particle at time $t$ with density $\rho > 0.64 \rhocloud$ and temperature $T < 0.9 T_{ambient}$ is associated with the cloud, and its mass accumulated to evaluate $M(t)$. \cref{fig:cloudfrac} illustrates the temporal behavior of the cloud mass fraction in both the 2 and 3D cases. We see in both scenarios all of the solvers agree on time-scales of order $t \lessapprox \tauKH$, i.e., before vortical shredding due to shock dynamics is prevalent, consistent with previous results (e.g. Figure 6 in \cite{Agertz2007}). Once entering the instability dominated regime $t \gtrapprox \tauKH$ the schemes beging to differ, with CRKSPH completely disrupting the cloud at roughly $t=6.5\tauKH$ in 2D and $3.5\tauKH$ in 3D, whereas compSPH, and to a lesser extent compSPH+Cullen, demonstrate retarded mixing, as was found above.

It is worth noting an issue that complicates code comparisons with other studies, particularly in the non-linear mixing regimes. As demonstrated in \cite{Agertz2007}, the early evolution is shock-driven with little linear instability growth. Thus, comparisons should (and do) agree reasonably during this early period, as we note here in our examples. However, for a rigorous analysis of the non-linear behavior, one must ensure the initial KH perturbations are standardized between schemes. For the case of \cite{Agertz2007}, a particle glass type IC was used, evoking a random perturbation, which is not trivially mapped to a mesh-based initial condition for a grid-code comparison, nor necessarily consistent for different resolutions or realizations of the point distribution. Our models start with nearly noiseless initial conditions -- the only perturbations at the cloud interface are due to the clipping of the background initial lattice used to create the cloud -- making the process of mapping these perturbation to a grid equally unclear.  This lack of rigorous well-defined seed perturbations as the basis for later amplification by instability growth makes quantitative comparisons at late-time difficult. A possible improvement to this test case would be to establish specified perturbations of the cloud-background material interface (as is done in \cite{Read2010}), where we would expect different models of the subsequent evolution to converge provided that the perturbation scale is resolved. For now we simply conclude that CRKSPH shows more evolution of the cloud material interface than either of the comparable compSPH models; the discrepancy is favorable toward CRKSPH, but it would be useful to have a more concrete specification in order to design a useful reference for comparison.

\subsubsection{Triple Point Shock Test with Vorticity}
\label{sec:triplepoint}
\begin{figure}[ht]
  \centering
  \includegraphics[width=0.65\textwidth]{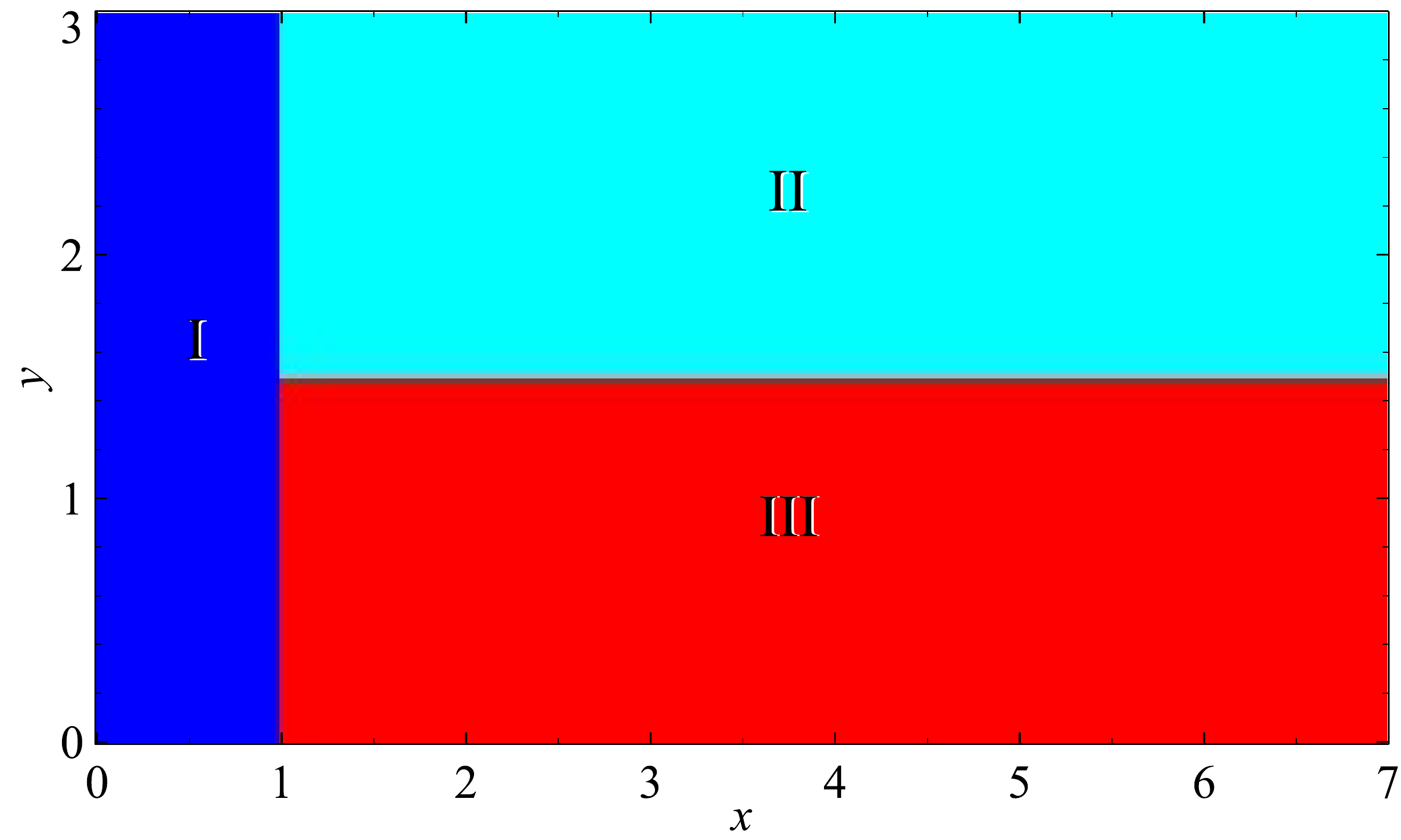}
  \caption{A plot of the three different regions in the initial condition of the triple-point test. The initial density and pressure of each region is defined to be $(\rhoI,\PI)=(1.0, 1.0)$, $(\rhoII,\PII)=(0.125, 0.1)$, and $(\rhoIII,\PIII)=(1.0, 0.1)$ for regions I, II and III, respectively. The resolution chosen in our evaluation populates the three domains with $(n_x, n_y)_I = (160, 480)$, $(n_x, n_y)_{II} = (320, 80)$ and $(n_x, n_y)_{III} = (960, 240)$ particles. }
  \label{fig:tripleIC}
\end{figure}
\begin{figure*}[ht]
  \centering
  \begin{tabular}{rl}
    \includegraphics[width=0.45\textwidth]{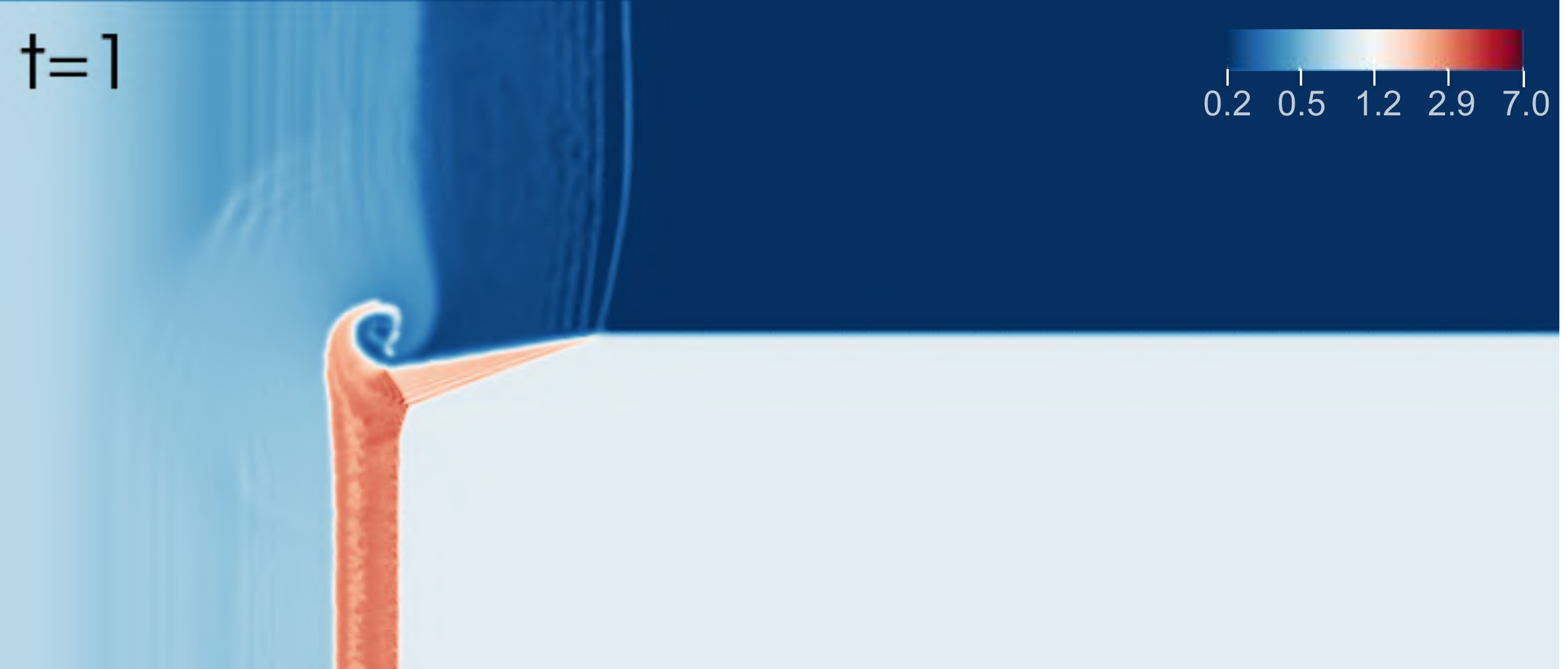} &
    \includegraphics[width=0.45\textwidth]{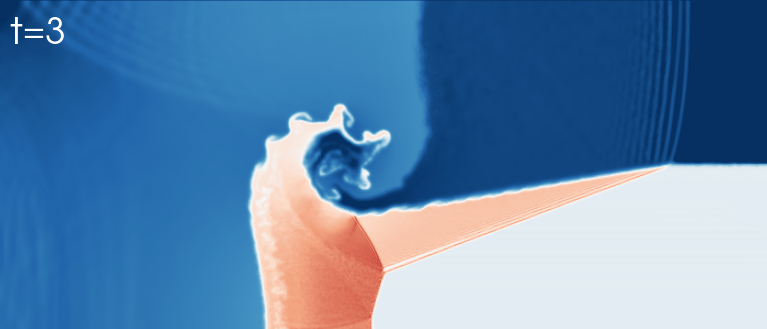} \\
    \includegraphics[width=0.45\textwidth]{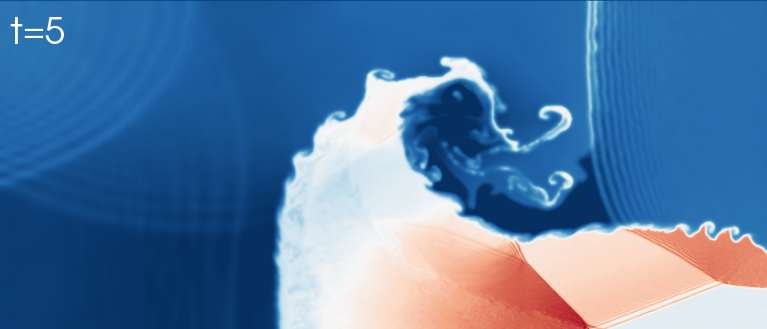} &
    \includegraphics[width=0.45\textwidth]{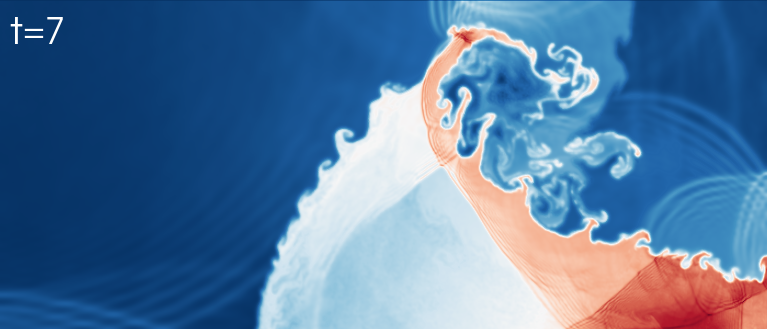}
  \end{tabular}
  \caption{Time sequence of the evolution of the mass density in the CRKSPH model of the triple-point problem using the initial conditions of \cref{fig:tripleIC}. The vertical density feature downstream of the vortex at $t=5,$ is a visual marker of the original shock after reflecting off the $x=7$ domain boundary. The complex reshocking and vortical interactions of the problem are clear at late times.}
  \label{fig:tripleSequence}
\end{figure*}

Our final example is a triple material Riemann problem, the initial conditions of which are depicted in \cref{fig:tripleIC}.  In this test case, a high-density, high-pressure material (region I) drives a shock in a direction parallel to a density discontinuity (between regions II \& III).  As the sound speed is larger in the upper region (II) compared with region III, the shock races ahead in region II vs.~region III, seeding vorticity that progressively rolls up the interface between these two regions. We visually illustrate the time-evolution of the shock roll-up in \cref{fig:tripleSequence} using the CRKSPH solver, where the initial conditions are described below.  As shown in the figure, this problem requires a numerical method that can handle both shock-hydrodynamics and vorticity treatments -- an ideal test of our goals with CRKSPH.  The triple-point problem has been examined previously in the context of reconnecting Lagrangian meshed methods, such as ReALE \cite{Loubere2010} and high-order finite-element Lagrangian schemes \cite{Kolev2009,Dobrev2012,Dobrev2013}.  In our example, we include the results of a ReALE calculation for comparison.  The presented ReALE methodology is based on the original study of \cite{Loubere2010}; the details of this ReALE implementation can be found in \cite{Rathkopf2000,Starinshak2014,Starinshak2016a,Starinshak2016b}.

We establish the initial conditions in a 2D rectangular region, $(x,y) \in ([0,7], [0,3])$, with reflecting boundaries.  The triple segmented domain consists of region I occupying $(x_{\text{I}}, y_{\text{I}}) \in ([0,1], [0,3])$, region II $(x_{\text{II}}, y_{\text{II}}) \in ([1,7], [1.5,3])$, and region III $(x_{\text{III}}, y_{\text{III}}) \in ([1,7], [0,1.5])$.  The density and pressure of each region is specified in \cref{fig:tripleIC}.  Regions I \& II use a $\gamma=1.5$ ideal gas, while region III is a $\gamma=1.4$ material; all domains are initialized with zero velocity.  We create initial lattices of points in each region according to: region I, $(n_x, n_y) = (160, 480)$; region II, $(n_x, n_y) = (320, 80)$; region III, $(n_x, n_y) = (960, 240)$.  For the ReALE comparison, we use an equivalent number of zones in each region, though drawing comparisons of the resolution of such methods vs.~meshfree techniques can be difficult to quantify.

\begin{figure}[ht]
  \centering
  \includegraphics[width=0.95\textwidth]{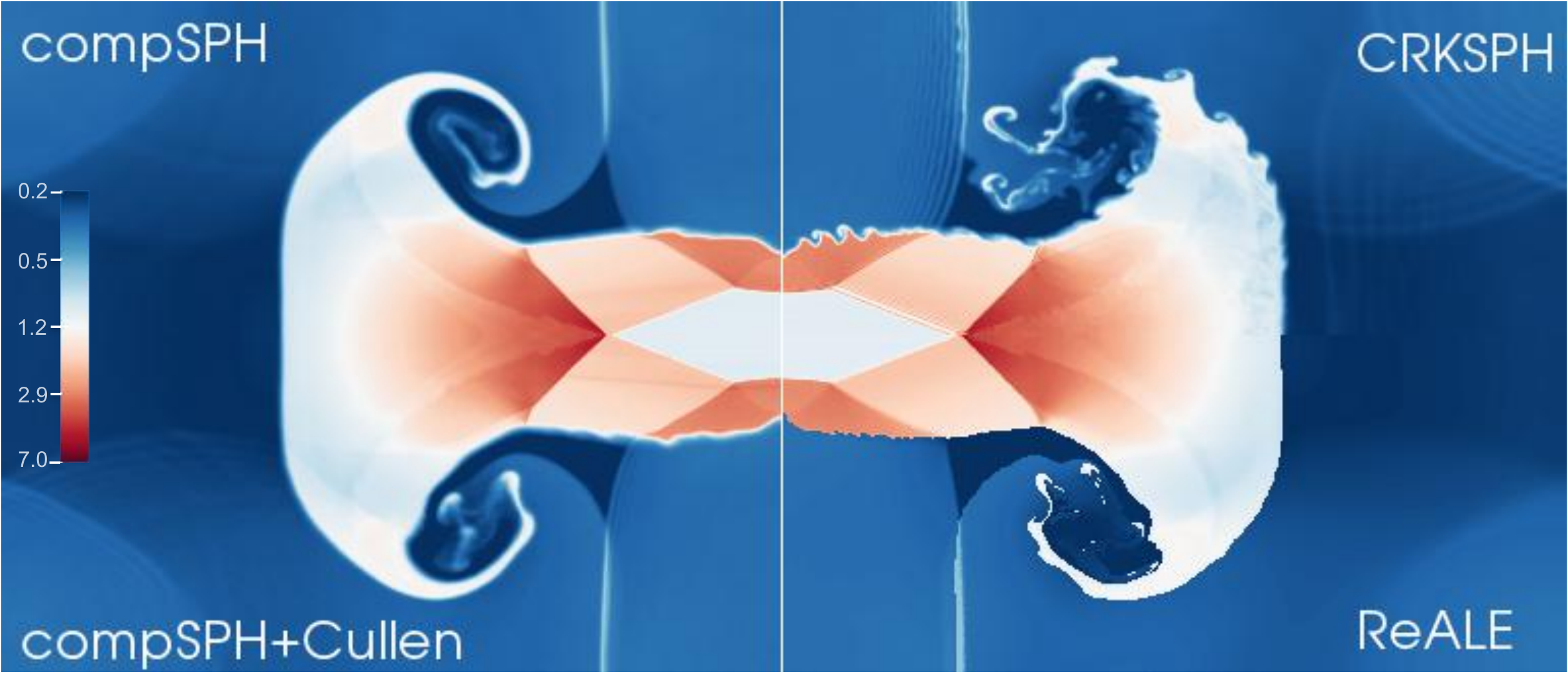}
  \caption{Pseudocolor plots in the logarithm of density at $t=5$ for the triple-point calculations.  The orientation of the compSPH calculation is unmodified -- the compSPH+Cullen, ReALE, and CRKSPH calculations have been reflected to the other three quadrants in order to line up for comparison of major features. The symmetric matching of the shock-lines at the panel boundaries illustrates the shock structure agreement within region III between the calculations. With respect to instability detail, compSPH shows the least mixing evolution that improves for each subsequent method (proceeding counter-clockwise between panels), where CRKSPH demonstrates the most structure.   }
  \label{fig:triple5}
\end{figure}
\Cref{fig:triple5}, shows the state of the mass density in models using compSPH (upper-left), compSPH+Cullen (lower-left), CRKSPH (upper-right), and ReALE (lower-right) at $t=5$, a time often used as the final state of this problem.  The 10-fold overpressure in region I drives a shock into the initially equilibrium regions II \& III.  The resulting evolution of this test is a complex interplay between strong shock-hydrodynamic and the growth of instabilities -- such as Kelvin-Helmholtz (due to the shear between regions II \& III) and Richtmyer-Meshkov -- as shocks repeatedly cross the unstable interfaces between these materials.  The reflecting boundaries cause the shocks to repeatedly reflect and interact, both with each other, and the material interfaces.  By $t=5,$ the main shock launched from region I has reflected off of the $x=7$ domain boundary, and is just reshocking the interface of regions I \& II; simultaneously, region I is expanding into the area formally occupied by region II and being wrapped around the evolving vortex.  In these plots, we can see this reshocking of the region I \& II interface, marked by the vertical density enhancement near $x=5,$ just downstream of the vortex (as noted in \cref{fig:tripleSequence}).  All four calculations agree well on the shock structure at this time, as can be seen in the shock transitions reflecting and interacting about in the ``stem'' of the problem in the region III material.  However, there is a definite progression of detail in the evolution of the fluid instabilities; the amount of structure evident at the material interfaces progressively increases as we step from compSPH $\to$ compSPH+Cullen $\to$ ReALE $\to$ CRKSPH, i.e., counter-clockwise around \cref{fig:triple5}.  This is suggestive of the amount of diffusion in each method, slowing the growth of the instabilities. The ordering of the compSPH calculations vs.~CRKSPH is consistent with the previous results in \crefrange{sec:KelvinHelmholtz}{sec:RayleighTaylor}, though the addition of ReALE is interesting in this case.

\begin{figure}[ht]
  \centering
  \includegraphics[width=0.95\textwidth]{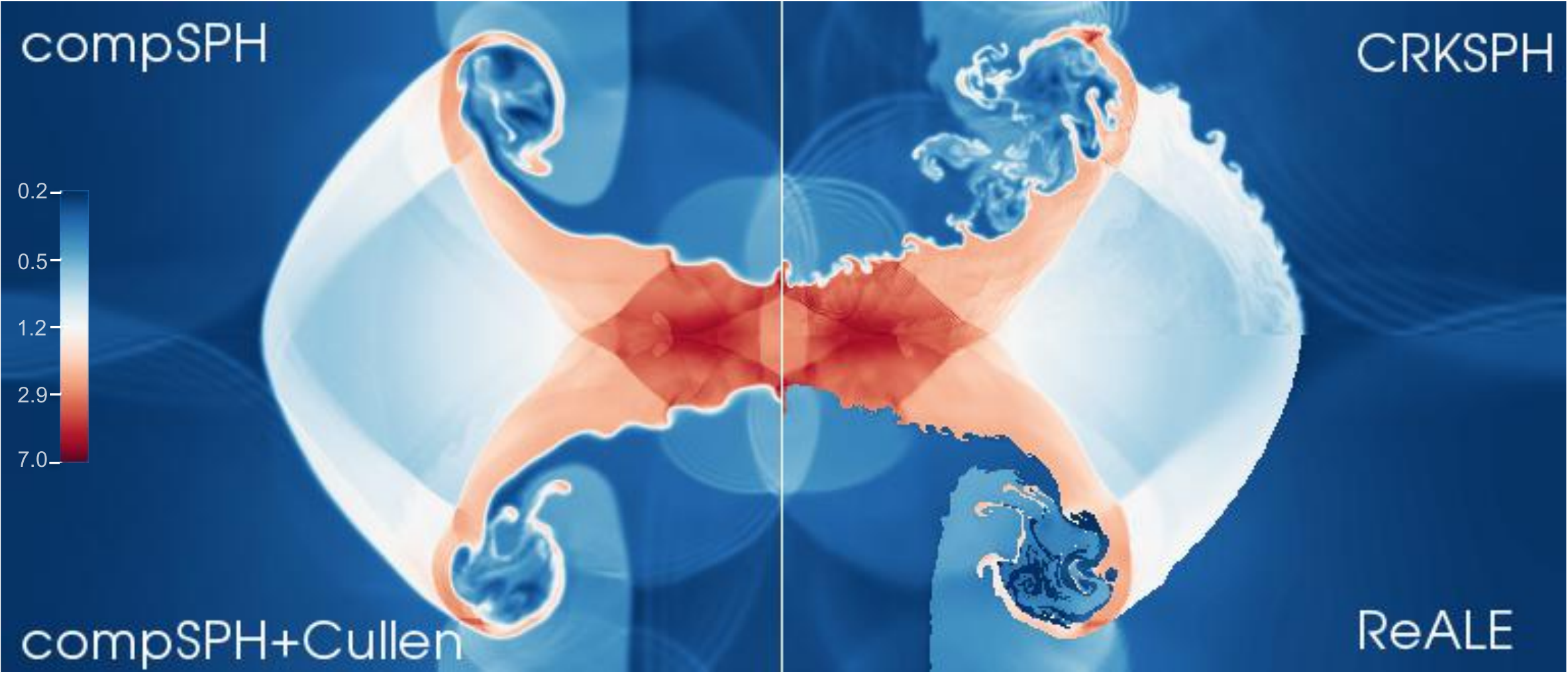}
  \caption{Same as \cref{fig:triple5} at $t=7$. Once again, we see agreement within the shock-dominated region III.  The growth of the unstable interfaces are now drastically different, however.  CRKSPH demonstrates the most evolution, marked by visible secondary KH instabilities on the regional interfaces, including the front-side of the expanding plume. These late-forming instabilities are shown to a lesser and lesser extent in the clock-wise panels starting from CRKSPH, and ending with compSPH, which has the most diffusion (and hence the least mixing).}
  \label{fig:triple7}
\end{figure}
In order to further examine the evolution of the fluid instabilities in the triple-point, \cref{fig:triple7} shows the state of these same four calculations at $t=7$.  By this time, the pressure from the reflected main shock is propagating back into the problem, significantly compressing the stem of region III, where that material is being further forced back into the growing vortex.  Once again, all methods agree quite nicely in the resulting shock structure at this time, though we do see some evidence for increased post-shock ringing in the CRKSPH model; this post-shock ringing suggests that the viscosity limiter may be overly aggressive in the post-shock flow. CRKSPH also shows significantly more growth of the secondary fluid instabilities relative to either of the compSPH models.  In particular, the main vortex is being significantly distorted by the growth of these secondary instabilities, and we see Kelvin-Helmholtz setting in on the front side of the expanding plume of region III material into region I.  The ReALE result is intermediate between the compSPH and CRKSPH models: it does show significant shredding of the main vortex, along with some amount of Kelvin-Helmholtz beginning at the interface of regions I \& III.  It is reasonable that ReALE would show more numerical dissipation than CRKSPH as this model progresses, since the ReALE methodology involves a significant degree of remapping, which will introduce advective diffusion -- less than a purely Eulerian method, but more than a truly meshfree scheme.

Overall, the results of this test are consistent with our previous examples; CRKSPH and compSPH agree well on the shock-dominated portion of the problem, but CRKSPH resolves significantly more evolution due to the onset of fluid instabilities.  As the test interfaces are hydrodynamically unstable to both Kelvin-Helmholtz and Richtmyer-Meshkov phenomena, an additional amount of structure is expected.  The ReALE comparison is suggestive but not conclusive: ReALE indicates there should be more evolution than either compSPH variant identifies, resulting in an answer intermediate between compSPH+Cullen and CRKSPH.  It is plausible that ReALE suffers some amount of numerical diffusion retarding fluid growth of the interface, and since (as in the blob test) we have not introduced perturbations on the interfaces of a known scale, it is difficult to predict exactly how much growth we should see.  Further investigations with a refined problem specification, as well as a trusted reference, could be fruitful.  For now, we can say that CRKSPH yields a reasonable answer to this problem, capturing shock phenomena well, due to its rigorously conservative nature, while also improving SPH's weakness of suppressing fluid instabilities from E0 errors and overactivity of the viscosity.

\subsection{Performance vs. Accuracy}
\label{sec:performance}
\begin{figure}[ht]
\centering
\includegraphics[width=0.45\textwidth]{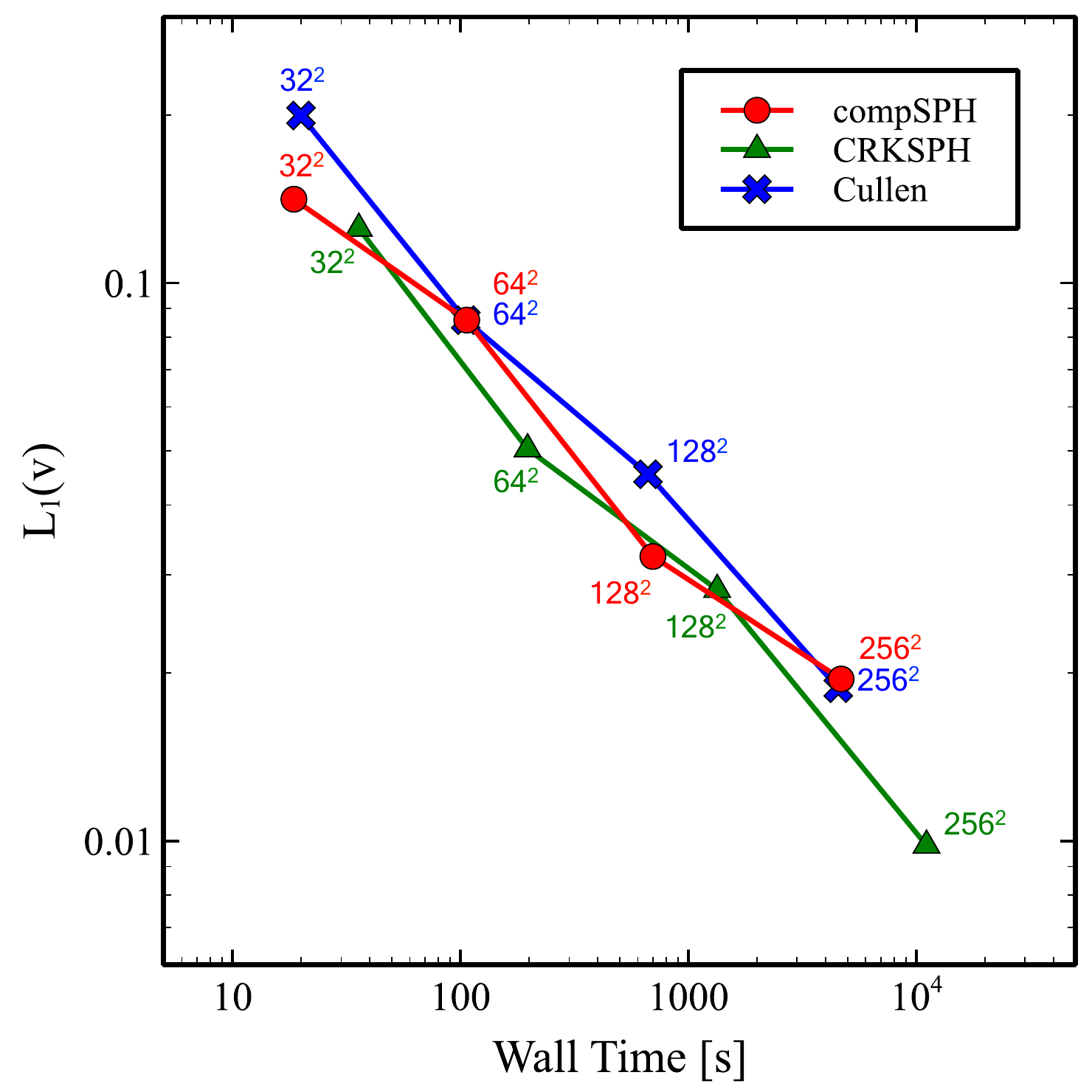}
\includegraphics[width=0.45\textwidth]{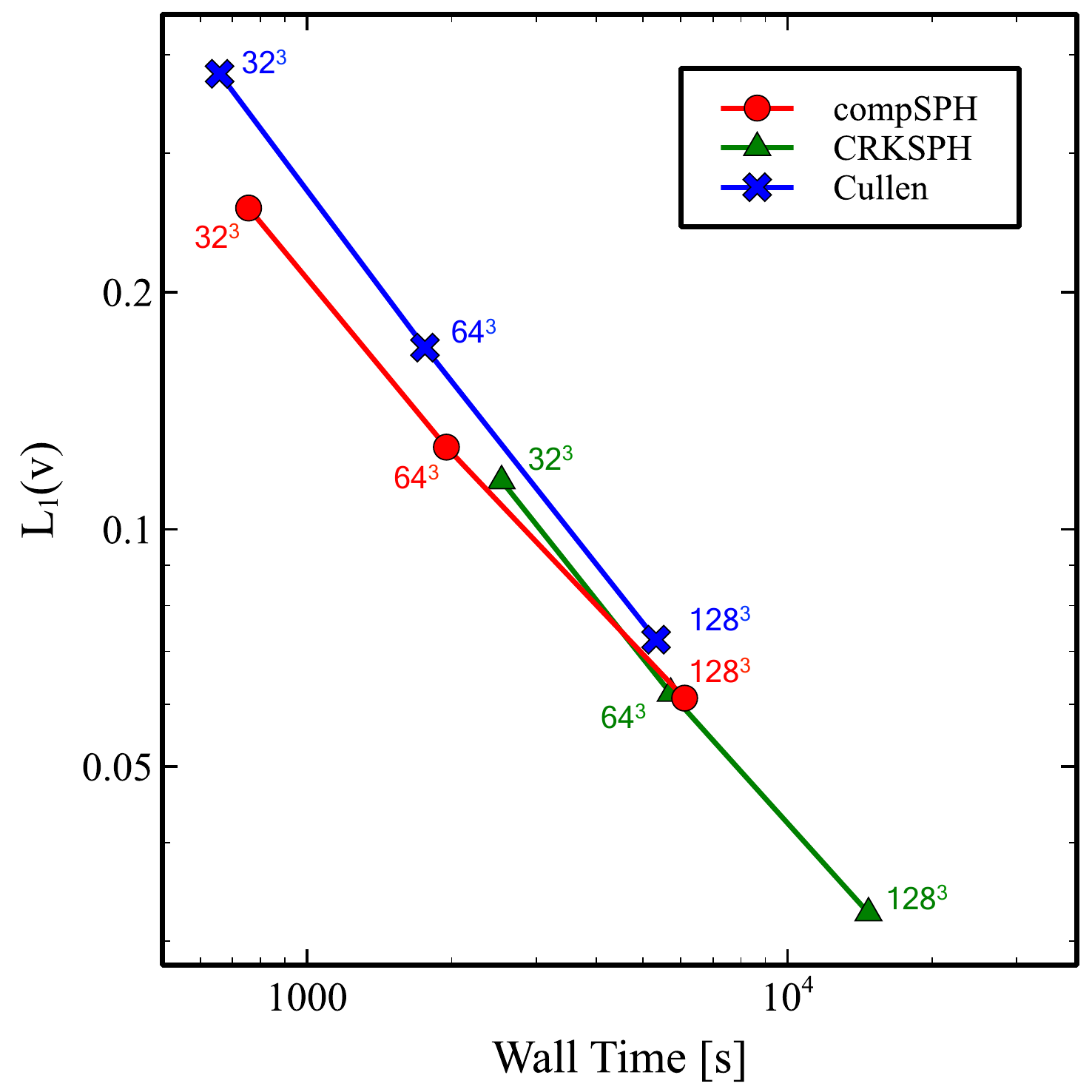}
\caption{Benchmark data: The left panel plots $L_1$ error in velocity $v$ vs. wall clock time for both CRKSPH and compSPH (using both unmodified and Cullen viscosity models), simulating the 2D Noh problem (\cref{sec:Noh}) to time $t=0.6$ at resolutions $N=32^2, 64^2, 128^2$, and $256^2$. The right panel is the equivalent study in 3D at resolutions of $N=32^3, 64^3,$ and $128^3.$ The 3D simulations were run in parallel, weak scaling the number of processors, whereas the 2D simulations were run serially. In both panels, the performance curves of all three methods lie on approximately the same line, indicating similar accuracy given the same wall-time running time. 
However, CRKSPH requires less resolution to achieve similar accuracy to SPH (compare, e.g., the $64^3$ CRKSPH run to the $128^3$ SPH data points).  This tells us SPH requires significantly more points (particularly in 3D) in order to achieve similar accuracy, implying a CRKSPH model will require much less memory (even given the extra storage requirements for the additional kernel parameters of RK) to achieve a given level of accuracy.  Of course CRKSPH also has the advantage of improved solutions on the wide-range of test-cases investigated in previous sections: here we are comparing the accuracy vs~cost of the methods on a problem where all three solvers perform well for fairness.  }
\label{fig:bench}
\end{figure}

Before concluding our evaluation, we briefly turn to the question of computational performance of the CRKSPH formalism compared to SPH. When evaluating a new algorithm such as this, it is important to measure if any additional computational expense is compensated by accuracy; in other words, is the extra work worth it. In this section we investigate how well CRKSPH compares in performance to SPH for a non-trivial multi-dimensional test problem with an analytical solution so we can measure accuracy.

In order to quantitatively compare the computational expense vs.~accuracy of CRKSPH and SPH, we re-examine the Noh test from \cref{sec:Noh} at multiple resolutions comparing the accuracy of the error in the velocity vs.~wall clock time; each model was run to time $t=0.6,$ as was done in our previous Noh examples. 
We select the Noh problem since it has both 2D and 3D configurations with an analytic solution, allowing us to precisely measure the error of both methods.
We also intentionally have chosen a test where ordinary SPH methods perform well in order to focus solely on computational performance, in contrast with the wider variety of test cases we have examined that have historically proved challenging for SPH solvers.

In \cref{fig:bench}, we plot the $L_1$ error in velocity vs.~time-to-solution for both 2D and 3D cases using CRKSPH and compSPH, marking each data-point with the corresponding resolution of the run ($N_{\text{2D}} = \{32^2, 64^2, 128^2$ and $256^2\}$ and $N_{\text{3D}} = \{32^3, 64^3,$ and $128^3\}$ ).
As the Noh test is a shock dominated problem, we also include results of compSPH using the Cullen viscosity model.  
An important practical detail to keep in mind is that while we are able to run the 2D problems serially, simplifying the comparison of relative computational expense, the higher-resolution 3D models are too large to fit in memory for a serial run.
We have therefore used a weak scaling approach to the 3D runs, i.e., scaling the number of processors ($N_{\text{p}}$) such that the number of points per parallel domain remains roughly constant.
Our test employs $N_{\text{P}}=2$ for $N_{\text{3D}}=32^3$, $N_{\text{P}}=16$ for $N_{\text{3D}}=64^3$, and $N_{\text{P}}=128$ processors for $N_{\text{3D}}=128^3.$
Given that we are maintaining 4 radial neighbors per point for all runs (equalizing resolution), as well as the fact that we use tabulated kernels in our implementations (equalizing kernel costs), the heads-up comparisons are highlighting the computational expense of the additional passes over the particle neighbor sets required in CRKSPH, as the matrix calculations for the reproducing kernels are comparatively minimal.  

As we can see in \cref{fig:bench}, the performance curve of both the 2D and 3D cases of CRKSPH and both SPH implementations fall roughly onto the same line, implying the methods achieve similar accuracy given the same compute time, and we see the additional particle-neighbor summations required in CRKSPH are being well compensated by improvements in accuracy.
However, as highlighted by the 3D results, the SPH implementations require higher resolution (by approximately a factor of 8 more particles in 3D) to achieve the same error level as CRKSPH, a rather steep memory cost in scaling. Although it is true that CRKSPH requires more memory per particle interaction to store the RK coefficients (which is typically at most a 60\% overhead depending heavily on one's SPH implementation and memory optimization), the required SPH memory to achieve similar accuracy eclipses this additional cost due to the number of SPH points required.
In the high-performance computing (HPC) realm, total machine memory is often the limiting constraint on the problem, where one tries to maximize the accuracy of the numerical solution given a fixed problem size (dictated by the biggest problem that can fit in main memory), favoring CRKSPH's improved fidelity for fixed resolution.  
Furthermore, the RK coefficients are not evolved quantities, and therefore, can be calculated and utilized on-the-fly -- a preferred work-load for accelerated systems, where reduced main memory algorithms that require more computational kernels are desired. In upcoming work, we investigate specific algorithmic approaches of CRKSPH targeting modern HPC architectures, and the various tricks therein to achieve further performance. We should also note that we intend on performing a wider variety of code-comparisons (including other Lagrangian and Eulerian methods for instance) in the future, wherein performance tests of this nature should be further illuminating.
Finally, it is worth pointing out the usual caveat with performance comparisons such as this: our implementations of CRKSPH will continue to evolve, and these measures will no doubt change.
Our current CRKSPH implementation has had little work done for optimization,  rather opting for explicitness and simplicity of implementation as we have developed the methodology.
There are many opportunities to improve on the current state of our performance, which we will be pursuing.

\section{Summary, Conclusions, and Future Directions}
\label{sec:conclusion}
We have presented and demonstrated the efficacy of a new meshfree method, Conservative Reproducing Kernel Smoothed Particle Hydrodynamics, or CRKSPH.  Our motivation in this study was to formulate an approach as close to standard SPH as possible, thereby leveraging the strengths and wealth of experience from the successful application of SPH to a variety of problems both within astrophysics, and elsewhere. Concurrently, we sought to improve what we view as the greatest weaknesses of SPH:  the poor interpolation properties of the underlying SPH approximation -- manifesting as ``E0-errors'' that cannot reproduce a constant field -- and the frequently excessive (unphysical) application of artificial viscosity. Towards that end, we replaced the standard SPH interpolation approach with the reproducing kernel (RK) formalism of \cite{Liu1995,Liu1998,Bonet2000}. We chose RK theory, as it represents a minimal augmentation of the ordinary SPH kernels, such that interpolation of fields to arbitrary order could be achieved. For this study, we employed linear reproducing kernels, implying functions up to linear-order are represented precisely. Although the RK formalism has been known for some time, its utilization in meshfree modeling has typically sacrificed the conservative properties of ordinary SPH, rendering such implementations ill-suited for applications involving strong shock compressible hydrodynamics. In order to maintain conservation we used the mathematical framework of \cite{Dilts1999,Dilts2000}, along with the compatible differencing methodology of \cite{Owen2014}, to construct hydrodynamic equations that rigorously maintain mass, linear momentum, and energy conservation to machine precision. These are the same major invariants as traditional formulations of SPH (compSPH, PSPH) with the exception of angular momentum, and entropy on the condition that the method in question employs an entropy-based discretization. Exact total angular momentum conservation can be restored by employing the zeroth order RK. However, in rigorously conserving linear momentum, we have sacrificed exact consistency in differencing the momentum equation, though rigorous consistency is maintained for other relations (see \cref{sec:consist} for a discussion of this trade-off). 

In an effort to address the excessive activation of artificial viscosity in SPH, we presented a simple method of limiting the treatment based on the work of \cite{Christensen1990}. Our implementation was derived solely to eliminate viscous interactions for any linear velocity field, rather than requiring complex shock-detectors or other physical prescriptions to switch the viscosity on and off. The new viscosity limiter relies on an accurate velocity gradient, which in our case is afforded through the use of reproducing kernels, and appears solely as a higher-order projection of the velocity difference in the standard Monaghan-Gingold viscosity \cite{Monaghan:1983dn}. Other than computing the velocity jump between points -- using a mid-point high-order difference -- the CRKSPH artificial viscosity is identical to the well-known pair-wise form due to \cite{Monaghan:1983dn}.  We titled our scheme CRKSPH, following our addition of these aforementioned elements; Conservative reformulation of the hydrodynamic equations, Reproducing Kernel interpolation, and our novel artificial viscosity limiter. 

In a series of increasingly complex tests, we have demonstrated that CRKSPH handles strong-shock physics as well as contemporary SPH based methods, if not better. In every case, we find our viscosity limiter improves the localization of the shock-jump condition, without considerable introduction of undue oscillations; notably, as demonstrated in the 2D and 3D Noh test case, CRKSPH reduces, or eliminates, the unphysical activation of the artificial viscosity in smoothly compressing flows.  We also demonstrated the applicability of CRKSPH in problems with hydrodynamically unstable interfaces, by addressing the unphysical ``artificial surface tension'' due to the E0 errors in SPH -- though we also acknowledge the CRKSPH inconsistency caveat described in \cref{sec:consist}. The improved interpolation of RK allows CRKSPH to model static surfaces, as well as hydrodynamically unstable interfaces, more effectively than ordinary SPH. Analyzing idealized mixing tests, e.g. Kelvin-Helmholtz and Rayleigh-Taylor, we illustrated how CRKSPH performs well at capturing the growth of such unstable interfaces. We also demonstrated how these benefits carry through to more complex realistic modeling, such as the so-called ``blob'' test of \cite{Agertz2007} and the triple-point test discussed in \cref{sec:triplepoint}.

While CRKSPH performs well on the tests presented here and is already a useful method, we believe there are still areas where the method can be expanded upon and improved.  One concern is that small-scale noise in the CRKSPH point field can grow and degrade the quality of the solution, i.e. manifestations of so-called ``hourglass'' error modes.  SPH also suffers from this problem, but the very reproducing/accurate nature of RK interpolation can make CRKSPH further susceptible.  Consider, for example, a set of points with a uniform pressure field: if the positions of those points are perturbed randomly, SPH will detect and react somewhat to such perturbations, albeit in an overly smoothed manner.  A strict RK method would explicitly be blind to these perturbations, allowing such small-scale noise to persist or even grow.  Since we partially sacrifice strict reproducibility for explicit conservation, enforced in the CRKSPH momentum relation, our formalism will not be completely oblivious to such perturbations (see the example in \cref{sec:consist}); however, it is very likely that the method would benefit from explicit treatment to remove perturbations below the resolution scale.  For now, we simply utilize our viscosity limiter -- in particular, the exponential term of \cref{eq:phi} -- to suppress high-frequency particle movement. There currently exist many possibilities for an improved correction, such as the regularization ideas of \cite{Borve2001,Borve2005}; however, it remains to be seen what the best approaches to this problem will be.

We are also interested in exploring the multi-material aspects of CRKSPH.  The treatment in this paper is largely appropriate for single-fluid calculations, though we demonstrate good results on a few simple multiple-material fluid problems, such as the box tension test (\cref{sec:HydroBox}), Kelvin-Helmholtz (\cref{sec:KelvinHelmholtz}), Rayleigh-Taylor (\cref{sec:RayleighTaylor}), ``blob'' test (\cref{sec:blob}), and  triple-point (\cref{sec:triplepoint}).  However, in problems with true surfaces, such as solids, CRKSPH will likely benefit from a more rigorous surface treatment.  For instance, the derivation of the CRKSPH relations in \cref{sec:crkmomderiv}, results in terms that involve integrals over the bounding surfaces of the discretized material, which we neglect in this work.  That choice is appropriate for the continuous fluids we examine here, but is not a valid assumption when dealing with solids and discrete surfaces delineating very different materials, where a more rigorous examination of these terms is warranted. 
 
Lastly, we reaffirm that although there is no doubt additional algorithmic work to be done, the CRKSPH formalism presented here represents a simple variation of SPH, yielding useful improvements on a wide class of fluid dynamic problems. CRKSPH is a relatively non-invasive modification of an existing SPH implementation, requiring a few additional pre-passes over the points and their neighbors to compute the kernel enhancement terms (\crefrange{eq:coefA}{eq:gradB}), improved density (\cref{eq:CRKrho}), and volume definition (\cref{eq:CRKvol}),  before evaluating the hydrodynamical relations of \crefrange{eq:CRKDvDt}{eq:CRKcompE2}. These extra passes over the connectivity represent the major additional cost of CRKSPH: computing the correction terms themselves only involves inverting a $2 \times 2$ (2D) or $3 \times 3$ (3D) matrix per point, which is essentially free compared with walking the topology.  Moreover, as discussed in \cref{sec:performance}, the accuracy gains of CRKSPH justify the additional computational effort, whereby CRKSPH achieves similar accuracy to SPH using significantly reduced particle counts.

We conclude by remarking that we intend to examine how CRKSPH performs on a variety of interesting astrophysical problems, especially compared to a collection of solvers (both meshless and Eulerian), including large-scale baryonic cosmological simulations, performed on current and future high-performance architectures (such as GPU and Xeon-Phi based machines). The CRKSPH methodologies described here can be found in the publicly available code Spheral\footnote{https://sourceforge.net/projects/spheral/}, and are currently being implemented in the cosmology N-body code HACC \cite{habib2016hacc}, specialized for supercomputing hardware;  algorithmic formulations of CRKSPH that are optimized for HPC architectures will be discussed in future work.

\section*{Acknowledgments}
NJF would like to acknowledge support from the Department of Energy Computational Science Graduate Fellowship (DOE-CSGF) program, in addition to support from the Nambu Fellowship provided by the University of Chicago. All work done by NJF at Argonne National Laboratory was supported under the U.S. Department of Energy Contract DE-AC02-06CH11357. In the case of CDR and JMO, this work was performed under the auspices of the U.S. Department of Energy by Lawrence Livermore National Laboratory under Contract DE-AC52-07NA27344. We would also like to acknowledge the many Bothans that died to bring us this information. 

\appendix
\setcounter{figure}{0} 
\section{Linear Reproducing Kernel Derivation and Validation}
\label{sec:RKDerv}
\begin{figure}[ht]
\centering
\includegraphics[width=0.4\textwidth]{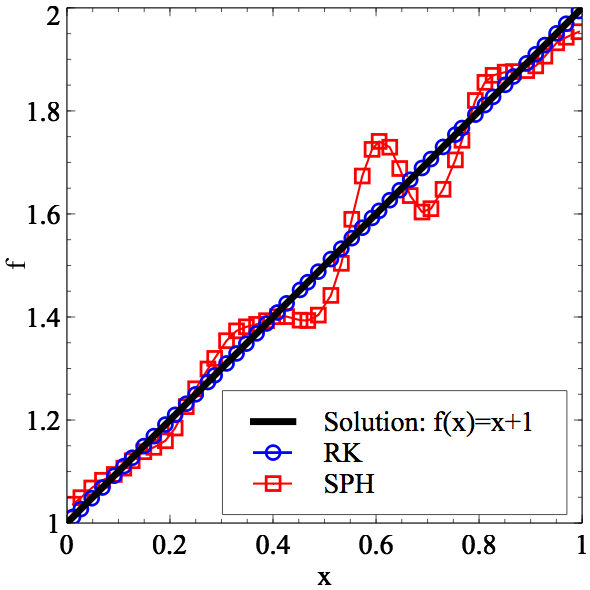}
\includegraphics[width=0.4\textwidth]{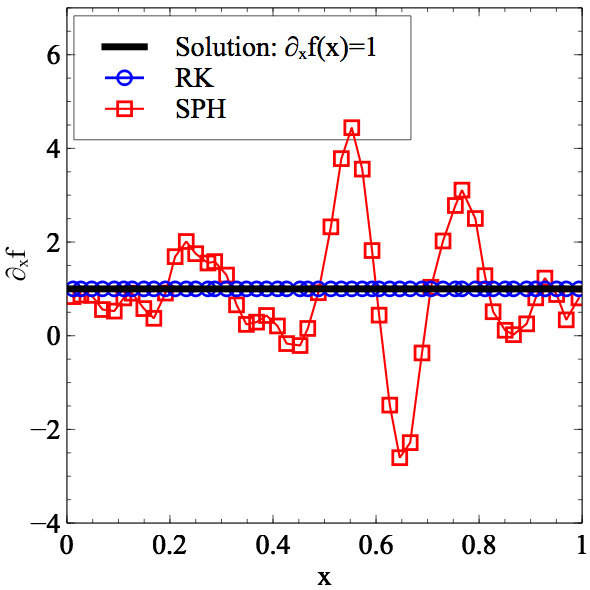}
\caption{Interpolation of a linear field and its gradient using RK and SPH kernels over $N=50$ particles randomly displaced by a small fraction of the interparticle spacing from a uniform distribution. The RK kernel exactly reproduces the field, while the irregular particle distribution causes SPH interpolation errors. We note that both kernels used seventh-order splines at a resolution of 4 radial neighbors.}
\label{fig:1dInterp}
\end{figure}
To explicitly derive the linear RK coefficients from  \cref{eq:coefA,eq:coefB}, we begin by substituting the definition of $\mathcal{W}^R$ from \cref{eq:wR} into the consistency relation of \cref{eq:Cons2}:
\begin{align}
 0 &= \sum_j x_{ij}^\alpha V_j\Wrj \nonumber \\
 &= \sum_j x_{ij}^\alpha V_j A_i \left(1+B_i^\beta x_{ij}^\beta\right)\Wj \nonumber \\
 &= A_i \sum_j x_{ij}^\alpha V_j \Wj + A_i B_i^\beta \sum_j x_{ij}^\alpha x_{ij}^\beta V_j \Wj \nonumber \\
 B_i^\alpha &= -\left(m_2^{-1}\right)^{\alpha\beta}m_1^\beta
\end{align}
where we used the moment definitions from \crefrange{eq:m0}{eq:m2}. To find the normalization coefficient $A_i$, we utilize the additional consistency relation from \cref{eq:Cons1}, together with our evaluation of $B_i^\alpha$, viz. 

\begin{align}
 1&=\sum V_j A_i \left(1+B_i^\beta x_{ij}^\beta\right)\Wj \nonumber \\
 &= A_i \sum V_j \Wj + A_i B_i^\beta\sum x_{ij}^\beta V_j \Wj \nonumber \\
 &= A_i \left(m_0 + B_i^\beta m_1^\beta\right) \nonumber \\
 &= A_i \left(m_0-\left(m_2^{-1}\right)^{\alpha\beta}m_1^\beta m_1^\alpha\right) \nonumber \\
 A_i^{-1} &= m_0-\left(m_2^{-1}\right)^{\alpha\beta}m_1^\beta m_1^\alpha.
\end{align}

The coefficients $(A,B^\alpha)$ now fully define the reproducing kernel $\mathcal{W}^R$. Evaluating the coefficient and kernel derivatives (\cref{eq:gradwR}, \cref{eq:gradA}, and \cref{eq:gradB}), one can interpolate an arbitrary field $F(x^\alpha)$ and its gradient using \cref{eq:rkinterp} and \cref{eq:rkgrad}. By construction, $\mathcal{W}^R$ is accurate to first-order.  To illustrate, \cref{fig:1dInterp} plots the interpolation of a linear 1D field $F(x)= y_0+ m_0x$, and its gradient, using both the SPH and RK kernels (replacing $\mathcal{W}^R$ with $W$ for SPH in \cref{eq:rkinterp} and \cref{eq:rkgrad}); both methods used seventh order splines at a resolution of four radial neighbor points. We sampled $N=50$ points using a displacement that is a random fraction of 0.2 from uniform spacing, with field constants $y_0=m_0 = 1.0$. As a result, the benefit of RK kernels becomes clear -- the SPH kernel displays significant noise given the irregular particle distribution, while the RK kernel exactly reproduces the linear field and its gradient to machine precision, regardless of the point geometry.  

\setcounter{figure}{0} 
\section{The Tradeoff Between Consistency and Conservation in CRKSPH}
\label{sec:consist}
\begin{figure}[ht]
\centering
\includegraphics[width=0.45\textwidth]{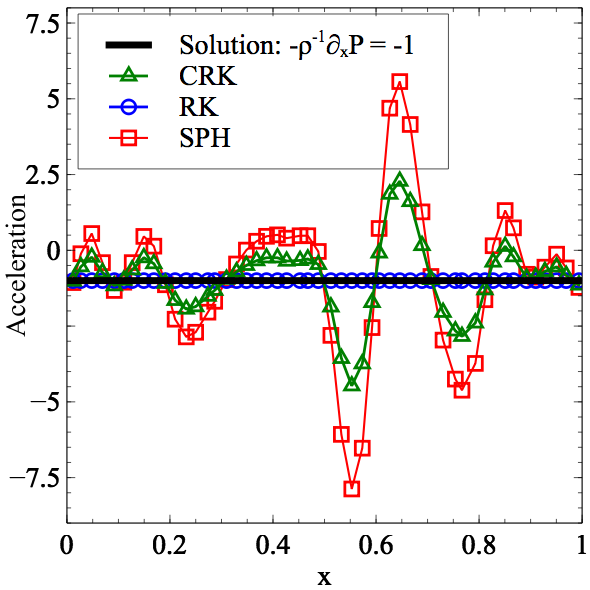}
\caption{Particle acceleration estimates using CRK, RK, and SPH evaluation of a 1D linear pressure profile with $N=50$ points, randomly displaced by a small fraction from a uniform distribution. As expected, RK exactly reproduces the field, owing to its first-order accuracy. However, the CRK momentum relation (\cref{eq:momeq2}) sacrifices exact reproduction of the linear acceleration field, in order to maintain conservation of momentum, though it is still more accurate than the SPH example. Similar to SPH, the inconsistency errors vanish when the particle spacing is uniform. We note that all three kernels used seventh-order splines at a resolution of 4 radial neighbors for a fair comparison.}
\label{fig:1dConsist}
\end{figure}
As discussed in \cref{sec:crkmomderiv}, simply replacing the SPH kernel with the accurate reproducing kernel ($W \rightarrow \mathcal{W}^R$) in the traditional formalism, results in fluid equations that are no longer conservative due to the non-symmetric nature of the RK kernels.  Therefore, a reformulation of the fluid equations is required to maintain conservation, yielding the CRK evolution relations \cref{eq:momeq2,eq:engeq2}. Unfortunately, enforcing pair-wise linear momentum conservation reintroduces a consistency error, which no longer guarantees reproducibility to the order of the underlying RK approximation. The extent of the error is dictated by the degree of irregularity in the underlying point distribution: the more regular the points, the more the inconsistency is reduced, fully vanishing if the particles are exactly uniformly spaced. As a 1D example, we initialize $N=50$ particles using a random pairwise displacement of up to 0.2 times the initial uniform particle spacing with unit density and linear pressure profile $P(x) =  y_0+ m_0x$. Setting $y_0=m_0 = 1.0$, \cref{fig:1dConsist} plots the analytical acceleration of the fluid (-$\partial_\alpha P/\rho$), as well as the estimated particle accelerations using the CRK formalism (\cref{eq:momeq2}), the RK interpolation of -$\partial_\alpha P/\rho$ (\cref{eq:rkgrad}), and the traditional SPH kernel interpolation for good measure (\cref{eq:SPHvel}, without viscosity terms). Note, all three solutions were generated using the same kernel (seventh-order spline) at a resolution of 4 radial neighbors, for a fair comparison. As expected, the RK interpolants exactly reproduce the analytically expected acceleration to machine precision. The CRK formalism, however, does not maintain the exact solution, illustrating the inconsistency error resulting from a non-regular particle geometry. The CRK solution remains more accurate than the SPH example, however, while maintaining conservation of linear momentum to machine precision. The RK solution, while formally more exact, is, in general, not conservative. These differences illustrate the trade-off between conservation and consistency in our equations.

\begin{figure}[ht]
\centering
\includegraphics[width=0.45\textwidth]{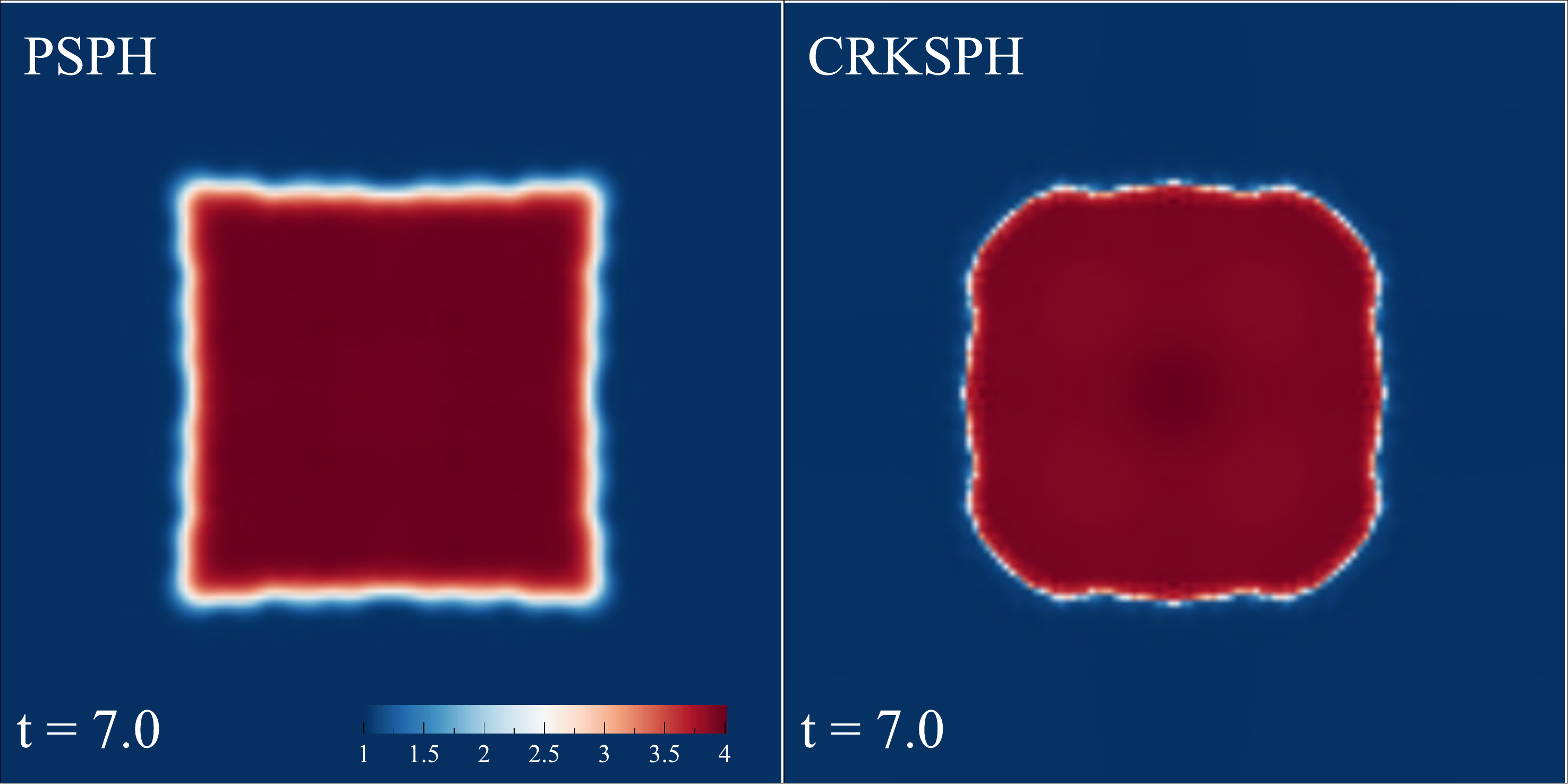}
\caption{Hydrostatic box with equal mass particles at $t=7$ (same as \cref{fig:box}) using both CRK and PSPH (without artificial conductivity) solvers. CRK no longer maintains exact equilibrium due to the inconsistency error derived from the irregular particle spacing at the boundary. PSPH also degrades slightly, illustrating that changing to a smooth discretization variable does not fully circumvent this problem. In general, density and volume estimates in multi-material phenomena, particularly with large density differences, require further development to be addressed in future work.}
\label{fig:equalMbox}
\end{figure}
We can examine some of the consequences of this inconsistency error by returning to the box tension test of \cref{sec:HydroBox}, using equal mass particles, rather than equally spaced as assigned originally.  This implies the points are packed into the box region four times more densely than compared to the surrounding medium, resulting in a box boundary that has a discontinuous jump in the particle spacing.  \Cref{fig:equalMbox} shows the resulting configuration at $t=7$ (just as in \cref{fig:box}) for PSPH (without conduction in this case) and CRKSPH.  Both methods now show some deviation from maintaining the perfect original square interface owing to this inconsistency error in the acceleration equation.  The PSPH result is very similar to that noted in \cite{Saitoh2013} for equal mass points; we see in \cref{fig:equalMbox} that CRKSPH shows more rounding of the corners but less diffusion of the interface compared with PSPH. We note the addition of artificial conductivity only worsens the diffusion in the PSPH case (compare e.g. \cite{Hopkins2015}).

Although we have sacrificed some of the underlying accuracy/consistency of the RK methodology by formulating CRKSPH in a conservative manner (as was clearly shown in these examples), we find that maintaining invariants of the continuum equations (like conservation of total momentum and energy) is the superior compromise; in particular, conservativeness is critical for obtaining accurate solutions in scenarios where strong shocks and/or highly compressible evolution dominate.  Moreover, we find that CRKSPH demonstrates significant improvement over SPH in these sorts of gaseous compressible problems, as outlined in the many tests of \cref{sec:crktests}, owing in large part to the improved accuracy of the RK interpolation theory, even though we have sacrificed that property to some degree in the name of conservation.  We note that regularization of particle geometry and/or boundary treatments are compelling areas of investigation that may further improve the inconsistency errors of CRKSPH. 

We should remark that the results in \cref{fig:equalMbox} are obtained by evaluating the density using the discretized continuity equation (\cref{eq:contEqn}), as opposed to our regular treatment of $\rho$ in \cref{eq:CRKrho}. As seen in \cref{sec:HydroBox}, our density definition is advantageous as it gets density discontinuities exactly correct when particles are equally spaced, unlike the traditional SPH summation definition in \cref{eq:SPHSum}, which smooths the density across the surface. However, for unequal particle separations such as we have in \cref{fig:equalMbox}, \cref{eq:CRKrho} will again incur an averaging error over the discontinuity, resulting in pressure errors that eclipse the inaccuracy attributed to the inconsistency of the solver that we wished to point out here. This is not to say that we advise running with the continuity equation in production for fluids; \cref{eq:contEqn} is merely convenient in this particular case as it correctly captures multi-material surfaces under static flow. In general, it is often advisable to use quantities derived from integral forms, such as the summation definitions of \cref{eq:CRKrho} or \cref{eq:SPHSum}, as opposed to discretized differential forms, which do not conserve mass exactly, as well as encounter issues with non-static discontinuities, where the solutions possess infinite derivatives (see e.g. \cite{Price2008}).


\setcounter{figure}{0} 
\section{Angular Momentum Conservation}
\label{sec:angularMomentum}
One of the major strengths of SPH is its explicit conservation of angular momentum -- a consequence of the fact that the SPH pair-wise forces between neighboring points are radially aligned (for spherical kernels interpolating scalar pressure forces).
However, the inclusion of any non-zero curl forces -- such as those arising from approximate gravity solvers or tensor strength forces -- will violate this constraint and introduce an error into the total angular momentum.
With respect to reproducing kernel theory, any kernel correction of order greater than zero no longer ensures accelerations are oriented along the pairwise separation vector.
In the aggregate, this results in a resolution-dependent error in the total angular momentum.
Recall, however, that the CRKSPH formalism does ensure equal and opposite pairwise forces, thus exactly preserving linear momentum. 

\begin{figure}[ht]
\centering
\includegraphics[width=0.45\textwidth]{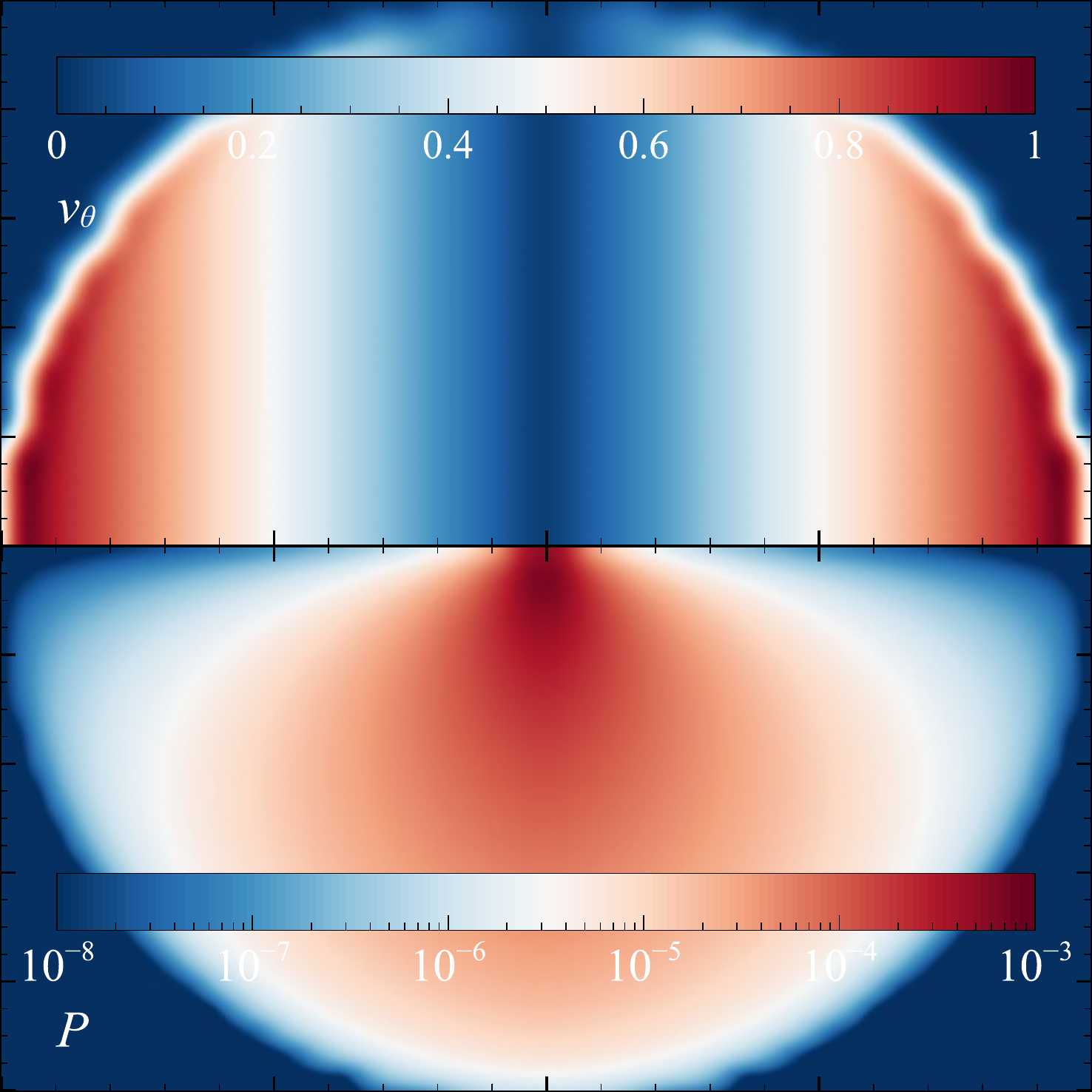}
\caption{Initial conditions for the pressure $P$ and velocity $v_\theta$ of the rotating gas cloud. }
\label{fig:Lsetup}
\end{figure}

In order to quantify the violation of angular momentum conservation, we examine the gravitational collapse of a gas cloud of radius $R$ and constant density $\rho.$
The gas is initially in solid body rotation about the $z$-axis with $\boldsymbol{\omega}=\sqrt{GM/R^3}\bold{\hat{z}}$ in the presence of a central gravitational source following a Plummer softened gravitational potential, $\Phi_i = -(GM)/\sqrt{r_i^2+a_p^2}$, where $a_p=0.1R$.
At $t=0$, the velocity is given by
\be
\bold{v}=\boldsymbol{\omega}\times \bold{r},
\ee
and the pressure by
\be
P=\rho G M\left(|\bold{r}|^{-1}-R^{-1}\right)\left(1-\sin\theta\right).
\ee
This problem is based on a similar prescription for a 2D rotating disk described in \cite{Raskin2016}, generalized for 3D and using a constant density.
The resulting cloud rapidly collapses to form a rotating disk with a hot, pressure-supported bulge in the center.
The initial configuration is shown in \cref{fig:Lsetup}.

\begin{figure}[ht]
\centering
\includegraphics[width=0.45\textwidth]{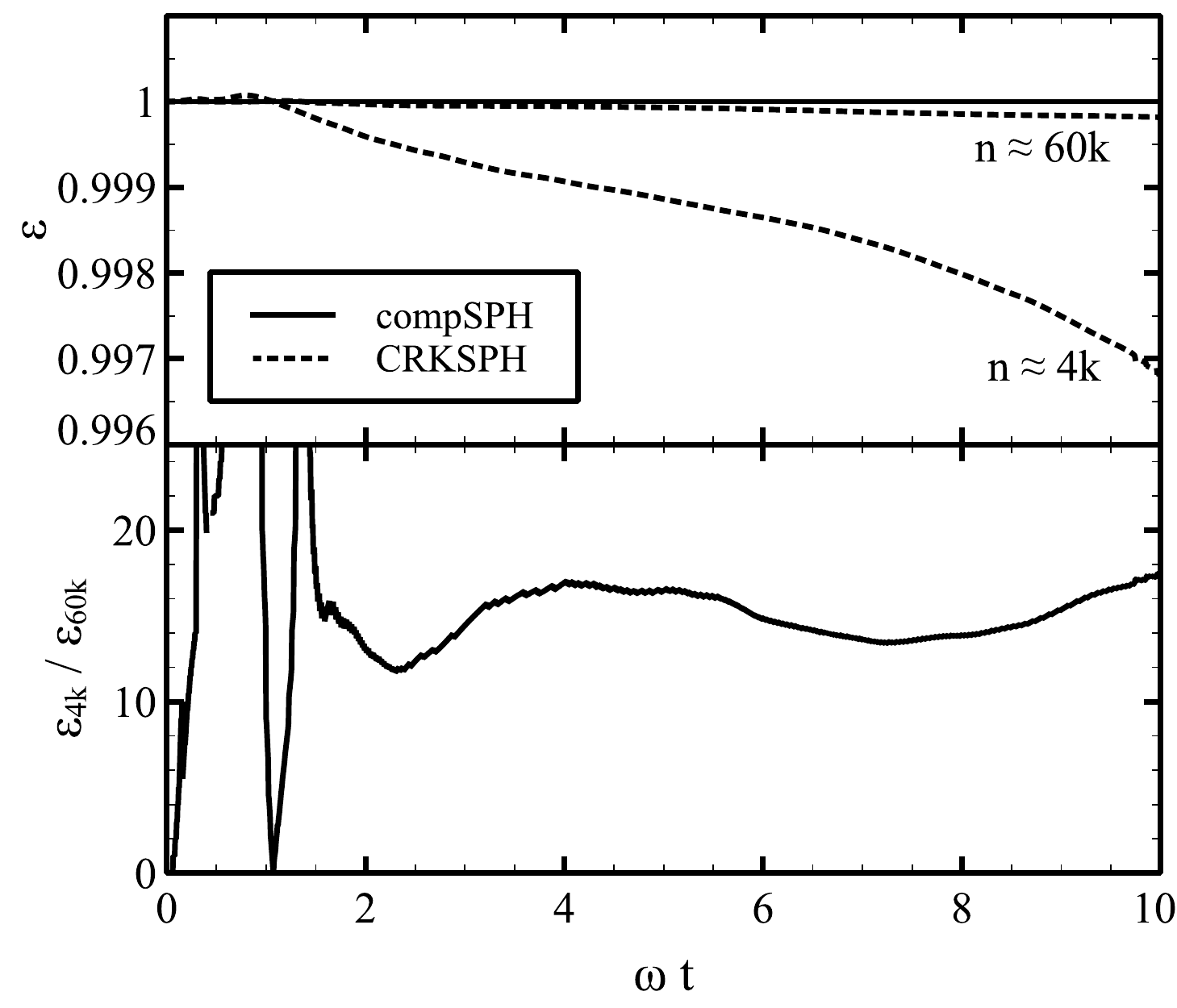}
\caption{The ratio ($\varepsilon$) of the measured angular momentum $L_z$ over the analytical value $L_0$, as a function of time for $n=4,000$ and $n=60,000$ particle simulations of the rotating gas cloud, using both compSPH and CRKSPH solvers. As the gravitational acceleration is calculated exactly in this idealized central potential problem, compSPH demonstrates machine-precision accuracy in this metric, as expected.  The angular momentum error for CRKSPH is resolution dependent, where the errors are $<0.5\%$ and $<0.05\%$ for the low and high resolution runs, respectively.  Lower Panel: The ratio of the measured CRKSPH $\varepsilon$ from both simulations (low/high) overtime shows a roughly $16\times$ reduction of the error for a $2.5\times$ increase in linear particle resolution.}
\label{fig:Lresult}
\end{figure}

In this test we use a gamma-law equation of state with $\gamma=5/3$, $R=GM=1$ and $\rho=10^{-4}$.
The particles are initially arranged in a simple cubical lattice clipped to form a spherical distribution.
In order to ascertain the angular momentum errors incurred in our linearly-corrected CRK formalism, we measure the evolution of the total angular momentum over the duration of this test ($\omega t=10$), shown in \cref{fig:Lresult}. Here, the figure of merit is the angular momentum error $\varepsilon$ defined to be the ratio of the measured angular momentum ($z-$component) $L_z$, divided by the initial (analytically constant) angular momentum $L_0,$ where we recall that $\textbf{L} = \sum_i (\textbf{r}_i \times m_i \textbf{v}_i)$ summing over the contributions of all particles. 

We initialized two distributions of points on lattices of $20^3$ and $50^3$ particles, resulting in $\approx 4,000$ and $\approx 60,000$ particles, respectively, after each distribution has been clipped to a sphere. We find the total angular momentum conservation error to be $\varepsilon <0.5\%$ at $\approx 4,000$ particles, so angular momentum is very nearly conserved. 
In the higher resolution study ($\approx 60,000$ particles), the error drops to $\varepsilon<0.05\%$, implying the error is converging rapidly with spatial resolution; indeed, taking the ratio of the low-resolution simulation error measurement ($\varepsilon_{\text{4k}}$) over the high-resolution result ($\varepsilon_{\text{60k}}$), as was done in the lower panel of \cref{fig:Lresult}, we measure a roughly $16\times$ reduction of the error for a $2.5\times$ increase in linear particle resolution.

As expected, the compSPH calculation exactly conserves angular momentum to machine precision in this test.
We remark, however, if this were a self-gravitating fluid where we modeled the gravitational force using ordinary N-body methods, such as a tree code or particle-mesh solver, there would again be angular momentum errors due to the gravitational term.
This idealized test uses an imposed central potential for which we exactly evaluate the acceleration, allowing us to examine just the error due to the hydrodynamics.
The CRKSPH conservation error is strongly resolution-dependent, whereas the error introduced by a tree-gravity approach is not simply resolution-dependent, thus, one might expect the angular momentum errors due to the gravitational forces to take precedence in practical astrophysical problems. Regardless, tracking the total angular momentum in rotation problems (similar to measuring total energy when using non-energy conserving solvers), is important to monitor effects from all sources of error to this quantity. 

Comparable sources of momentum violation aside, a final important point (as mentioned in \cref{sec:crkmomderiv}) is the fact that although CRKSPH does not preserve total angular momentum precisely, the benefit of accurately simulating momentum transport drastically improves numerical solutions of gravitational disk phenomena -- classic problems which are heavily dependent on the proper treatment of angular momentum. For a more detailed examination of how CRKSPH fares and compares with traditional SPH in a generalized Keplerian problem, see \cite{Raskin2016}. 

\setcounter{figure}{0} 
\section{The choice of interpolation kernel in CRKSPH}
\label{sec:Wchoice}
\begin{figure}[ht]
  \centering
  \includegraphics[width=0.75\textwidth]{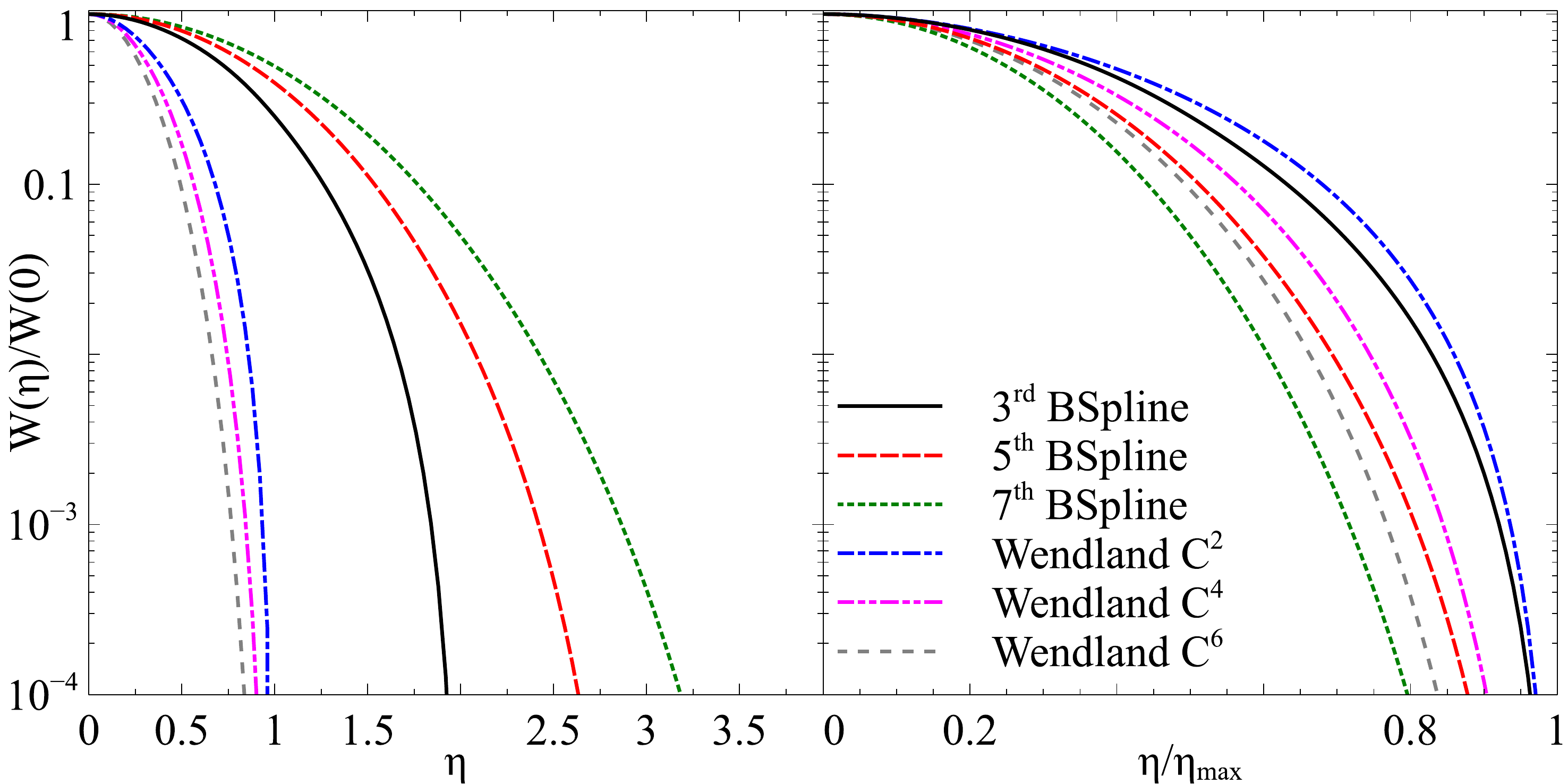}
  \caption{Profiles of the six kernels considered here: \nth{3}, \nth{5}, and \nth{7}-order B-spline, and the C$^2$, C$^4$, and C$^6$ Wendland kernels drawn as a function of $\eta$ (left), and normalized to $\eta/\eta_{\text{max}}$ (right).}
  \label{fig:Wprofiles}
\end{figure}
\begin{table}[h]
  \begin{tabular}{r|cccc}
    Kernel           & $\eta_{\max}$ & $n_h$ & $C_l$ & $C_q$ \\ \hline
    \nth{3}-order B-spline & 2           & 2.0  & 1.0   & 0.5  \\
    \nth{5}-order B-spline & 3           & 1.35 & 1.5   & 0.75 \\
    \nth{7}-order B-spline & 4           & 1.0  & 2.0   & 1.0  \\
    Wendland C$^2$         & 1           & 4.0  & 0.5   & 0.25 \\
    Wendland C$^4$         & 1           & 4.0  & 0.5   & 0.25 \\
    Wendland C$^6$         & 1           & 4.0  & 0.5   & 0.25 \\
  \end{tabular}
  \caption{Kernel extents ($\eta_{\max}$) and the corresponding settings for important numerical parameters in order to fairly compare the resulting calculations.}
  \label{tab:Wparams}
\end{table}
Just as in ordinary SPH the choice of interpolation kernel (i.e, $W(\eta)$) for CRKSPH is a free parameter.  There have been many studies searching for optimal kernel selections for SPH (e.g. \cite{WalterAly2012,Read2010}), and the introduction of a new formulation like CRKSPH opens up the possibility of a different optimization of this choice. Given that the CRKSPH kernel $\Wrij$ (\cref{eq:wR}) applies correction terms ($A_i$ and $B_i^\alpha$), the schema may be less sensitive to the exact form of the underlying $W(\eta)$; it is certainly true, for example, that the explicit enforcement of zeroth-order consistency in $\Wrij$ (encapsulated in the $A_i$ coefficient), renders the volume normalization term ordinarily applied to SPH interpolation kernels irrelevant.

In this section, we briefly compare the impact of varying the interpolation kernel on selected test cases considered in this paper.  For this purpose we consider the \nth{3}, \nth{5}, and \nth{7}-order B-spline kernels (\cref{eq:nbspline}), as well as the C$^2$, C$^4$, and C$^6$ Wendland kernels \cite{Wendland1995,Wendland2005,WalterAly2012}, which are given as (appropriate for 2D or 3D)
\begin{align}
  \label{eq:wendland}
  W_{\text{C2}}(\eta) &= \left(1 - \eta\right)^4_+ (1 + 4\eta) \\
  W_{\text{C4}}(\eta) &= \left(1 - \eta\right)^6_+ (1 + 6\eta + \frac{35}{3} \eta^2) \\
  W_{\text{C6}}(\eta) &= \left(1 - \eta\right)^8_+ (1 + 8\eta + 25\eta^2 + 32\eta^3).
\end{align}
\Cref{fig:Wprofiles} plots the shapes of these kernels, both as a function of $\eta = x/h$ as well as normalized to the same radial extent $\eta/\eta_{\max}$.  The figure immediately highlights one aspect to consider when varying the interpolation kernel: in general, such kernels do not necessarily have the same spatial extent.
This difference in $\eta_{\max}$ implies we need to adjust relevant numerical parameters used in our calculations in order to fairly compare the results.  We choose to maintain the same total number of neighbors per CRKSPH particle, regardless of kernel choice (thereby keeping the same computational expense for each calculation); namely, in these test cases we maintain a total radial number of 4 neighbors.  In our implementation, this is controlled by adjusting the effective number of neighbors per smoothing scale $n_h$, the value of which is summarized in the \nth{3} column of \cref{tab:Wparams} for each kernel.  An additional consideration is the spatial scale of the dissipation of the artificial viscosity.  Inspection of the viscous $\mu_i$ term in \cref{eq:unlimmu,eq:limmu} reveals $\mu \propto \eta^{-1} \propto h$, so $\mu$ scales as $h$.  Therefore, in order to keep roughly the same spatial dissipation in our comparisons we can adjust the viscous coefficients in response to the different $(\eta_{\max}, n_h)$ values as shown in \cref{tab:Wparams}: note the quantities in the row for the \nth{7}-order B-spline correspond to our CRKSPH default values outlined in \cref{sec:crkeqs}, and used throughout our evaluation in \cref{sec:crktests}. 

\begin{figure}[ht]
  \centering
  \includegraphics[width=0.9\textwidth]{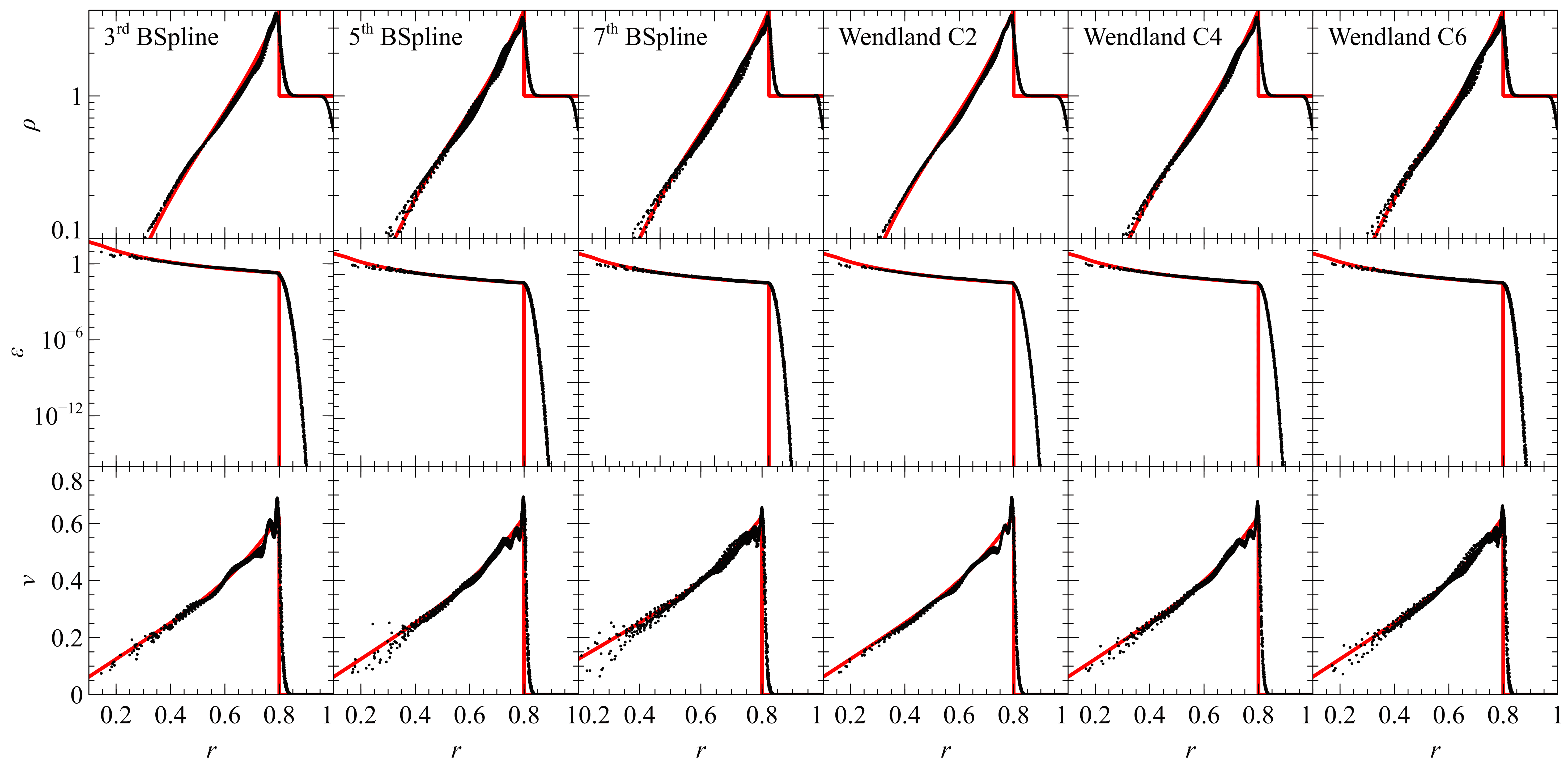}
  \caption{Comparison of the 2D cylindrical Sedov problem (\cref{sec:Sedov}) for each of the six different choices of interpolation kernel in CRKSPH. The simulations are displayed when the shock is predicted to be at $r_s=0.8$ using $100^2$ particles for each realization. The results are quite similar, where we see a trend of slightly more scatter as we go to higher order kernels, and the Wendlend kernels partially more dissipative than the B-splines. }
  \label{fig:Sedov_Wprofiles}
\end{figure}
In order to ascertain how our adjusted parameters handle a strong shock problem, we revisit the 2D cylindrical Sedov test case from \cref{sec:Sedov}.  \Cref{fig:Sedov_Wprofiles} plots the radial profiles of this problem for CRKSPH using each of our six different kernel choices (for comparison see the left panel of \cref{fig:sedov-planar}). As before, we use $100^2$ points in the positive $(x,y)$ quadrant initially placed on a lattice, enforce reflecting boundaries along $x=0$ and $y=0$, and place all the initial energy on the central-most point.  We find the results are largely indifferent to the kernel choice: the shock transition is resolved roughly the same in each calculation, and the fits to the post-shock analytic solutions are about equivalent.  There is some evidence that the higher-order kernels show a bit more scatter in the profiles at the core of the expanding bubble, with the Wendland kernels showing a bit less scatter in this region than the corresponding same order B-splines.  Overall it appears our adjustments to the artificial viscosity coefficients are working reasonably.

\begin{figure}[ht]
  \centering
  \includegraphics[width=0.9\textwidth]{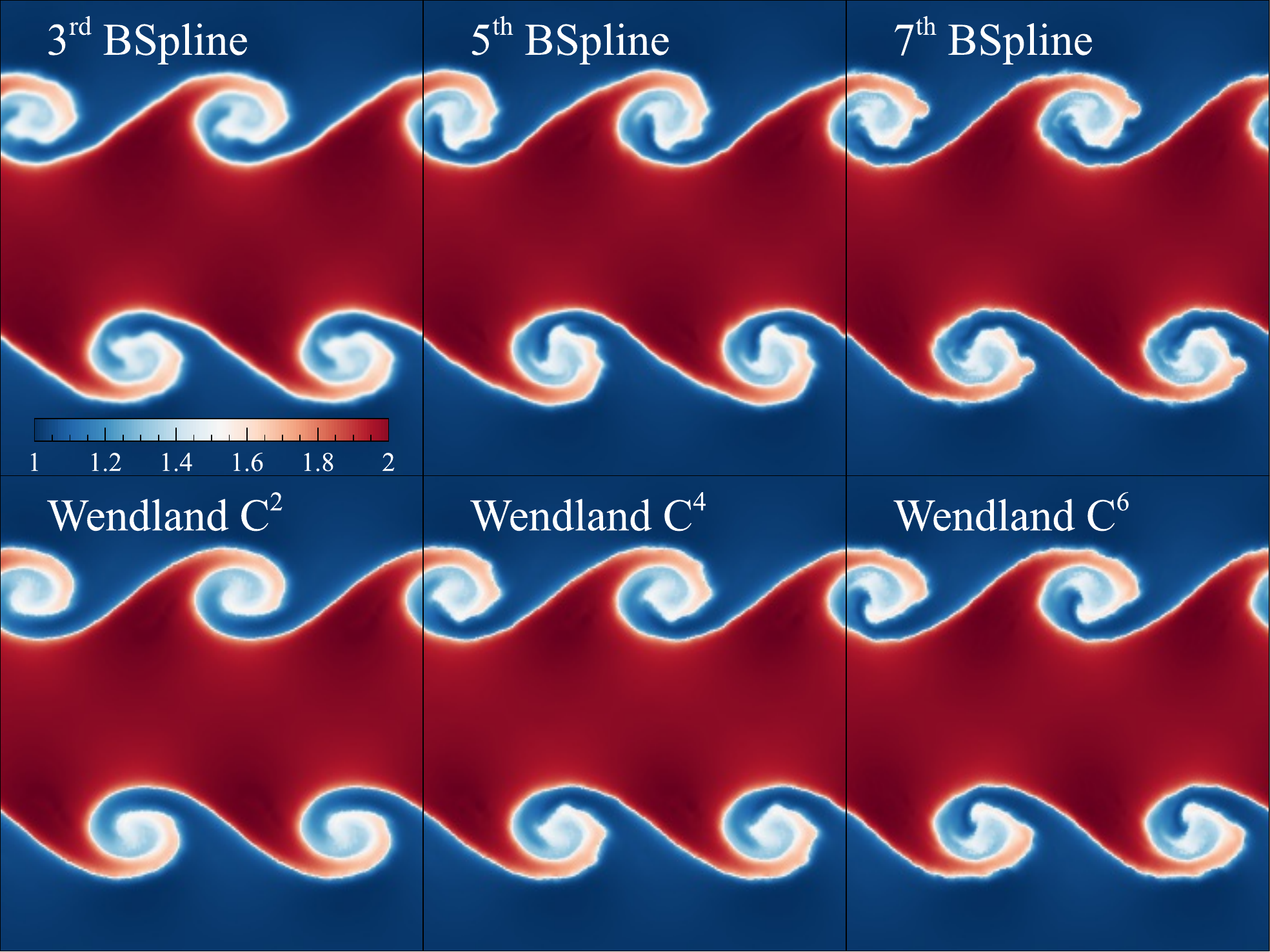}
  \caption{Comparison of the six different interpolation kernels in CRKSPH on the Kelvin-Helmholtz problem (\cref{sec:KelvinHelmholtz}). Presented is the final state of the mass density at $t=2$  using $N=256^2$ particles. The results are qualitatively quite similar, with slightly more substructure in the higher-order kernels. }
  \label{fig:KH_Wimages}
\end{figure}
\begin{figure}
  \centering
  \includegraphics[width=0.45\textwidth]{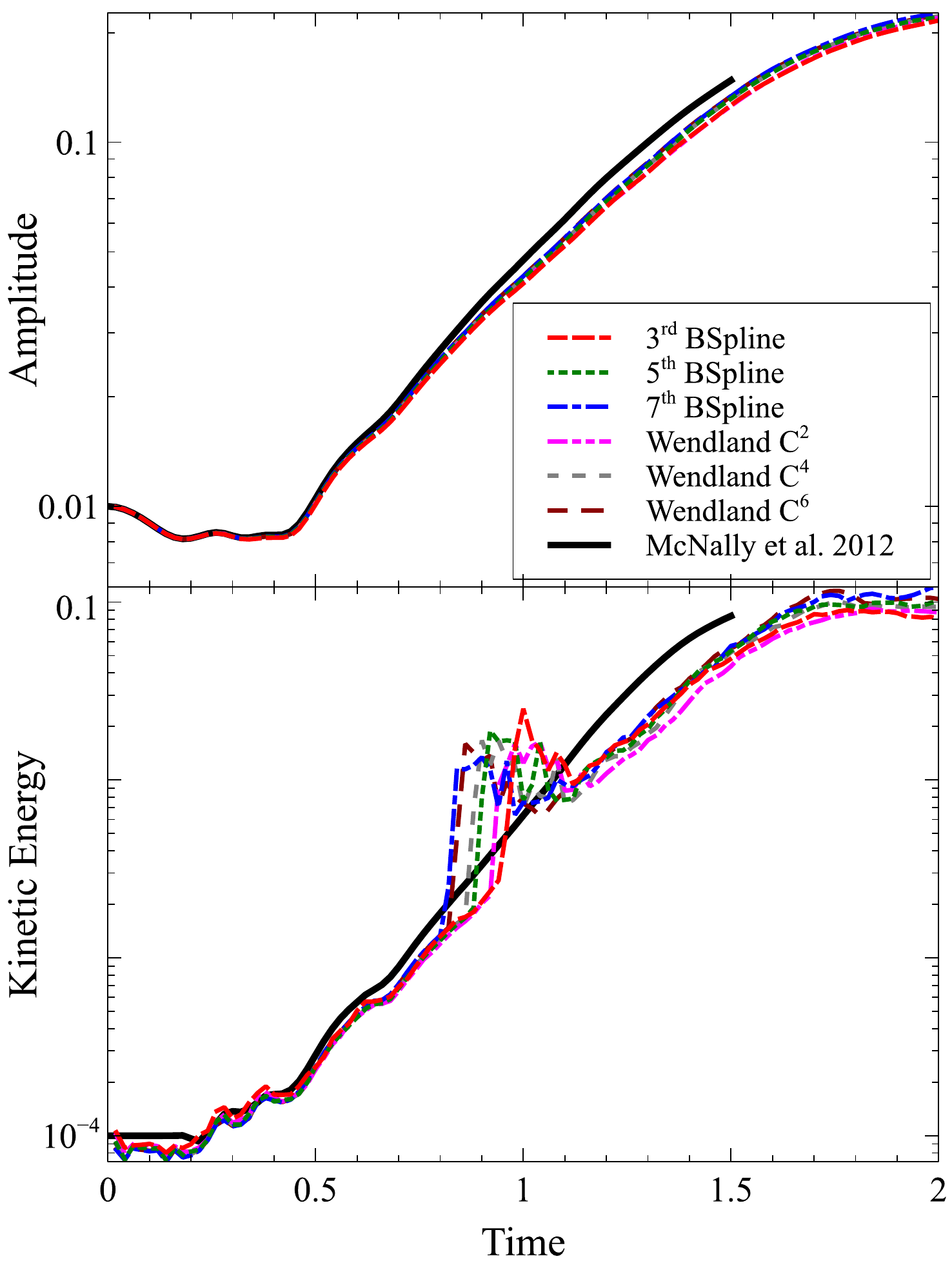}
  \caption{Time evolution of the mixing scale (top) and maximum kinetic energy (bottom) for the CRKSPH models of the Kelvin-Helmholtz problem using different interpolation kernels. The results show a more quantitative similarity between the kernels for the KH test (reinforcing the visual results of \cref{fig:KH_Wimages}). The \nth{7}-order B-spline demonstrates the most mixing, but the variation is small.}
  \label{fig:KH_Wmixing}
\end{figure}
Next we turn our attention to shockless hydrodynamic mixing problems, where we might expect the biggest differences due to their inherent instability.  First we consider the Kelvin-Helmholtz problem described in \cref{sec:KelvinHelmholtz}.  We rerun this problem using our six kernel choices, implementing the same initialization procedure outlined in \cref{sec:KelvinHelmholtz} on $256^2$ points.  \Cref{fig:KH_Wimages} shows the final state of the mass density at $t=2$, well into the regime when we expect the Kelvin-Helmholtz driven roll-up of the fluid interface to be present.  \Cref{fig:KH_Wmixing} plots the time evolution of the scale of mixing (top panel) as well as the maximum $y$-component kinetic energy (bottom panel) as was done in \cref{fig:KHMixing}, both compared with the reference solution of \cite{McNally2012}.  We can see that the extent of the mixing region is nearly identical, regardless of kernel choice, though the \nth{7}-order B-spline shows marginally the most mixing (albeit the variations are tiny).  Similarly the B-spline kernels tend to show more substructure developing within the Kelvin-Helmholtz whirls compared with the equivalent Wendland kernels: both series show a trend for more structure with higher-order kernel.  Such difference are minor though, and in general we find fairly consistent results regardless of the kernel choice.

\begin{figure}[ht]
  \centering
  \includegraphics[width=0.45\textwidth]{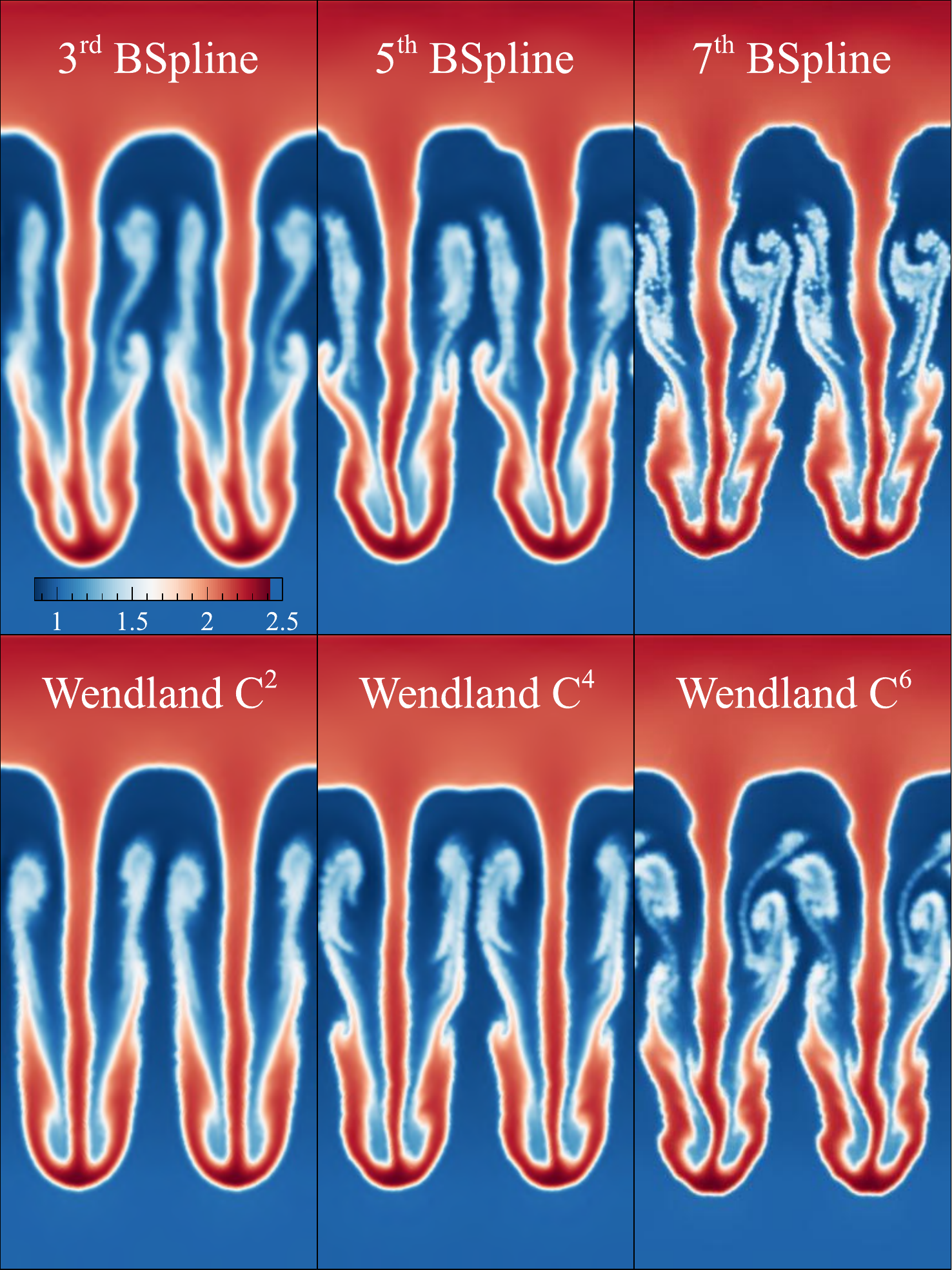}
  \caption{Comparison of the six different interpolation kernels in CRKSPH on the Rayleigh-Taylor problem (\cref{sec:RayleighTaylor}) evolved to $t=4$ using $N=128\times256$ particles. Once again, the different kernels produce very similar evolutions, where we see slightly increasing secondary instabilities from low to high-order kernels. }
  \label{fig:RT_Wimages}
\end{figure}
Finally, we revisit the Rayleigh-Taylor test outlined in \cref{sec:RayleighTaylor}.  \Cref{fig:RT_Wimages} shows the final state of these models at $t=4$, corresponding to our earlier comparisons in \cref{fig:RT}.  As in the Kelvin-Helmholtz results, we find the CRKSPH models are quite consistent (relative to comparisons with other techniques such as compSPH or PSPH from \cref{fig:RT}).  Again we also see evidence that secondary instabilities set in more readily with the higher-order kernels, mostly evident by increasing Kelvin-Helmholtz distortions of the trailing edges of the mushroom plumes as they descend.  There also is further evidence that the Wendland kernels show more dissipation than their corresponding B-spline counterparts.  However, these differences are relatively minor -- by significant metrics, such as the scale of the mixing layer, these calculations are very similar.

In conclusion, we find that although there are interesting minor differences between the results, the CRKSPH models are relatively insensitive to our choice of basis interpolation kernel.  The trends we do note are the increasing presence of secondary structures with increasing order of kernel, and that results using the Wendland kernels may be a bit more dissipative than the B-splines.  Examining the normalized kernel shapes in the right panel of \cref{fig:Wprofiles} suggests that this slight dissipative nature in the Wendland kernels could be due to the fact that those functions fall off less rapidly with $\eta/\eta_{\max}$ than the similar B-splines, and therefore, effectively the B-splines are ``sharper'', i.e., more strongly dominated by local particle contributions to the average values (for equivalent total numbers of neighbors).  The flip-side of this trend is that results using the B-splines also tend to be less stable than those based on the Wendland kernels, showing more rapid onset of secondary instabilities.  In this paper we settled on the \nth{7}-order B-spline as our default by a tiny margin, but as we see here, CRKSPH is not greatly sensitive to this choice.  In future work we may revisit this issue and delve more deeply into the implications of the choice of $W(\eta),$ as well as a wider parameter optimization consideration for each kernel.  This has been an area of study for several papers in SPH, and we have only begun to similarly explore the corresponding choices implied by CRKSPH.

\setcounter{figure}{0} 
\section{Compatible Smoothed Particle Hydrodynamics}
\label{sec:compSPH}
Our compatible SPH discretization is taken from \cite{Owen2014}, with the exception that we use the entropy weighted distribution of the pair-wise work described in \cref{sec:compwork}.  The evolution equations are 
\begin{align}
  \label{eq:SPHrho}
  \rho_i &= \sum_j m_j W_{i} \\
  \label{eq:SPHvel}
  \frac{Dv_i^\alpha}{Dt} &= -\sum_j m_j \left[ \left(\Omega_i^{-1} \frac{P_i}{\rho_i^2} + \half \Pi_i \right) \partial_\alpha W_i + \left(\Omega_j^{-1} \frac{P_j}{\rho_j^2} + \half \Pi_j \right) \partial_\alpha W_j \right] \\
  \label{eq:SPHeng}
  \frac{Du_i}{Dt} &= \sum_j m_j \left(\Omega_i^{-1} \frac{P_i}{\rho_i^2} + \half \Pi_i\right) \vij^\alpha\partial_\alpha W_i \\
  \label{eq:SPHDvDx}
  \partial_\beta v_i^\alpha &= -\left(M_i^{-1}\right)^{\phi \beta} \sum_j m_j \vij^\alpha \partial_\phi \Wi \\
  \label{eq:GradHOmg}
   \Omega_i &= 1-\frac{\partial h_i}{\partial \rho_i}\sum_j m_j \frac{\partial {W}_i}{\partial h_i} = -\frac{1}{\nu \rho_i}\sum_j m_j \eta_i \frac{\partial \Wi}{\partial \eta_i} \\
  \label{eq:SPHM}
  M_i^{\alpha \beta} &= -\sum_j m_j \xij^\alpha \partial_\beta\Wi
\end{align}
where $\rho$ is the mass density, $m$ the mass, $v^\alpha$ the velocity, $P$ the pressure, $u$ the specific thermal energy.  Note in these relations we use subscript $i$ and $j$ to indicate what smoothing scale is used for each term: $\Wi = W(\xij, h_i)$, $\Wj = W(\xij, h_j)$, $\Pi_i$ is the viscosity using $h_i$, $\Pi_j$ using $h_j$, etc. We also use the standard (but sometimes confusing) SPH convention that $ij$ on physical quantities indicates a difference: $\xij = x_i - x_j$, $\vij = v_i - v_j$. The $\Omega_i$ terms represent the so called ``grad-h'' corrections, resulting from a Lagrangian formulation of the SPH equations as described in \cite{Springel:2002wh, Monaghan:2002uy}, given here appropriately for $\nu$ dimensions.

The $\Pi$ term in \cref{eq:SPHvel,eq:SPHeng} is the artificial viscosity, for which we use the Monaghan-Gingold form \cite{Monaghan:1983dn}:
\begin{align}
  \label{eq:MGPi}
  \Pi_i &= \rho_i^{-1} \left(-C_l c_i \mu_i + C_q \mu_i^2 \right) \\
  \mu_i &= \min\left(0, \frac{\vij^\alpha \eta_i^\alpha}{\eta_i^\alpha \eta_i^\alpha + \epsilon^2}\right) \\
  \eta_i^\alpha &= \xij^\alpha/h_i
\end{align}
where $\vij^\alpha \equiv v_i^\alpha - v_j^\alpha$, $\xij^\alpha \equiv x_i^\alpha - x_j^\alpha$, $(C_l, C_q)$ are the viscous linear and quadratic coefficients, $c_i$ is the sound speed, and $\epsilon=0.1$ is a small number to avoid division by zero.  Using our subscript convention, $\Pi_j$ is obtained by using $h_j$ in the above relations. \Cref{eq:SPHDvDx} is the SPH estimate of the velocity gradient.  Following \cite{Randles1996} we apply the correction $M^{\alpha \beta}$ from \cref{eq:SPHM} that makes this gradient exact for linear velocity fields. 

The major distinction of the compatible SPH formalism is that we advance the specific thermal energy with the compatible formalism described in \cref{sec:compwork}; the time evolution equation for $u_i$ (\cref{eq:SPHeng}) is only used to compute intermediate values of $u_i$ during the time advancement cycle.

\setcounter{figure}{0} 
\section{Cullen-Dehnen Modified Viscosity Model}
\label{sec:CDHvisc}
In the tests employing the Cullen-Dehnen viscosity, we use the Hopkins modified form \citep{Cullen2010,Hopkins2015}.  This algorithm evolves the coefficients used in the viscosity $(C_l, C_q)$, replacing them with pair-wise values 
\begin{align}
  {C_l}_{ij} &= \half (\alpha_i + \alpha_j) C_l \\
  {C_q}_{ij} &= \half (\alpha_i + \alpha_j) C_q.
\end{align}

The point-wise time dependent multiplier $\alpha_i$ is evolved according to
\begin{align}
  \label{eq:CDalpha}
  \alpha_i &= \max\left( \alphamin,
    \frac{ |\beta_\xi \xi_i^4 \partial_\phi v_i^\phi |^2 }
         { |\beta_\xi \xi_i^4 \partial_\phi v_i^\phi |^2 + S^{\phi \psi} S^{\psi \phi} } \alphaz_i(t) \right) \\
  \xi_i &= 1 - \rho_i^{-1} \sum_j \sgn(\partial_\phi v_i^\phi) m_j \Wi
\end{align}
\begin{equation}
  \alphatmp_i = \left\{ \begin{array}{l@{\quad}l}
    0 & \partial_t\left(\partial_\phi v_i^\phi\right) \ge 0 \;\text{or}\; \partial_\phi v_i^\phi \ge 0 \\
    \frac{\alphamax |\partial_t\left(\partial_\phi v_i^\phi\right)|}
         {\alphamax |\partial_t\left(\partial_\phi v_i^\phi\right)| + \beta_c c_i^2 (\fkern h_i)^{-2}} & \text{otherwise} \\
  \end{array} \right.
\end{equation}
\begin{equation}
  \alphaz_i(t + \Delta t) = \left\{ \begin{array}{l@{\quad}l}
    \alphatmp_i & \alphatmp_i \ge \alphaz_i(t) \\
    \alphatmp_i + \left(\alphaz_i(t) - \alphatmp_i\right) e^{-\beta_d \Delta t \, \vsig_i /(2 \fkern h_i)} & \text{otherwise} \\
  \end{array}\right.
\end{equation}
where $\partial_\phi v_i^\phi$ is evaluated by \cref{eq:SPHDvDx} (note \cite{Cullen2010} used an alternative form for the linearly corrected velocity gradient than presented here), and $S$ is the shear tensor described in \cite{Cullen2010}.  We adopt the values for the constants from \citep{Hopkins2015}: $\alphamin = 0.02$, $\alphamax = 2$, $\beta_c = 0.7$, $\beta_d = 0.05$, and $\fkern = 1/3$.

\setcounter{figure}{0} 
\section{Pressure-based Smoothed Particle Hydrodynamics}
\label{sec:PSPH}
Our PSPH examples follow the pressure-energy description of \cite{Hopkins2015}, which is, in turn, based on density independent SPH (DISPH) of \cite{Saitoh2013}.  In PSPH, the pressure is defined by a summation relation rather than equation of state lookups using the density and energy, and the weighting per point is a function of the pressure rather than mass density.  Although the mass density, therefore, does not play a direct role in the hydrodynamical equations, it can also be found via summation.  The pressure, mass density, and number density are given as
\begin{align}
  \label{eq:PSPHpbar}
  P_i &= \sum_j (\gamma - 1) m_j u_j \Wi \\
  \label{eq:PSPHrhobar}
  \rho_i &= \sum_j m_j \Wi \\
  \label{eq:PSPHnbar}
  n_i &= \sum_j \Wi
\end{align}

The PSPH hydrodynamical equations are
\begin{align}
  \label{eq:PSPHDvDt}
  \frac{Dv_i^\alpha}{Dt} &= -\sum_j m_j \left[ (\gamma - 1)^2 u_i u_j \left( \frac{f_{ij}}{P_i} \partial_\alpha \Wi + \frac{f_{ji}}{P_j} \partial_\alpha \Wj \right)  + \qacc \right] \\
  \label{eq:PSPHDEDt}
  \frac{DE_i}{Dt} &= m_i v_i^\alpha \frac{Dv_i^\alpha}{Dt} +
    \sum_j m_i m_j \left[ (\gamma - 1)^2 u_i u_j \frac{f_{ij}}{P_i} \vij^\alpha \partial_\alpha \Wi + \vij^\alpha \qacc \right] \\
  \label{eq:PSPHqacc}
  \qacc &= \half (\rho_i \Pi_i + \rho_j \Pi_j) \frac{\partial_\alpha \Wi + \partial_\alpha \Wj}{\rho_i + \rho_j} \\
  \label{eq:PSPHfij}
  f_{ij} &= 1 - \left( \frac{h_i}{\nu (\gamma - 1) n_i m_j u_j} \frac{\partial P_i}{\partial h_i} \right) \left(1 + \frac{h_i}{\nu n_i} \frac{\partial n_i}{\partial h_i} \right)^{-1} \\
  \frac{\partial n_i}{\partial h_i} &= -\sum_j h_i^{-1} \left( \nu \Wi + \eta_i \frac{\partial W}{\partial \eta}(\eta_i) \right) \\
  \frac{\partial P_i}{\partial h_i} &= -\sum_j (\gamma - 1) m_j u_j h_i^{-1} \left( \nu \Wi + \eta_i \frac{\partial W}{\partial \eta}(\eta_i) \right)
\end{align}
where $\eta_i \equiv \xij/h_i$.  For PSPH, we choose to evolve the total rather than specific thermal energy via \cref{eq:PSPHDEDt}, as this seems to be a common practice.  For all of our PSPH comparisons, the artificial viscosity used in \cref{eq:PSPHqacc} is always the Cullen-Dehnen modification (\cref{sec:CDHvisc}) of the Monaghan-Gingold viscosity (\cref{eq:MGPi}).  Our PSPH implementation also uses the linearly corrected velocity gradient (\cref{eq:SPHDvDx,eq:SPHM}) described in the compatible SPH discussion, which yields better behavior in combination with the Cullen-Dehnen viscosity model.

For PSPH examples, we also incorporate the artificial conductivity term described in \cite{Price2008, Hopkins2015, Read2012}, which adds additional diffusion to the energy equation according to
\begin{align}
  \frac{DE_i}{Dt} &= \alpha_C \sum_j m_i m_j \alpha_{ij} \tilde{v}_s (u_i - u_j) \frac{|P_i - P_j|}{P_i + P_j}
    \frac{\partial_\alpha \Wi + \partial_\alpha \Wj}{\rho_i + \rho_j} \\
  \tilde{v}_s &\equiv c_i + c_j - 3 \vij^\alpha \xij^\alpha / |\xij| \\
  \alpha_{ij} &\equiv \half \left( \alpha_i + \alpha_j \right)
\end{align}
when $\tilde{v}_s > 0$. The Cullen-Dehnen coefficients from \cref{eq:CDalpha} are $(\alpha_i, \alpha_j)$, and $\alpha_C = 0.25$ a constant.

Finally, again for consistency with prior published results in our PSPH examples, we use the specialized quintic kernel described in \cite{WalterAly2012} that has been rescaled to terminate at $\eta=1$, namely
\begin{equation}
  \label{eq:Wfive}
  W(\eta) = \left\{ \begin{array}{l@{\quad}l}
    \left(1 - \eta\right)^5 - 6 \left(2/3 - \eta\right)^5 + 15 \left(1/3 - \eta\right)^5 & \eta \in [0, 1/3) \\
    \left(1 - \eta\right)^5 - 6 \left(2/3 - \eta\right)^5                                & \eta \in [1/3, 2/3) \\
    \left(1 - \eta\right)^5                                                              & \eta \in [2/3, 1] \\
    0                                                                                   & \eta > 1.
    \end{array} \right.
\end{equation}
This is functionally the same as the quintic kernel derived from \cref{eq:nbspline}, only with a different spatial extent.

\bibliographystyle{apalike}
\bibliography{bib}
\end{document}